\documentclass[10pt, a4paper, onecolumn]{article} % 10pt font size (11 and 12 also possible), A4 paper (letterpaper for US letter) and two column layout
\usepackage{lscape}
\usepackage{hyperref}
\usepackage{amsfonts,amssymb}
\usepackage{graphicx}
\usepackage{float}
\usepackage{amsmath}
\usepackage{cuted}
\usepackage{geometry}
\usepackage{rotating}
\usepackage{pdflscape}
\usepackage{blindtext}

\newcommand{\ket}[1]{\left|#1\right\rangle}
\newcommand{\bra}[1]{\left\langle#1\right|}

\title{Analytical view on non-invasive measurement of moving charge \\ by various topologies of position-based semiconductor qubit}
\author{ Krzysztof Pomorski$^{1,2,3} $ \\ \\ %\textsuperscript{1,2,3}
		\newline\newline % Space before institutions
    1: University College Dublin, School of Computer Science, Ireland %}
    \\ % Institution 1
     2: Department of Microelectronics and Computer Science, Lodz University of Technology, Poland %}
     \\ % Institution 2
    3: Quantum Hardware Systems:\texttt{www.quantumhardwaresystems.com} %} % Institution 3
}
\begin{document}
\maketitle

\begin{abstract}
Detection of moving charge in free space is presented in the framework of single electron CMOS devices.
It opens the perspective for construction of new type detectors for beam diagnostic in accelerators.
%and moving charge entanglement is presented in the framework of single electron CMOS devices.
General phenomenological model of noise acting on position based qubit implemented in semiconductor quantum dots is given in the framework of simplistic tight-binding model.
\end{abstract}

\section{Motivation for weak measurement of moving charged particles}
In nature matter has the attribute of having electric charge.Interaction between charged particles is the foundation base for atoms and molecules.
Currently various experiments are conducted with charged particles as present in CERN and DESY accelerators. The controlled movement of charged particles as protons, electrons, light and heavy ions or other elementary particles takes place under static and time-dependent electric field and magnetic field generated in well organized pattern that is consequence of Maxwell equations. In particular one uses the magnetic focussing to keep the accelerator beam confined to certain finite geometrical space. Moving charges generate electric and magnetic field what is reflected in time-dependent electric field and time-dependent vector potential field. Such time-dependent electric and magnetic field can be sensed by various types of detectors. If movement of the charged particle is traced only by time-dependent fields that are generated by observed particle one can deal with weak non-invasive measurement. Moving particle will always bring some action on the detector. On another hand the detector will respond to the action of external time-dependent magnetic and electric field.
This detector response will generate counter electric and magnetic field that will try to compensate the effect of field trying to change the state of detector. Therefore one has mutual interaction between moving charged particles and detector.
However if the speed of moving particles is very high this interaction will take very short time and will only slightly deflect the trajectory of moving charged particle that is under observation. Therefore one deals with weak measurement that is changing the physical state of object under observation in
perturbative way [1]. Now we will concentrate on the description of the detector of moving charged particles. One can choice various types of detectors as measuring device for example: superconducting SQUIDs, Josephson junctions, NV-diamond sensors or single electron semiconductor devices.
Because of rapid developments in cryogenic CMOS technology and scalability of those detectors we will concentrate on single electron semiconductor devices as most promising detectors for massive use described by [1], [3], [4] and [8].
 % and is placed in the accelerating electric field with magnetic lenses.
%We have $10^{10}$ protons in moving beam.

\section{Single electron devices as detectors of moving charged particles}
\begin{figure}
    \centering
    \includegraphics[scale=0.4]{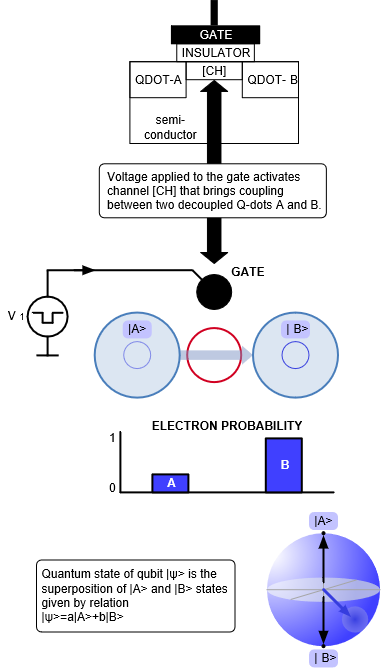}
    \includegraphics[scale=0.5]{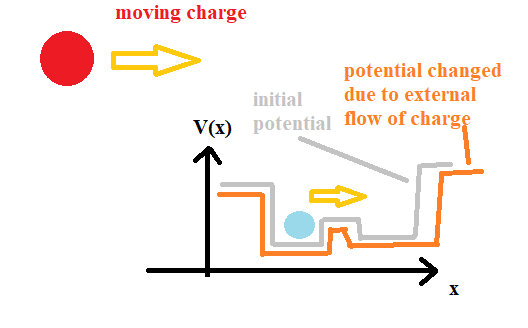}
   \includegraphics[scale=1.3]{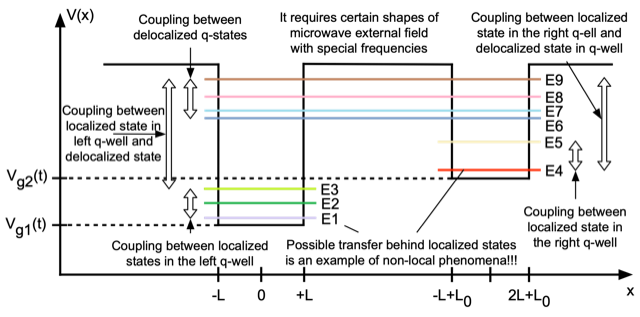}
   \includegraphics[scale=0.2]{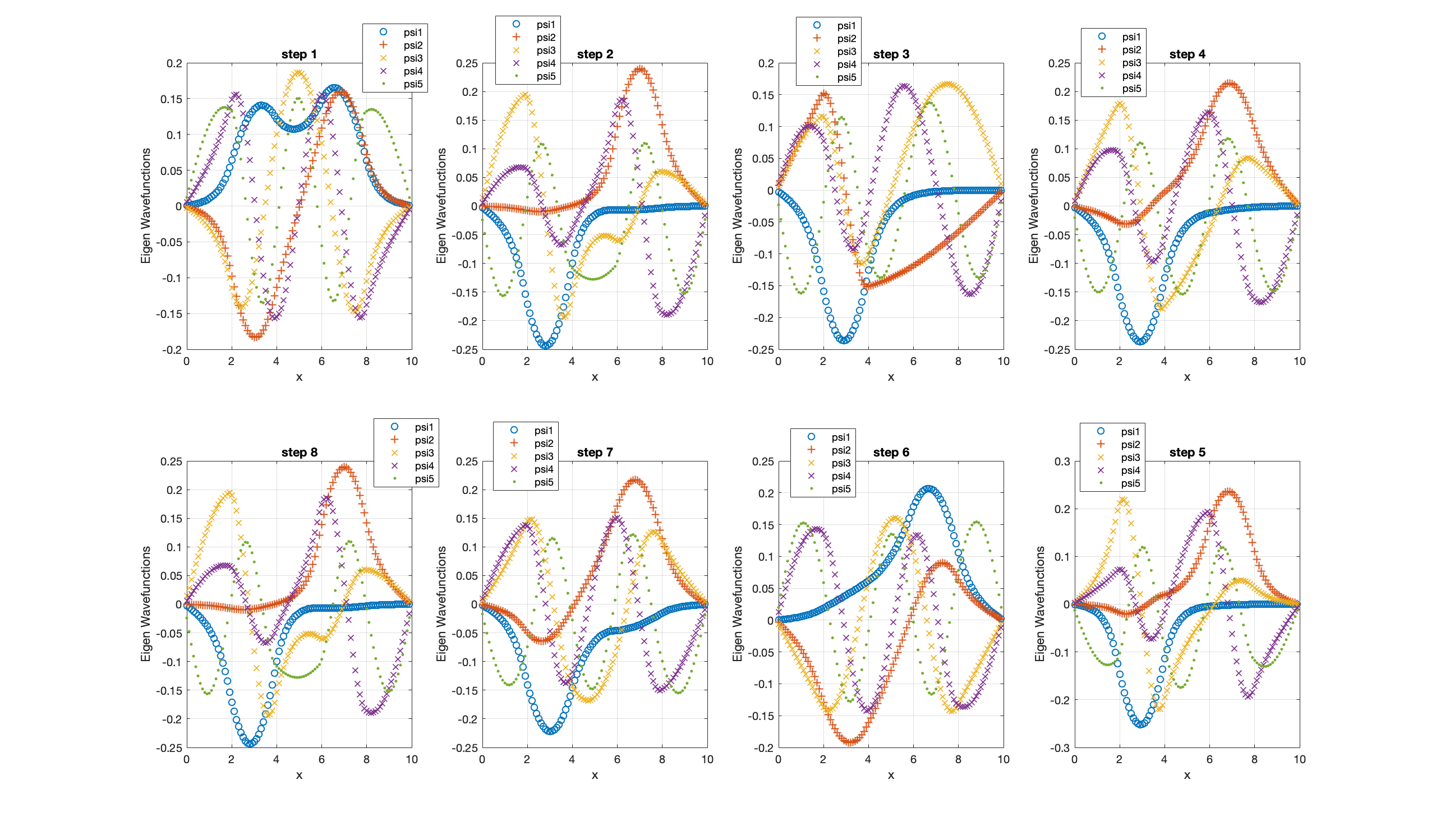}
    \caption{(Top left): Scheme of position based qubit as given by [6] and act of weak measurement by external charged probe [3]; (Top right): Act of passage of charged particle in the proximity of position based qubit and renormalization of qubit confining potential due to the external perturbation; (Middle): Scheme of various energy levels present in qubit [4]; (Bottom): Different qubit eiegenergy levels for different confining potential cases. It is worth mentioning that passing electric charge can induce quantum system transitions between many energetic levels.}
  \label{CMOSSET}
\end{figure}

%We can construct similar of electrons in single electron devices sensing the movement of proton beam.
Quite recently it has been proposed by Fujisawa and Petta to use single electron devices for classical and quantum information processing. This technology relies on the chain of coupled quantum dots that can be implemented in various semiconductors. In particular one can use CMOS transistor with source and drain with channel in-between that is controlled by external polarizing gate as it is depicted in Fig.\ref{CMOSSET}.  Recent CMOS technologies allow for fabrication of transistor with channel length from 22nm to 3nm. If one can place one electron in source-channel-drain system (S-CH-D) than one can approximate the physical system by two coupled quantum dots. It is convenient to use tight-binding formalism to describe electron dynamics with time. In such case instead of wavefunction of electron it is usefull to use maximum localized wavefunctions (Wannier functions) of that electron on the left and right quantum dot that are denoted by $\ket{1,0}$ and $\ket{0,1}$.
One obtains the following simplistic Hamiltonian of position-based qubit given as
\begin{equation}
H=E_p(1)\ket{1,0}\bra{1,0}+E_p(2)\ket{0,1}\bra{0,1}+|t|_{1 \rightarrow 2}\ket{1,0}\bra{0,1}+|t|_{2 \rightarrow 1}\ket{0,1}\bra{1,0}.
\end{equation}

Here $E_p(1)$ or $E_p(2)$  has the meaning of minima of confining potential on the left or right quantum dot.
It can be recognized as localized energy on the left or right quantum dot. The tunneling process between left and right quantum dot or classical movement electron between left and right quantum dot can be accounted by the term $|t|_{1 \rightarrow 2}$ that has the meaning of delocalized energy (energy participating in particle transfer between quantum dots). If electron kinetic energy is much beyond the potential barrier separating left and right quantum dot that one can assign the meaning of kinetic energy to the term $|t|_{1 \rightarrow 2}$ or $|t|_{2 \rightarrow 1}$.
The quantum state of position based qubit is given as superposition of presence on the left and right quantum dot and is expressed by the formula
\begin{eqnarray}
\ket{\psi}=\alpha_t \ket{1,0} + \beta_t \ket{0,1},
\end{eqnarray}
where $\ket{1,0}=w_L(x)$, $\ket{0,1}=w_R(x)$ are maximum localized functions on the left and right side of position based qubit.
In case of position dependent qubit we have
\begin{equation}
    \label{simple_equation}
    i \hbar \frac{d}{dt}\ket{\psi} =
    \begin{pmatrix}
    E_{p1} & t_{s12} \\
    t_{s12}^{*} & E_{p2} \\
    \end{pmatrix}\ket{\psi}=
    E(t) \ket{\psi}.
\end{equation}
For simplicity we consider $E_{p1}=E_{p2}=E_{p}, t_{s12}=t_s$.
We have two eigenergies $E_1=E_p-t_s$ and $E_2=E_p+t_s$ and eigenstates are as follows
\begin{eqnarray}
\ket{E_1}=\frac{1}{\sqrt{2}}(\ket{1,0}-\ket{0,1})=\frac{1}{\sqrt{2}}
    \begin{pmatrix}
     +1 \\
      -1 \\
    \end{pmatrix}, \nonumber \\
\ket{E_2}=\frac{1}{\sqrt{2}}(\ket{1,0}+\ket{0,1})=\frac{1}{\sqrt{2}}
    \begin{pmatrix}
      +1 \\
      +1 \\
    \end{pmatrix}.
\end{eqnarray}
In general case we have superposition of energy levels $E_1$ and $E_2$ as $\ket{\psi}=c_{E1}e^{i \phi_{E1}}\ket{E1}_t+c_{E2}e^{i \phi_{E2}}\ket{E2}_t$ and in details we have
\begin{eqnarray}
\ket{\psi}=e^{i \phi_{E1}}
\begin{pmatrix}
+c_{E1}e^{\frac{E_1}{i\hbar}t} \\
 -c_{E1}e^{\frac{E_1}{i\hbar}t} \\
\end{pmatrix} + e^{i \phi_{E2}}
\begin{pmatrix}
+c_{E2}e^{\frac{E_2}{i\hbar}t} \\
+c_{E2}e^{\frac{E_2}{i\hbar}t} \\
\end{pmatrix}=
\begin{pmatrix}
+e^{i \phi_{E1}}c_{E1}e^{\frac{E_1}{i\hbar}t}+e^{i \phi_{E2}}c_{E2}e^{\frac{E_2}{i\hbar}t} \\
- e^{i \phi_{E1}}c_{E1}e^{\frac{E_1}{i\hbar}t}+e^{i \phi_{E2}}c_{E2}e^{\frac{E_2}{i\hbar}t} \\
\end{pmatrix}=
\begin{pmatrix}
\alpha(t) \\
\beta(t) \\
\end{pmatrix}
\end{eqnarray}
where $|c_{E1}|^2+|c_{E2}|^2=1$ (sum of occupancy probability of energy level 1 and 2) and $|\alpha(t)|^2+|\beta(t)|^2=1$ (sum of probability of occupancy of left and right side by electron).
Under influence of very quickly moving charge we have
\begin{equation}
    \label{simple_equation1}
    i \hbar \frac{d}{dt}\ket{\psi} =
    \begin{pmatrix}
    E_{p1}+f_1(t) & t_{s12}+f_3(t) \\
    t_{s12}^{*}+f_3(t)^{*} & E_{p2}+f_2(t) \\
    \end{pmatrix}\ket{\psi}=
    E(t) \ket{\psi}.
\end{equation}
More exactly we have
\begin{equation}
    \label{simple_equation2}
    i \hbar \frac{d}{dt}
    \begin{pmatrix}
    \alpha(t) \\
    \beta(t)
    \end{pmatrix}
    =
    \begin{pmatrix}
    E_{peff1}(t) & t_{eff-s12}(t) \\
    t_{eff-s12}(t)^{*} & E_{peff2}(t)  \\
    \end{pmatrix}
    \begin{pmatrix}
    \alpha(t) \\
    \beta(t)
    \end{pmatrix}
    =
    \begin{pmatrix}
    E_{p1}+f_1(t) & t_{s12}+f_3(t) \\
    t_{s12}^{*}+f_3(t)^{*} & E_{p2}+f_2(t) \\
    \end{pmatrix}
    \begin{pmatrix}
    \alpha(t) \\
    \beta(t)
    \end{pmatrix}
    =
    E(t) \begin{pmatrix}
    \alpha(t) \\
    \beta(t)
    \end{pmatrix}
\end{equation}
Single proton movement in proximity of position based qubit
\begin{eqnarray}
f_1(t)= \sum_{k=1}^{Nprotons}V_1(k) \delta(t-t_1(k)),
f_2(t)= \sum_{k=1}^{Nprotons}V_2(k) \delta(t-t_1(k)),
f_3(t)= \sum_{k=1}^{Nprotons}(V_3(k)+iV_4(k))\delta(t-t_1(k)).
\end{eqnarray}
In general case one shall have effective values of $E_{peff1}(t)$, $E_{peff2}(t)$, $t_{eff-s12}(t)$ and $t_{eff-s21}(t)$ given by formulas
\begin{eqnarray}
E_{peff1}(t)=\int_{-\infty}^{+\infty}dx w_L^{*}(x)(-\frac{\hbar^2}{2m_e}\frac{d^2}{dx^2}+V_{pol}(x)+V_p(t))w_L(x), \nonumber \\
E_{peff1}(t)=\int_{-\infty}^{+\infty}dx w_R^{*}(x)(-\frac{\hbar^2}{2m_e}\frac{d^2}{dx^2}+V_{pol}(x)+V_p(t))w_R(x),  \nonumber \\ t_{eff-s12}(t)=\int_{-\infty}^{+\infty}dx w_R^{*}(x)(-\frac{\hbar^2}{2m_e}\frac{d^2}{dx^2}+V_{pol}(x)+V_p(t))w_L(x),  \nonumber \\ t_{eff-s21}(t)=\int_{-\infty}^{+\infty}dx w_L^{*}(x)(-\frac{\hbar^2}{2m_e}\frac{d^2}{dx^2}+V_{pol}(x)+V_p(t))w_R(x),
\end{eqnarray}
where $w_L(x)$ and $w_R(x)$ are maximum localized states (Wannier functions) in the left and right quantum dots and where $V_{pol}(x)$ is the qubit polarizing electrostatic potential with $V_p(t)$ as electrostatic potential coming from proton moving in the accelerator beam.
For simplicity let us consider 3 terms perturbing single electron qubit Hamiltonian
\begin{eqnarray}
f_1(t)= V_1 \delta(t-t_1), f_2(t)= V_2\delta(t-t_2), %%\nonumber \\
%%%$f_2(t)= V_2\delta(t-t_2)$,
f_3(t)=(V_3+iV_4)\delta(t-t_2)
\end{eqnarray}
and we obtain the modified Hamiltonian of qubit as
\begin{equation}
    \label{simple_equation2}
    i \hbar \frac{d}{dt}
    \begin{pmatrix}
    \alpha(t) \\
    \beta(t)
    \end{pmatrix}
    =
    \begin{pmatrix}
    E_{p1}+V_1 \delta(t-t_1) & t_{s12}+(V_3+iV_4) \delta(t-t_1) \\
    t_{s12}^{*}+ (V_3- i V_4) \delta(t-t_1) & E_{p2}+V_2 \delta(t-t_1) \\
    \end{pmatrix}
    \begin{pmatrix}
    \alpha(t) \\
    \beta(t)
    \end{pmatrix}
    =
    E(t) \begin{pmatrix}
    \alpha(t) \\
    \beta(t)
    \end{pmatrix}
\end{equation}
and is the system of 2 coupled differential equations
\begin{eqnarray}
    \label{simple_equation4}
    i \hbar \frac{d}{dt}\alpha(t)=(E_{p1}+V_1 \delta(t-t_1))\alpha(t) + (t_{s12}+(V_3+iV_4)\delta(t-t_1))\beta(t), \nonumber \\
    i \hbar \frac{d}{dt}\beta(t)=(E_{p2}+V_2 \delta(t-t_1))\beta(t) + (t_{s12}^{*}+(V_3-iV_4)\delta(t-t_1))\alpha(t),
\end{eqnarray}
that can be rewritten in discrete form as %and we can have finite state representation
\begin{eqnarray}
    \label{simple_equation5}
    i \hbar \frac{1}{2\delta t}(\alpha(t+\delta t)-\alpha(t-\delta t))=(E_{p1}+V_1 \delta(t-t_1))\alpha(t) + (t_{s12}+(V_3+iV_4)\delta(t-t_1))\beta(t), \nonumber \\
    i \hbar \frac{1}{2\delta t}(\beta(t+\delta t)-\beta(t-\delta t))=(E_{p2}+V_2 \delta(t-t_1))\beta(t) + (t_{s12}^{*}+(V_3-iV_4)\delta(t-t_1))\alpha(t),
\end{eqnarray}

Applying operator $\int_{t_1-\delta t}^{t_1-\delta t}dt'$ to both sides of previous equations with very small $\delta t \rightarrow 0$ we obtain
two algebraic relations as
\begin{eqnarray}
    \label{simple_equation6}
    i \hbar (\alpha(t_1^{+})-\alpha(t_1^{-}))=V_1 \alpha(t_1^{+}) + (V_3+iV_4)\beta(t_1^{+}), \nonumber \\
    i \hbar (\beta (t_1^{+})- \beta(t_1^{-}))=V_2 \beta(t_1^{+})  + (V_3-iV_4)\alpha(t_1^{+}).
\end{eqnarray}
Linear combination of quantum states of qubit before the measurement is expressed by quantum states of qubit after weak measurement that was due to the interaction of qubit with external passing charged particle so we obtain
\begin{eqnarray}
    \label{simple_equation7}
    (i \hbar- V_1)\alpha(t_1^{+})-(V_3+iV_4)\beta(t_1^{+})  = i \hbar \alpha(t_1^{-}) , \nonumber \\
    (i \hbar- V_2)\beta (t_1^{+})- (V_3-iV_4)\alpha(t_1^{+})=i \hbar \beta(t_1^{-}).
\end{eqnarray}
Last equations can be written in the compact form as
\begin{eqnarray}
    \label{simple_equation8}
\begin{pmatrix}
(i \hbar- V_1) & -(V_3+iV_4) \\
- (V_3-iV_4) & (i \hbar- V_2) \\
\end{pmatrix}
\begin{pmatrix}
\alpha(t_1^{+}) \\
\beta(t_1^{+}) \\
\end{pmatrix}=i \hbar
\begin{pmatrix}
\alpha(t_1^{-}) \\
\beta(t_1^{-}) \\
\end{pmatrix}
\end{eqnarray}
or equivalently
\begin{eqnarray}
    \label{simple_equation9}
\begin{pmatrix}
\alpha(t_1^{+}) \\
\beta(t_1^{+}) \\
\end{pmatrix}=i \hbar
\begin{pmatrix}
(i \hbar- V_1) & -(V_3+iV_4) \\
- (V_3-iV_4) & (i \hbar- V_2) \\
\end{pmatrix}^{-1}
\begin{pmatrix}
\alpha(t_1^{-}) \\
\beta(t_1^{-}) \\
\end{pmatrix}
\end{eqnarray}
and it implies that quantum state after weak measurement is obtained as the linear transformation of the quantum state before the measurement so
\begin{eqnarray}
    \label{simple_equation10}
\begin{pmatrix}
\alpha(t_1^{+}) \\
\beta(t_1^{+}) \\
\end{pmatrix}= \frac{\hbar}{(\hbar + i V_1) (\hbar + i V_2) + V_3^2 + V_4^2}
\begin{pmatrix}
 (\hbar + i V_2) & (-i V_3+V_4) \\
(-iV_3 - V_4) & (\hbar + i V_1) \\
\end{pmatrix}
\begin{pmatrix}
\alpha(t_1^{-}) \\
\beta(t_1^{-}) \\
\end{pmatrix}
\end{eqnarray}
and hence
\begin{eqnarray}
    \label{simple_equation11}
    \begin{pmatrix}
\alpha(t_1^{+}) \\
\beta(t_1^{+}) \\
\end{pmatrix} %%%= \nonumber \\
    =\hat{M}
    \begin{pmatrix}
\alpha(t_1^{-}) \\
\beta(t_1^{-}) \\
\end{pmatrix}
\begin{pmatrix}
\alpha(t_1^{+}) \\
\beta(t_1^{+}) \\
\end{pmatrix} %%%= \nonumber \\
=\begin{pmatrix}
M_{1,1} & M_{1,2} \\
M_{2,1} & M_{2,2}
\end{pmatrix}
\begin{pmatrix}
\alpha(t_1^{-}) \\
\beta(t_1^{-}) \\
\end{pmatrix} = \nonumber \\
\frac{1}{(\hbar^4 + (-V_1 V_2 + V_3^2 +  V_4^2)^2 + \hbar^2 (V_1^2 + V_2^2 + 2 (V_3^2 + V_4^2)))}  %%%%\times \nonumber \\
\times \begin{pmatrix}
M_{r(1,1)}+iM_{i(1,1)}  & M_{r(1,2)}+ i M_{i(1,2)} \\
M_{r(2,1)}+ i M_{i(2,1)} & M_{r(2,2)} +iM_{i(2,2)}
\end{pmatrix}.
\end{eqnarray}

%%= \frac{1}{(\hbar^4 + (-V_1 V_2 + V_3^2 +
%%   V_4^2)^2 + \hbar^2 (V_1^2 + V_2^2 + 2 (V_3^2 + V_4^2)))} \times \nonumber \\
%%\begin{pmatrix}
%%(\hbar^2 (\hbar^2 + V_2^2 + V_3^2 + V_4^2)) + i (\hbar (-\hbar^2 V_1 + V_2 (-V_1 V_2 + V_3^2 + V_4^2)) ) & M_{1,2} \\
%%M_{2,1} & M_{2,2}
%%\end{pmatrix}
%%\begin{pmatrix}
%%\alpha(t_1^{-}) \\
%%\beta(t_1^{-}) \\
%%\end{pmatrix}
%%\end{eqnarray}
Now we identify diagonal parts of matrix $\hat{M}$ as $M_{1,1}$ %xxxxxxxxxxxxxxxxxxxxxxxxxxxxxxxxxxxxxxxxxxxxxxxxxxxxxxxxxxxxxxxxxxxxxxxxxxxxxxxxxxxxxxxxxxxxaaaaaaaaaaaaaaaaaaaaaaaaaaaaaaaaaaaaaaaaaaaaaaaaaaaaaaaaaaaaaaaaaaaaaaaaaaaaaaaaaaaaaaaaaaaaaaaaaaaaaaaaaaaaaaaaaaaaaaaaaaaaaaaaaaaaaaaaaaaaaaaaaaaaaaaaaaaaaaaaaaaaaaaaaaaaaaaaaaaaaaaaaaaaaaaaaaaaaaaaaaaaaaaaaaaaaaaaaaaaaaaaaaaaaaaaaaaaaaaaaaaaaaaaaaaaaaaaaaaaaaaaaaaaaaaaaaaaaaaaaaaaaaaaaaaaaaaaaaaaaaaaaaaaaaaaaaaaaaaaaaaaaaaaaaaaaaaaaaaaaaaaaaaaaaaaaaaaaaaaaaaaaaaaaaaaaaaaaaaaaaaaaaaaaaaaaaaaaaaaaaaaaaaaaaaaaaaaaaaaaaaaaaaaaaaaaaaaaaaaaaaaaaaaaaaaaaaaaaaaaaaaaaaaaaaaaaaaaaaaaaaaaaaaaaaaaaaaaaaaaaaaaaaaaaaaaaaaaaaaaaaaaaaaaaaaaaaaaaaaaaaaaaaaaaaaaaaaaaaaaaaaaaaaaaaaaaaaaaaaaaaaaaaaaaaaaaaaaaaaaaaaaaaaaaaaaaaaaaaaaaaaaaaaaaaaaaaaaaaaaaaaaaaaaaaaaaaaaaaaaaaaaaaaaaaaaaaaaaaaaaaaaaaaaaaaaaaaaaaaaaaaaaaaaaaaaaaaaaaaaaaaaaaaaaaaaaaaaaaaaaaaaaaaaaaaaaaaaaaaaaaaaaaaaaaaaaaaaaaaaaaaaaaaaaaaaaaaaaaaaaaaaaaaaaaaaaaaaaaaaaaaaaaaaaaaaaaaaaaaaaaaaaaaaaaaaaaaaaaaaaaaaaaaaaaaaaaaaaaaaaaaaaaaaaaaaaaaaaaaaaaaaaaaaaaaaaaaaaaaaaaaaaaaaaaaaaaaaaaaaaaaaaaaaaaaaaaaaaaaaaaaaaaaaaaaaaaaaaaaaaaaaaaaaaaaaaaaaaaaaaaaaaaaaaaaaaaaaaaaaaaaaaaaaaaaaaaaaaaaaaaaaaaaaaaaaaaaaaaaaaaaaaaaaaaaaaaaaaaaaaaaaaaaaaaaaaaaaaaaaaaaaaaaaaaaaaaaaaaaaaaaaaaaaaaaaaaaaaaaaaaaaaaaaaaaaaaaaaaaaaaaaaaaaaaaaaaaaaaaaaaaaaaaaaaaaaaaaaaaaaaaaaaaaaaaaaaaaaaaaaaaaaaaaaaaaaaaaaaaaaaaaaaaaaaaaaaaaaaaaaaaaaaaaaaaaaaaaaaaaaaaaaaaaaaaaaaaaaaaaaaaaaaaaaaaaaaaaaaaaaaaaaaaaaaaaaaaaaaaaaaaaaaaaaaaaaaaaaaaaaaaaaaaaaaaaaaaaaaaaaaaaaaaaaaaaaaaaaaaaaaaaaaaaaaaaaaaaaaaaaaaaaaaaaaaaaaaaaaaaaaaaaaaaaaaaaaaaaaaaaaaaaaaaaaaaaaaaaaaaaaaaaaaaaaaaaaaaaaaaaaaaaaaaaaaaaaaaaaaaaaaaaaaaaaaaaaaaaaaaaaaaaaaaaaaaaaaaaaaaaaaaaaaaaaaaaaaaaaaaaaaaaaaaaaaaaaaaaaaaaaaaaaaaaaaaaaaaaaaaaaaaaaaaaaaaaaaaaaaaaaaaaaaaaaaaaaaaaaaaaaaaaaaaaaaaaaaaaaaaaaaaaaaaaaaaaaaaaaaaaaaaaaaaaaaaaaaaaaaaaaaaaaaaaaaaaaaaaaaaaaaaaaaaaaaaaaaaaaaaaaaaaaaaaaaaaaaaaaaaaaaaaaaaaaaaaaaaaaaaaaaaaaaaaaaaaaaaaaaaaaaaaaaaaaaaaaaaaaaaaaaaaaaaaaaaaaaaaaaaaaaaaaaaaaaaaaaaaaaaaaaaaaaaaaaaaaaaaaaaaaaaaaaaaaaaaaaaaaaaaaaaaaaaaaaaaaaaaaaaaaaaaaaaaaaaaaaaaaaaaaaaaaaaaaaaaaaaaaaaaaaaaaaaaaaaaaaaaaaaaaaaaaaaaaaaaaaaaaaaaaaaaaaaaaaaaaaaaaaaaaaaaaaaaaaaaaaaaaaaaaaaaaaaaaaaaaaaaaaaaaaaaaaaaaaaaaaaaaaaaaaaaaaaaaaaaaaaaaaaaaaaaaaaaaaaaaaaaaaaaaaaaaaaaaaaaaaaaaaaaaaaaaaaaaaaaaaaaaaaaaaaaaaaaaaaaaaaaaaaaaaaaaaaaaaaaaaaaaaaaaaaaaaaaaaaaaaaaaaaaaaaaaaaaaaaaaaaaaaaaaaaaaaaaaaaaaaaaaaaaaaaaaaaaaaaaaaaaaaaaaaaaaaaaaaaaaaaaaaaaaaaaaaaaaaaaaaaaaaaaaaaaaaaaaaaaaaaaaaaaaaaaaaaaaaaaaaaaaaaaaaaaaaaaaaaaaaaaaaaaaaaaaaaaaaaaaaaaaaaaaaaaaaaaaaaaaaaaaaaaaaaaaaaaaaaaaaaaaaaaaaaaaaaaaaaaaaaaaaaaaaaaaaaaaaaaaaaaaaaaaaaaaaaaaaaaaaaaaaaaaaaaaaaaaaaaaaaaaaaaaaaaaaaaaaaaaaaaaaaaaaaaaaaaaaaaaaaaaaaaaaaaaaaaaaaaaaaaaaaaaaaaaaaaaaaaaaaaaaaaaaaaaaaaaaaaaaaaaaaaaaaaaaaaaaaaaaaaaaaaaaaaaaaaaaaaaaaaaaaaaaaaaaaaaaaaaaaaaaaaaaaaaaaaaaaaaaaaaaaaaaaaaaaaaaaaaaaaaaaaaaaaaaaaaaaaaaaaaaaaaaaaaaaaaaaaaaaaaaaaaaaaaaaaaaaaaaaaaaaaaaaaaaaaaaaaaaaaaaaaaaaaaaaaaaaaaaaaaaaaaaaaaaaaaaaaaaaaaaaaaaaaaaaaaaaaaaaaaaaaaaaaaaaaaaaaaaaaaaaaaaaaaaaaaaaaaaaaaaaaaaaaaaaaaaaaaaaaaaaaaaaaaaaaaaaaaaaaaaaaaaaaaaaaaaaaaaaaaaaaaaaaaaaaaaaaaaaaaaaaaaaaaaaaaaaaaaaaaaaaaaaaaaaaaaaaaaaaaaaaaaaaaaaaaaaaaaaaaaaaaaaaaaaaaaaaaaaaaaaaaaaaaaaaaaaaaaaaaaaaaaaaaaaaaaaaaaaaaaaaaaaaaaaaaaaaaaaaaaaaaaaaaaaaaaaaaaaaaaaaaaaaaaaaaaaaaaaaaaaaaaaaaaaaaaaaaaaaaaaaaaaaaaaaaaaaaaaaaaaaaaaaaaaaaaaaaaaaaaaaaaaaaaaaaaaaaaaaaaaaaaaaaaaaaaaaaaaaaaaaaaaaaaaaaaaaaaaaaaaaaaaaaaaaaaaaaaaaaaaaaaaaaaaaaaaaaaaaaaaaaaaaaaaaaaaaaaaaaaaaaaaaaaaaaaaaaaaaaaaaaaaaaaaaaaaaaaaaaaaaaaaaaaaaaaaaaaaaaaaaaaaaaaaaaaaaaaaaaaaaaaaaaaaaaaaaaaaaaaaaaaaaaaaaaaaaaaaaaaaaaaaaaaaaaaaaaaaaaaaaaaaaaaaaaaaaaaaaaaaaaaaaaaaaaaaaaaaaaaaaaaaaaaaaaaaaaaaaaaaaaaaaaaaaaaaaaaaaaaaaaaaaaaaaaaaaaaaaaaaaaaaaaaaaaaaaaaaaaaaaaaaaaaaaaaaaaaaaaaaaaaaaaaaaaaaaaaaaaaaaaaaaaaaaaaaaaaaaaaaaaaaaaaaaaaaaaaaaaaaaaaaaaaaaaaaaaaaaaaaaaaaaaaaaaaaaaaaaaaaaaaaaaaaaaaaaaaaaaaaaaaaaaaaaaaaaaaaaaaaaaaaaaaaaaaaaaaaaaaaaaaaaaaaaaaaaaaaaaaaaaaaaaaaaaaaaaaaaaaaaaaaaaaaaaaaaaaaaaaaaaaaaaaaaaaaaaaaaaaaaaaaaaaaaaaaaaaaaaaaaaaaaaaaaaaaaaaaaaaaaaaaaaaaaaaaaaaaaaaaaaaaaaaaaaaaaaaaaaaaaaaaaaaaaaaaaaaaaaaaaaaaaaaaaaaaaaaaaaaaaaaaaaaaaaaaaaaaaaaaaaaaaaaaaaaaaaaaaaaaaaaaaaaaaaaaaaaaaaaaaaaaaaaaaaaaaaaaaaaaaaaaaaaaaaaaaaaaaaaaaaaaaaaaaaaaaaaaaaaaaaaaaaaaaaaaaaaaaaaaaaaaaaaaaaaaaaaaaaaaaaaaaaaaaaaaaaaaaaaaaaaaaaaaaaaaaaaaaaaaaaaaaaaaaaaaaaaaaaaaaaaaaaaaaaaaaaaaaaaaaaaaaaaaaaaaaaaaaaaaaaaaaaaaaaaaaaaaaaaaaaaaaaaaaaaaaaaaaaaaaaaaaaaaaaaaaaaaaaaaaaaaaaaaaaaaaaaaaaaaaaaaaaaaaaaaaaaaaaaaaaaaaaaaaaaaaaaaaaaaaaaaaaaaaaaaaaaaaaaaaaaaaaaaaaaaaaaaaaaaaaaaaaaaaaaaaaaaaaaaaaaaaaaaaaaaaaaaaaaaaaaaaaaaaaaaaaaaaaaaaaaaaaaaaaaaaaaaaaaaaaaaaaaaaaaaaaaaaaaaaaaaaaaaaaaaaaaaaaaaaaaaaaaaaaaaaaaaaaaaaaaaaaaaaaaaaaaaaaaaaaaaaaaaaaaaaaaaaaaaaaaaaaaaaaaaaaaaaaaaaaaaaaaaaaaaaaaaaaaaaaaaaaaaaaaaaaaaaaaaaaaaaaaaaaaaaaaaaaaaaaaaaaaaaaaaaaaaaaaaaaaaaaaaaaaaaaaaaaaaaaaaaaaaaaaaaaaaaaaaaaaaaaaaaaaaaaaaaaaaaaaaaaaaaaaaaaaaaaaaaaaaaaaaaaaaaaaaaaaaaaaaaaaaaaaaaaaaaaaaaaaaaaaaaaaaaaaaaaaaaaaaaaaaaaaaaaaaaaaaaaaaaaaaaaaaaaaaaaaaaaaaaaaaaaaaaaaaaaaaaaaaaaaaaaaaaaaaaaaaaaaaaaaaaaaaaaaaaaaaaaaaaaaaaaaaaaaaaaaaaaaaaaaaaaaaaaaaaaaaaaaaaaaaaaaaaaaaaaaaaaaaaaaaaaaaaaaaaaaaaaaaaaaaaaaaaaaaaaaaaaaaaaaaaaaaaaaaaaaaaaaaaaaaaaaaaaaaaaaaaaaaaaaaaaaaaaaaaaaaaaaaaaaaaaaaaaaaaaaaaaaaaaaaaaaaaaaaaaaaaaaaaaaaaaaaaaaaaaaaaaaaaaaaaaaaaaaaaaaaaaaaaaaaaaaaaaaaaaaaaaaaaaaaaaaaaaaaaaaaaaaaaaaaaaaa
\begin{eqnarray}
M_{1,1}=\frac{(\hbar^2 (\hbar^2 + V_2^2 + V_3^2 + V_4^2))}{(\hbar^4 + (-V_1 V_2 + V_3^2 +
   V_4^2)^2 + \hbar^2 (V_1^2 + V_2^2 + 2 (V_3^2 + V_4^2)))}+ \nonumber \\
 + i  ( \frac{ ( \hbar (-\hbar^2 V_1 + V_2 (-V_1 V_2 + V_3^2 + V_4^2)) )}{(\hbar^4 + (-V_1 V_2 +
   V_3^2 + V_4^2)^2 + \hbar^2 (V_1^2 + V_2^2 + 2 (V_3^2 + V_4^2)))} )= \nonumber \\
=\frac{1}{(\hbar^4 + (-V_1 V_2 + V_3^2 +
   V_4^2)^2 + \hbar^2 (V_1^2 + V_2^2 + 2 (V_3^2 + V_4^2)))} \times \nonumber \\
\times  [(\hbar^2 (\hbar^2 + V_2^2 + V_3^2 + V_4^2)) + i (\hbar (-\hbar^2 V_1 + V_2 (-V_1 V_2 + V_3^2 + V_4^2)) )]= \nonumber \\
\frac{1}{(\hbar^4 + (-V_1 V_2 + V_3^2 +
   V_4^2)^2 + \hbar^2 (V_1^2 + V_2^2 + 2 (V_3^2 + V_4^2)))} [ M_{r(1,1)}+iM_{i(1,1)}  ]
\end{eqnarray}
and  $M_{2,2}$
\begin{eqnarray}
M_{2,2}=\frac{ (\hbar^2 (\hbar^2 + V_1^2 + V_3^2 + V_4^2)) }{(\hbar^4 + (-V_1 V_2 + V_3^2 +
   V_4^2)^2 + \hbar^2 (V_1^2 + V_2^2 + 2 (V_3^2 + V_4^2)))}+ \nonumber \\
+i ( \frac{ (\hbar (-\hbar^2 V_2 + V_1 (-V_1 V_2 + V_3^2 + V_4^2)))  }{(\hbar^4 + (-V_1 V_2 +
   V_3^2 + V_4^2)^2 + \hbar^2 (V_1^2 + V_2^2 + 2 (V_3^2 + V_4^2)))} ) = \nonumber \\
=\frac{1}{(\hbar^4 + (-V_1 V_2 + V_3^2 +
   V_4^2)^2 + \hbar^2 (V_1^2 + V_2^2 + 2 (V_3^2 + V_4^2)))} \times \nonumber \\
\times [ (\hbar^2 (\hbar^2 + V_1^2 + V_3^2 + V_4^2)) + i (\hbar (-\hbar^2 V_2 + V_1 (-V_1 V_2 + V_3^2 + V_4^2)))  ]
= \nonumber \\
\frac{1}{(\hbar^4 + (-V_1 V_2 + V_3^2 +
   V_4^2)^2 + \hbar^2 (V_1^2 + V_2^2 + 2 (V_3^2 + V_4^2)))} [ M_{r(2,2)}+iM_{i(2,2)}  ].
\end{eqnarray}
Non-diagonal parts of matrix are given as
%%$
\begin{eqnarray}
M_{1,2}=\frac{(\hbar (-\hbar (V_1 + V_2) V_3 + (\hbar^2 - V_1 V_2 + V_3^2) V_4 +
   V_4^3))}{( \hbar^4 + (-V_1 V_2 + V_3^2 + V_4^2)^2 +
 \hbar^2 (V_1^2 + V_2^2 + 2 (V_3^2 + V_4^2))) } \nonumber \\
 - i ( \frac{ (\hbar (\hbar^2 V_3 + \hbar (V_1 + V_2) V_4 +
   V_3 (-V_1 V_2 + V_3^2 + V_4^2)))  }{(\hbar^4 + (-V_1 V_2 + V_3^2 + V_4^2)^2 +
 \hbar^2 (V_1^2 + V_2^2 + 2 (V_3^2 + V_4^2)))}) = \nonumber \\
=\frac{1}{(\hbar^4 + (-V_1 V_2 + V_3^2 +
   V_4^2)^2 + \hbar^2 (V_1^2 + V_2^2 + 2 (V_3^2 + V_4^2)))} \times \nonumber \\
\times [ ( \hbar (-\hbar (V_1 + V_2) V_3 + (\hbar^2 - V_1 V_2 + V_3^2) V_4 +
   V_4^3) ) + \nonumber \\  - i ( (\hbar (\hbar^2 V_3 + \hbar (V_1 + V_2) V_4 +
   V_3 (-V_1 V_2 + V_3^2 + V_4^2))) )  ] = \nonumber \\
\frac{1}{(\hbar^4 + (-V_1 V_2 + V_3^2 +
   V_4^2)^2 + \hbar^2 (V_1^2 + V_2^2 + 2 (V_3^2 + V_4^2)))} [ M_{r(1,2)}+iM_{i(1,2)}  ]
%%$
\end{eqnarray}

and
%%$
\begin{eqnarray}
M_{2,1}=( -(  \frac{ (  \hbar (\hbar (V_1 + V_2) V_3 + ( \hbar^2 - V_1 V_2 + V_3^2) V_4 + V_4^3) )  }{(
 \hbar^4 + (-V_1 V_2 + V_3^2 + V_4^2)^2 +
  \hbar^2 (V_1^2 + V_2^2 + 2 (V_3^2 + V_4^2)) )} ) ) + \nonumber \\
+i ( \frac{(\hbar (-\hbar^2 V_3 + \hbar (V_1 + V_2) V_4 -
   V_3 (-V_1 V_2 + V_3^2 + V_4^2)))}{(\hbar^4 + (-V_1 V_2 + V_3^2 + V_4^2)^2 +
 \hbar^2 (V_1^2 + V_2^2 + 2 (V_3^2 + V_4^2)))} )= \nonumber \\
=\frac{1}{(\hbar^4 + (-V_1 V_2 + V_3^2 +
   V_4^2)^2 + \hbar^2 (V_1^2 + V_2^2 + 2 (V_3^2 + V_4^2)))} \times \nonumber \\
\times [ -(  \hbar (\hbar (V_1 + V_2) V_3 + ( \hbar^2 - V_1 V_2 + V_3^2) V_4 + V_4^3) ) + \nonumber \\ + i ( \hbar (-\hbar^2 V_3 + \hbar (V_1 + V_2) V_4 -
   V_3 (-V_1 V_2 + V_3^2 + V_4^2)) )   ]  \nonumber \\
= \nonumber \\
\frac{1}{(\hbar^4 + (-V_1 V_2 + V_3^2 +
   V_4^2)^2 + \hbar^2 (V_1^2 + V_2^2 + 2 (V_3^2 + V_4^2)))} [ M_{r(2,1)}+iM_{i(2,1)}  ].
\end{eqnarray}
We recognize energy transfer during proton movement in close proximity to position dependent qubit so the quantum state after weak measurement is given as
\begin{eqnarray}
\ket{\psi}_{m}=e^{i \phi_{E1m}}c_{E1m}e^{\frac{E_1}{i\hbar}t}\ket{E_1}+e^{i \phi_{E2m}}c_{E2m}e^{\frac{E_2}{i\hbar}t}\ket{E_2}= \nonumber \\
=\begin{pmatrix}
+e^{i \phi_{E1m}}c_{E1m}e^{\frac{E_1}{i\hbar}t}+e^{i \phi_{E2m}}c_{E2m}e^{\frac{E_2}{i\hbar}t} \\
- e^{i \phi_{E1m}}c_{E1m}e^{\frac{E_1}{i\hbar}t}+e^{i \phi_{E2m}}c_{E2m}e^{\frac{E_2}{i\hbar}t} \\
\end{pmatrix}
=
\begin{pmatrix}
M_{1,1} & M_{1,2} \\
M_{2,1} & M_{2,2} \\
\end{pmatrix}
\begin{pmatrix}
+e^{i \phi_{E1}}c_{E1}e^{\frac{E_1}{i\hbar}t}+e^{i \phi_{E2}}c_{E2}e^{\frac{E_2}{i\hbar}t} \\
- e^{i \phi_{E1}}c_{E1}e^{\frac{E_1}{i\hbar}t}+e^{i \phi_{E2}}c_{E2}e^{\frac{E_2}{i\hbar}t} \\
\end{pmatrix}
\end{eqnarray}
We obtain the quantum state after weak measurement in the form as
\begin{eqnarray}
e^{i \phi_{E1m}}c_{E1m}=\frac{e^{-\frac{E_1}{i\hbar}t}}{\sqrt{2}}
\begin{pmatrix}
1 & -1 \\
\end{pmatrix}
\begin{pmatrix}
M_{1,1} & M_{1,2} \\
M_{2,1} & M_{2,2} \\
\end{pmatrix}
\begin{pmatrix}
+e^{i \phi_{E1}}c_{E1}e^{\frac{E_1}{i\hbar}t}+e^{i \phi_{E2}}c_{E2}e^{\frac{E_2}{i\hbar}t} \\
- e^{i \phi_{E1}}c_{E1}e^{\frac{E_1}{i\hbar}t}+e^{i \phi_{E2}}c_{E2}e^{\frac{E_2}{i\hbar}t} \\
\end{pmatrix}
\end{eqnarray}
and
\begin{eqnarray}
e^{i \phi_{E2m}}c_{E2m}=\frac{ e^{-\frac{E_2}{i\hbar}t}}{\sqrt{2}}
\begin{pmatrix}
1 & 1 \\
\end{pmatrix}
\begin{pmatrix}
M_{1,1} & M_{1,2} \\
M_{2,1} & M_{2,2} \\
\end{pmatrix}
\begin{pmatrix}
+e^{i \phi_{E1}}c_{E1}e^{\frac{E_1}{i\hbar}t}+e^{i \phi_{E2}}c_{E2}e^{\frac{E_2}{i\hbar}t} \\
- e^{i \phi_{E1}}c_{E1}e^{\frac{E_1}{i\hbar}t}+e^{i \phi_{E2}}c_{E2}e^{\frac{E_2}{i\hbar}t} \\
\end{pmatrix}
\end{eqnarray}
We have
\begin{eqnarray}
e^{i \phi_{E1m}}c_{E1m}=\frac{1}{\sqrt{2}} (c_{E2} e^{(
    i ( \phi_{E2} + (E_1 - E_2) t) )}   (M_{1,1} +M_{1,2} -M_{2,1} - M_{2,2}) + \nonumber \\
  + c_{E1} e^{(i \phi_{E1} )} (M_{1,1} -  M_{1,2} - M_{2,1} + M_{2,2}))=
\end{eqnarray}

and \newline $ $
\begin{eqnarray}
e^{i \phi_{E2m}}c_{E2m}=\frac{1}{\sqrt{2}} (c_{E2} e^{(
    i ( \phi_{E1} + (-E_1 + E_2) t) )} (M_{1,1} -  M_{1,2}  + M_{2,1} - M_{2,2}) + \nonumber \\
   + c_{E2} e^{(i \phi_{E2})} (M_{1,1} + M_{1,2}+ M_{2,1}+ M_{2,2} ))=
\end{eqnarray}
Last expression are given with parameters of weak measurement as \newline \newline
\begin{eqnarray}
e^{i \phi_{E1m}}c_{E1m}=\frac{1}{2 (\hbar^4 + (-V_1V_2+V_3^2 +
        V_4^2)^2 +
      \hbar^2 (V_1^2 + V_2^2 +
         2 (V_3^2 + V_4^2)))} \times \nonumber \\
\times \Bigg[\Big[\hbar (c_{E1}| \hbar (2 \hbar^2 + V_1^2 + V_2^2 + 2 V_1 V_3 + 2 V_3 (V_2+V_3)+
         2 V_4^2) Cos[\phi_{E1}] + \nonumber \\
+ c_{E2} (\hbar (-V_1^2+V_2^2)+2(\hbar^2 - V_1 V_2 + V_3^2) V_4 +
         2V_4^3) Cos[\phi_{E2}+(E_1-E_2)t]+ \nonumber \\
          +
      c_{E1} (\hbar^2 (V_1+V_2-2V_3) + (V_1+V_2+2V_3) (V_1V_2-V_3^2 -
            V_4^2)) Sin[\phi_{E1}] + \nonumber \\
    + c_{E2} ((V_1-V_2) (\hbar^2-V_1 V_2+V_3^2) +
         2 \hbar (V_1+V_2)V_4 + (V_1-V_2) V_4^2) Sin[
        \phi_{E2} + (E_1-E_2)t]))\Big]+ \nonumber \\ +i[( \hbar (c_{E1} (-\hbar^2 (V_1 + V_2 -
            2 V_3) %%\nonumber \\
            %%\end{eqnarray}
              - (V_1 + V_2 + 2 V_3) (V_1V_2 - V_3^2 - V_4^2)) Cos[
        \phi_{E1}]  \nonumber \\ %\\
      %%  \end{eqnarray}
       % \begin{eqnarray}
      -c_{E2} ((V_1 - V_2) (\hbar^2 - V_1 V_2 + V_3^2) +
         2 \hbar (V_1 + V_2)V_4 + (V_1 - V_2) V_4^2) Cos[
        \phi_{E2} + (E_1-E_2)t] + \nonumber  \\
      c_{E1} \hbar (2\hbar^2 + V_1^2 + V_2^2 + 2 V_1 V_3 + 2 V_3 (V_2 + V_3) +
         2 V_4^2) Sin[\phi_{E1}] + \nonumber \\
    %     \end{eqnarray}
    + c_{E2} (\hbar (-V_1^2 + V_2^2) + 2 (\hbar^2 - V_1 V_2 + V_3^2) V_4 +
         2 V_4^3) Sin[\phi_{E2} + (E_1 - E_2)t]))\Big]\Bigg] %\times \frac{1}{(2 (\hbar^4 + (-V_1 V_2 +
   %     V_3^2 + V_4^2)^2 + \hbar^2 (V_1^2 + V_2^2 + 2 (V_3^2 + V_4^2))))}
   \end{eqnarray}
and consequently we have
\begin{eqnarray}
e^{i \phi_{E2m}}c_{E2m}=
\frac{1}{ 2(\hbar^4 + (-V_1 V_2 + V_3^2 +
        V_4^2)^2 + \hbar^2 (V_1^2 + V_2^2 + 2 (V_3^2 + V_4^2)))  } \times
\end{eqnarray}
$ \times
 \Bigg[ \Big[(\hbar (c_{E2} \hbar (2 \hbar^2 + V_1^2 +
         V_2^2 - 2 (V_1 + V_2) V_3 + 2 V_3^2 + 2 V_4^2) Cos[\phi_{E2}]+$ \newline  $ -
      c_{E1} (\hbar (V_1 - V_2) (V_1 + V_2) + 2 (\hbar^2 - V_1 V_2 + V_3^2) V_4 +
         2 V_4^3) Cos[\phi_{E1} + (-E_1 + E_2)t] +$ \newline
      $+c_{E2} (\hbar^2 (V_1 + V_2 + 2 V_3) + (V_1 + V_2 - 2 V_3) (V_1 V_2 - V_3^2 -
            V_4^2)) Sin[\phi_{E2}] +$ \newline
      $+c_{E1} ((V_1 - V_2) (\hbar^2 - V_1 V_2 + V_3^2) -
         2 \hbar (V_1 + V_2) V_4 + (V_1 - V_2) V_4^2) Sin[
        \phi_{E1}+ (-E_1 + E_2)t]) )\Big]+$ \newline  $+i\Big[\hbar (-c_{E2} (\hbar^2 (V_1 + V_2 + 2 V_3) + (V_1 + V_2 - 2 V_3) (V_1 V_2 -
            V_3^2 - V_4^2)) Cos[\phi_{E2}] + $ \newline $
      +c_{E1} (-(V_1 - V_2) (\hbar^2 - V_1 V_2 + V_3^2) +
         2 \hbar (V_1 + V_2) V_4 + (-V_1 + V_2) V_4^2) Cos[
        \phi_{E1} -(E_1-E_2)t ] +$ \newline
      $+c_{E2} \hbar (2 \hbar^2 + V_1^2 + V_2^2 - 2 (V_1 + V_2) V_3 + 2 V_3^2 +
         2 V_4^2) Sin[\phi_{E2}]$ \newline $ -
      c_{E1} (\hbar (V_1 - V_2) (V_1 + V_2) + 2 (\hbar^2 - V_1 V_2 + V_3^2) V_4 +
         2 V_4^3) Sin[\phi_{E1} - (E_1-E_2)t ])\Big] \Bigg] $ \newline
\newline
\newline
It is quite straightforward to obtain probability of occupancy of energy $E_1$ by electron in position based qubit after weak measurement (one interaction with passing charge particle) and it is given as  \newline \newline
\begin{eqnarray}
(c_{E1m})^2=\frac{\hbar^2}{(4 (\hbar^4 + (-V_1 V_2 +
       V_3^2 + V_4^2)^2 + \hbar^2 (V_1^2 + V_2^2 + 2 (V_3^2 + V_4^2))))}\times \nonumber \\ \times [(c_{E1}^2 (4 \hbar^2 + (V_1 + V_2 + 2 V_3)^2) +
     c_{E2}^2 ((V_1 - V_2)^2 + 4 V_4^2) +  \nonumber \\
     +2 c_{E1} c_{E2}(-(V_1-V_2)(V_1 + V_2 + 2 V_3) + 4 \hbar V_4) Cos[
       \phi_{E1} - \phi_{E2} - (E_1-E_2) t_1] \nonumber \\
        -
     4 c_{E1} c_{E2}[\hbar(V_1-V_2)+(V_1 + V_2 + 2 V_3)V_4]Sin[
       \phi_{E1} - \phi_{E2} - (E_1-E_2)t_1])]  %\frac{1}{(4 (\hbar^4 + (-V_1 V_2 +
       %V_3^2 + V_4^2)^2 + \hbar^2 (V_1^2 + V_2^2 + 2 (V_3^2 + V_4^2))))}
       \end{eqnarray}
\newline \newline
In similar fashion we obtain the probability of occupancy of energy level $E_2$ by position dependent qubit that is given as %\newline
\begin{eqnarray}
(c_{E2m})^2=\frac{\hbar^2}{4 (\hbar^4 + (-V_1 V_2 +
       V_3^2 + V_4^2)^2 + \hbar^2 (V_1^2 + V_2^2 + 2 (V_3^2 + V_4^2)))} \times \nonumber \\ \times \Bigg[c_{E2}^2 (4 \hbar^2 + (V_1 + V_2 - 2 V_3)^2) +
     c_{E1}^2 ((V_1 - V_2)^2 + 4 V_4^2)\nonumber \\ -
     2 c_{E1} c_{E2} ((V_1 - V_2) (V_1 + V_2 - 2 V_3) + 4 \hbar V_4) Cos[
       \phi_{E1} - \phi_{E2} - (E_1-E_2)t] + \nonumber \\
     +4 c_{E1} c_{E2} (\hbar (V_1 - V_2) - (V_1 + V_2 - 2 V_3) V_4) Sin[
       \phi_{E1} - \phi_{E2} - (E_1-E_2)t]\Bigg]
\end{eqnarray}
\newline
\newline
Consequently we obtain phase imprint on energy eigenstate $E_1$ given by the relation \newline
\begin{eqnarray}
e^{i \phi_{E1m}}=\Bigg[\frac{1}{2 (\hbar^4 + (-V_1V_2+V_3^2 +
        V_4^2)^2 +
      \hbar^2 (V_1^2 + V_2^2 +
         2 (V_3^2 + V_4^2)))} \times \nonumber \\
\times \Bigg[\Big[\hbar (c_{E1}| \hbar (2 \hbar^2 + V_1^2 + V_2^2 + 2 V_1 V_3 + 2 V_3 (V_2+V_3)+
         2 V_4^2) Cos[\phi_{E1}] + \nonumber \\
+ c_{E2} (\hbar (-V_1^2+V_2^2)+2(\hbar^2 - V_1 V_2 + V_3^2) V_4 +
         2V_4^3) Cos[\phi_{E2}+(E_1-E_2)t]+ \nonumber \\
          +
      c_{E1} (\hbar^2 (V_1+V_2-2V_3) + (V_1+V_2+2V_3) (V_1V_2-V_3^2 -
            V_4^2)) Sin[\phi_{E1}] + \nonumber \\
    + c_{E2} ((V_1-V_2) (\hbar^2-V_1 V_2+V_3^2) +
         2 \hbar (V_1+V_2)V_4 + (V_1-V_2) V_4^2) Sin[
        \phi_{E2} + (E_1-E_2)t]))\Big]+ \nonumber \\ +i[( \hbar (c_{E1} (-\hbar^2 (V_1 + V_2 -
            2 V_3) %%\nonumber \\
            %%\end{eqnarray}
              - (V_1 + V_2 + 2 V_3) (V_1V_2 - V_3^2 - V_4^2)) Cos[
        \phi_{E1}]  \nonumber \\ %\\
      %%  \end{eqnarray}
       % \begin{eqnarray}
      -c_{E2} ((V_1 - V_2) (\hbar^2 - V_1 V_2 + V_3^2) +
         2 \hbar (V_1 + V_2)V_4 + (V_1 - V_2) V_4^2) Cos[
        \phi_{E2} + (E_1-E_2)t] + \nonumber  \\
      c_{E1} \hbar (2\hbar^2 + V_1^2 + V_2^2 + 2 V_1 V_3 + 2 V_3 (V_2 + V_3) +
         2 V_4^2) Sin[\phi_{E1}] + \nonumber \\
    %     \end{eqnarray}
    + c_{E2} (\hbar (-V_1^2 + V_2^2) + 2 (\hbar^2 - V_1 V_2 + V_3^2) V_4 +
         2 V_4^3) Sin[\phi_{E2} + (E_1 - E_2)t]))\Big]\Bigg]\Bigg] \times \nonumber \\
\times   \Bigg[\frac{\hbar^2}{(4 (\hbar^4 + (-V_1 V_2 +
       V_3^2 + V_4^2)^2 + \hbar^2 (V_1^2 + V_2^2 + 2 (V_3^2 + V_4^2))))}\times \nonumber \\ \times [(c_{E1}^2 (4 \hbar^2 + (V_1 + V_2 + 2 V_3)^2) +
     c_{E2}^2 ((V_1 - V_2)^2 + 4 V_4^2) +  \nonumber \\
     +2 c_{E1} c_{E2}(-(V_1-V_2)(V_1 + V_2 + 2 V_3) + 4 \hbar V_4) Cos[
       \phi_{E1} - \phi_{E2} - (E_1-E_2) t] \nonumber \\
        -
     4 c_{E1} c_{E2}[\hbar(V_1-V_2)+(V_1 + V_2 + 2 V_3)V_4]Sin[
       \phi_{E1} - \phi_{E2} - (E_1-E_2)t])] \Bigg]^{-\frac{1}{2}}
          %\times \frac{1}{(2 (\hbar^4 + (-V_1 V_2 +
   %     V_3^2 + V_4^2)^2 + \hbar^2 (V_1^2 + V_2^2 + 2 (V_3^2 + V_4^2))))}
   \end{eqnarray}
and phase imprint on energy eigenstate $E_2$ given by the relation \newline
\begin{eqnarray}
e^{i \phi_{E2m}}= \frac{1}{ \hbar \sqrt{(\hbar^4 + (-V_1 V_2 +
         V_3^2 + V_4^2)^2 +
      \hbar^2 (V_1^2 + V_2^2 +
         2 (V_3^2 + V_4^2)))}}  \times \nonumber \\
\Bigg[ i\Big[ \hbar (-c_{E2} (\hbar^2 (V_1 + V_2 + 2 V_3) + (V_1 + V_2 - 2 V_3) (V_1 V_2 -
            V_3^2 - V_4^2)) Cos[\phi_{E2}] + \nonumber \\
% \end{eqnarray}
      +c_{E1} (-(V_1 - V_2) (\hbar^2 - V_1 V_2 + V_3^2) +
         2 \hbar (V_1 + V_2) V_4 + (-V_1 + V_2) V_4^2) Cos[
        \phi_{E1} - (E_1-E_2)t] + \nonumber \\
      +c_{E2} \hbar (2 \hbar^2 + V_1^2 + V_2^2 - 2 (V_1 + V_2) V_3 + 2 V_3^2 +
         2 V_4^2) Sin[\phi_{E2}]+ \nonumber \\
      -c_{E1} (\hbar (V_1 - V_2) (V_1 + V_2) + 2 (\hbar^2 - V_1 V_2 + V_3^2) V_4 +
         2 V_4^3) Sin[\phi_{E1} - (E_1-E_2)t]) \Big] +  \nonumber \\
         + \Big[(\hbar (c_{E2} \hbar (2 \hbar^2 + V_1^2 +
         V_2^2 - 2 (V_1 + V_2) V_3 + 2 V_3^2 + 2 V_4^2) Cos[\phi_{E2}]+ \nonumber \\
         -c_{E1} (\hbar (V_1 - V_2) (V_1 + V_2) + 2 (\hbar^2 - V_1 V_2 + V_3^2) V_4 +
         2 V_4^3) Cos[\phi_{E1} + (-E_1 + E_2)t] + \nonumber \\
      +c_{E2} (\hbar^2 (V_1 + V_2 + 2 V_3) + (V_1 + V_2 - 2 V_3) (V_1 V_2 - V_3^2 -
            V_4^2)) Sin[\phi_{E2}] + \nonumber \\
      +c_{E1} ((V_1 - V_2) (\hbar^2 - V_1 V_2 + V_3^2) -
         2 \hbar (V_1 + V_2) V_4 + (V_1 - V_2) V_4^2) Sin[
        \phi_{E1}+ (-E_1 + E_2) t]) )\Big] \Bigg] \nonumber \\
    %  \end{eqnarray}
 % $ $
%\newline
        \times \Bigg[ %%%\frac{\hbar^2}{4(\hbar^4 + (-V_1 V_2 +
     %%  V_3^2 + V_4^2)^2 + \hbar^2 (V_1^2 + V_2^2 + 2 (V_3^2 + V_4^2)))} \times \nonumber \\
          ( (c_{E2}^2 (4 \hbar^2 + (V_1 + V_2 - 2 V_3)^2) +
     c_{E1}^2 ((V_1 - V_2)^2 + 4 V_4^2) \nonumber \\ -
     2 c_{E1} c_{E2} ((V_1 - V_2) (V_1 + V_2 - 2 V_3) + 4 \hbar V_4) Cos[
       \phi_{E1} - \phi_{E2} - (E_1-E_2)t] + \nonumber \\ +
     4 c_{E1} c_{E2} (\hbar (V_1 - V_2) - (V_1 + V_2 - 2 V_3) V_4) Sin[
       \phi_{E1} - \phi_{E2} - (E_1-E_2)t ]))\Bigg]^{-\frac{1}{2}}.
       \end{eqnarray}
%$
%\subsection{Subsection Heading Here}
%Write your subsection text here.
In general case of the position based qubit with complex values of hopping coefficients $t_{s12}=t_{sr}+i t_{si}$ brings eigenenergies
\begin{eqnarray}
E_1=\frac{1}{2} \left(-\sqrt{(E_{p1}-E_{p2})^2+4 \left(t_{si}^2+t_{sr}^2\right)}+(E_{p1}+E_{p2})\right), \nonumber \\
%and %\newline
E_2=\frac{1}{2} \left(\sqrt{(E_{p1}-E_{p2})^2+4 \left(t_{si}^2+t_{sr}^2\right)}+(E_{p1}+E_{p2})\right)
\end{eqnarray}

that are referred to the eigevectors
\begin{equation}
\ket{E_1}=
%%\left\{
\begin{pmatrix}
-\frac{i (t_{si}-i t_{sr}) \left(\sqrt{(E_{p1}-E_{p2})^2+4 \left(t_{si}^2+t_{sr}^2\right)}-E_{p1}+E_{p2}\right)}{2
   \sqrt{t_{si}^2+t_{sr}^2} \sqrt{\frac{1}{4} \left(\sqrt{(E_{p1}-E_{p2})^2+4
   \left(t_{si}^2+t_{sr}^2\right)}-E_{p1}+E_{p2}\right)^2+t_{si}^2+t_{sr}^2}}, \nonumber \\
   \frac{\sqrt{t_{si}^2+t_{sr}^2}}{\sqrt{\frac{1}{4}
   \left(\sqrt{(E_{p1}-E_{p2})^2+4 \left(t_{si}^2+t_{sr}^2\right)}-E_{p1}+E_{p2}\right)^2+t_{si}^2+t_{sr}^2}}
\end{pmatrix}
%%   \right\}
\end{equation}
and
\begin{equation}
\ket{E_2}=
%\left\{
\begin{pmatrix}
\frac{i (t_{si}-it_{sr}) \left(\sqrt{(E_{p1}-E_{p2})^2+4 \left(t_{si}^2+t_{sr}^2\right)}+E_{p1}-E_{p2}\right)}{2
   \sqrt{t_{si}^2+t_{sr}^2} \sqrt{\frac{1}{4} \left(\sqrt{(E_{p1}-E_{p2})^2+4
   \left(t_{si}^2+t_{sr}^2\right)}+E_{p1}-E_{p2}\right)^2+t_{si}^2+t_{sr}^2}}, \nonumber \\
   \frac{\sqrt{t_{si}^2+t_{sr}^2}}{\sqrt{\frac{1}{4}
   \left(\sqrt{+(E_{p1}-E_{p2})^2+4 \left(t_{si}^2+t_{sr}^2\right)}+E_{p1}-E_{p2}\right)^2+t_{si}^2+t_{sr}^2}}
\end{pmatrix}.
   %\right\}
\end{equation}
The quantum state before weak measurement is given in the form as
\begin{equation}
\ket{\psi_t}=c_{E1}e^{i \phi_{E1}}e^{\frac{E_1 t}{\hbar i}}\ket{E_1}+c_{E2}e^{i \phi_{E2}}e^{\frac{E_2 t}{\hbar i}}\ket{E_2}=
\begin{pmatrix}
\alpha(t), \nonumber \\
\beta(t),
\end{pmatrix}_x
\end{equation}
and the quantum state after measurement is given in the form as
\begin{equation}
\ket{\psi_t}=c_{E1m}e^{i \phi_{E1m}}e^{\frac{E_1 t}{\hbar i}}\ket{E_1}+c_{E2m}e^{i \phi_{E2m}}e^{\frac{E_2 t}{\hbar i}}\ket{E_2}=
\begin{pmatrix}
\alpha_m(t), \nonumber \\
\beta_m(t),
\end{pmatrix}_x .
\end{equation}
and for the weak measurement taking part at time instant $t_1$ we have
\begin{equation}
\begin{pmatrix}
\alpha_m(t_1^{+}), \nonumber \\
\beta_m(t_1^{+}),
\end{pmatrix}_x =
\begin{pmatrix}
M_{1,1} & M_{1,2} \\
M_{2,1} & M_{2,2} \\
\end{pmatrix}
\begin{pmatrix}
\alpha(t_1^{-}), \nonumber \\
\beta(t_1^{-}),
\end{pmatrix}_x.
\end{equation}
or equivalently
\begin{equation}
\begin{pmatrix}
\alpha_m(t_1^{+})^{*},\beta_m(t_1^{+})^{*},
\end{pmatrix}_x =
\begin{pmatrix}
\alpha(t_1^{-})^{*}, \beta(t_1^{-})^{*},
\end{pmatrix}_x
\begin{pmatrix}
M_{1,1}^{*} & M_{2,1}^{*} \\
M_{1,2}^{*} & M_{2,2}^{*} \\
\end{pmatrix}
\end{equation}
Last two relations implies
\begin{equation}
1=
\begin{pmatrix}
\alpha_m(t_1^{+})^{*},\beta_m(t_1^{+})^{*},
\end{pmatrix}_x
\begin{pmatrix}
\alpha_m(t_1^{+}), \nonumber \\
\beta_m(t_1^{+})
\end{pmatrix}_x =
\begin{pmatrix}
\alpha(t_1^{-})^{*}, \beta(t_1^{-})^{*},
\end{pmatrix}_x
\begin{pmatrix}
M_{1,1}^{*} & M_{2,1}^{*} \\
M_{1,2}^{*} & M_{2,2}^{*} \\
\end{pmatrix}
\begin{pmatrix}
M_{1,1} & M_{1,2} \\
M_{2,1} & M_{2,2} \\
\end{pmatrix}
\begin{pmatrix}
\alpha(t_1^{-}), \nonumber \\
\beta(t_1^{-})
\end{pmatrix}=1
\end{equation}
that simply implies
\begin{equation}
\begin{pmatrix}
M_{1,1}^{*} & M_{2,1}^{*} \\
M_{1,2}^{*} & M_{2,2}^{*} \\
\end{pmatrix}
\begin{pmatrix}
M_{1,1} & M_{1,2} \\
M_{2,1} & M_{2,2} \\
\end{pmatrix}
=\hat{1}=
\begin{pmatrix}
1 & 0 \\
0 & 1 \\
\end{pmatrix}.
\end{equation}
Such reasoning can be conducted also for many particle states.
Basing on the conducted analysis we obtain the parameters of the quantum state in eigenenergy representation after weak measurement at time $t_1$ in the form as
\begin{eqnarray}
e^{i \phi_{E1m}}c_{E1m}=e^{-\frac{E_1}{i\hbar}t_1} \nonumber \\
\begin{pmatrix}
\frac{-i (t_{si}-i t_{sr}) \left(\sqrt{(E_{p1}-E_{p2})^2+4 \left(t_{si}^2+t_{sr}^2\right)}-E_{p1}+E_{p2}\right)}{2
   \sqrt{t_{si}^2+t_{sr}^2} \sqrt{\frac{1}{4} \left(\sqrt{(E_{p1}-E_{p2})^2+4
   \left(t_{si}^2+t_{sr}^2\right)}-E_{p1}+E_{p2}\right)^2+t_{si}^2+t_{sr}^2}} &
\frac{\sqrt{t_{si}^2 + t_{sr}^2}}{\sqrt{t_{si}^2 + t_{sr}^2 +
 \frac{1}{4}(-E_{p1} + E_{p2} + \sqrt{(E_{p1} - E_{p2})^2 + 4 (t_{si}^2 + t_{sr}^2)})^2}},
\end{pmatrix} \times \nonumber \\ \times
\begin{pmatrix}
M_{1,1} & M_{1,2} \\
M_{2,1} & M_{2,2} \\
\end{pmatrix}
\begin{pmatrix}
\alpha(t_1^{-}), \nonumber \\
\beta(t_1^{-}),
\end{pmatrix}= \nonumber \\
=e^{-\frac{E_1}{i\hbar}t_1}
\begin{pmatrix}
\frac{-i (t_{si}-i t_{sr}) \left(\sqrt{(E_{p1}-E_{p2})^2+4 \left(t_{si}^2+t_{sr}^2\right)}-E_{p1}+E_{p2}\right)}{2
   \sqrt{t_{si}^2+t_{sr}^2} \sqrt{\frac{1}{4} \left(\sqrt{(E_{p1}-E_{p2})^2+4
   \left(t_{si}^2+t_{sr}^2\right)}-E_{p1}+E_{p2}\right)^2+t_{si}^2+t_{sr}^2}} &
\frac{\sqrt{t_{si}^2 + t_{sr}^2}}{\sqrt{t_{si}^2 + t_{sr}^2 +
 \frac{1}{4}(-E_{p1} + E_{p2} + \sqrt{(E_{p1} - E_{p2})^2 + 4 (t_{si}^2 + t_{sr}^2)})^2}},
\end{pmatrix} \times \nonumber \\ \times
\begin{pmatrix}
M_{1,1}\alpha(t_1^{-}) + M_{1,2}\beta(t_1^{-}) \\
M_{2,1}\alpha(t_1^{-}) + M_{2,2} \beta(t_1^{-}) \\
\end{pmatrix}
=e^{-\frac{\frac{1}{2} \left(-\sqrt{(E_{p1}-E_{p2})^2+4 \left(t_{si}^2+t_{sr}^2\right)}+(E_{p1}+E_{p2})\right)}{i\hbar} t_1}\times \nonumber \\
\Bigg[ \frac{-i (t_{si}-i t_{sr}) \left(\sqrt{(E_{p1}-E_{p2})^2+4 \left(t_{si}^2+t_{sr}^2\right)}-E_{p1}+E_{p2}\right)}{2
   \sqrt{t_{si}^2+t_{sr}^2} \sqrt{\frac{1}{4} \left(\sqrt{(E_{p1}-E_{p2})^2+4
   \left(t_{si}^2+t_{sr}^2\right)}-E_{p1}+E_{p2}\right)^2+t_{si}^2+t_{sr}^2}}(M_{1,1}\alpha(t_1^{-}) + M_{1,2}\beta(t_1^{-}))+ \nonumber \\
   +\frac{\sqrt{t_{si}^2 + t_{sr}^2}}{\sqrt{t_{si}^2 + t_{sr}^2 +
 \frac{1}{4}(-E_{p1} + E_{p2} + \sqrt{(E_{p1} - E_{p2})^2 + 4 (t_{si}^2 + t_{sr}^2)})^2}}(M_{2,1}\alpha(t_1^{-}) + M_{2,2}\beta(t_1^{-}))\Bigg]=e^{i \phi_{E1m}}c_{E1m}.
\end{eqnarray}
and
\begin{eqnarray}
e^{i \phi_{E2m}}c_{E2m}=e^{-\frac{E_2}{i\hbar}t_1}
\nonumber \\
\begin{pmatrix}
\frac{+i (t_{si}-i t_{sr}) \left(\sqrt{(E_{p1}-E_{p2})^2+4 \left(t_{si}^2+t_{sr}^2\right)}-E_{p2}+E_{p1}\right)}{2
   \sqrt{t_{si}^2+t_{sr}^2} \sqrt{\frac{1}{4} \left(\sqrt{(E_{p1}-E_{p2})^2+4
   \left(t_{si}^2+t_{sr}^2\right)}-E_{p2}+E_{p1}\right)^2+t_{si}^2+t_{sr}^2}} &
\frac{\sqrt{t_{si}^2 + t_{sr}^2}}{\sqrt{t_{si}^2 + t_{sr}^2 +
 \frac{1}{4}(-E_{p2} + E_{p1} + \sqrt{(E_{p1} - E_{p2})^2 + 4 (t_{si}^2 + t_{sr}^2)})^2}},
\end{pmatrix} \times \nonumber \\  \times
\begin{pmatrix}
M_{1,1} & M_{1,2} \\
M_{2,1} & M_{2,2} \\
\end{pmatrix}
\begin{pmatrix}
\alpha(t_1^{-}), \nonumber \\
\beta(t_1^{-})
\end{pmatrix}= \nonumber \\
e^{-\frac{E_2}{i\hbar}t}
\begin{pmatrix}
\frac{+i (t_{si}-i t_{sr}) \left(\sqrt{(E_{p1}-E_{p2})^2+4 \left(t_{si}^2+t_{sr}^2\right)}-E_{p2}+E_{p1}\right)}{2
   \sqrt{t_{si}^2+t_{sr}^2} \sqrt{\frac{1}{4} \left(\sqrt{(E_{p1}-E_{p2})^2+4
   \left(t_{si}^2+t_{sr}^2\right)}-E_{p2}+E_{p1}\right)^2+t_{si}^2+t_{sr}^2}} &
\frac{\sqrt{t_{si}^2 + t_{sr}^2}}{\sqrt{t_{si}^2 + t_{sr}^2 +
 \frac{1}{4}(-E_{p2} + E_{p1} + \sqrt{(E_{p1} - E_{p2})^2 + 4 (t_{si}^2 + t_{sr}^2)})^2}},
\end{pmatrix} \times \nonumber \\  \times
\begin{pmatrix}
M_{1,1}\alpha(t_1^{-})+M_{1,2}\beta(t_1^{-}), \nonumber \\
M_{2,1}\alpha(t_1^{-})+ M_{2,2}\beta(t_1^{-})
\end{pmatrix}= \nonumber \\
=e^{-\frac{\frac{1}{2} \left(\sqrt{(E_{p1}-E_{p2})^2+4 \left(t_{si}^2+t_{sr}^2\right)}+(E_{p1}+E_{p2})\right)}{i\hbar}t}\times \nonumber \\
\Bigg[\frac{+i (t_{si}-i t_{sr}) \left(\sqrt{(E_{p1}-E_{p2})^2+4 \left(t_{si}^2+t_{sr}^2\right)}-E_{p2}+E_{p1}\right)}{2
   \sqrt{t_{si}^2+t_{sr}^2} \sqrt{\frac{1}{4} \left(\sqrt{(E_{p1}-E_{p2})^2+4
   \left(t_{si}^2+t_{sr}^2\right)}-E_{p2}+E_{p1}\right)^2+t_{si}^2+t_{sr}^2}}(M_{1,1}\alpha(t_1^{-})+M_{1,2}\beta(t_1^{-})))+ \nonumber \\
   +\frac{\sqrt{t_{si}^2 + t_{sr}^2}}{\sqrt{t_{si}^2 + t_{sr}^2 +
 \frac{1}{4}(-E_{p2} + E_{p1} + \sqrt{(E_{p1} - E_{p2})^2 + 4 (t_{si}^2 + t_{sr}^2)})^2}}(M_{2,1}\alpha(t_1^{-})+M_{2,2}\beta(t_1^{-})))\Bigg]
 \nonumber =e^{i \phi_{E2m}}c_{E2m}.
\end{eqnarray}

\newpage
\section{Dynamic of two qubit electrostatic entanglement under influence of weak measurement}
We have the following two interacting qubit Hamiltonian for isolated quantum system given in the form
\begin{eqnarray}
H=
\begin{pmatrix}
E_{p1} + E_{p1'}+\frac{q^2}{d_{1,1'}} & t_{s2} & t_{s1} & 0 \\
t_{s2}^{*} & E_{p1}+E_{p2'}+\frac{q^2}{d_{1,2'}} & 0 & t_{s1}  \\
t_{s1}^{*} & 0 & E_{p2}+E_{p1'}+\frac{q^2}{d_{2,1'}} & t_{s2} \\
0   & t_{s1}^{*} & t_{s2}^{*} &  E_{p2}+E_{p2'}+\frac{q^2}{d_{2,2'}}
%%%%%0 & t_{s2}^{*} & t_{s1} & 0 & 0 & E_{p2} + E_{p2'} \\
\end{pmatrix}.
\end{eqnarray}
Placement of external probing particle affecting only qubit 1 modifies this Hamiltonian
\begin{eqnarray*}
H=
\begin{pmatrix}
E_{p1}+f_1(t) + E_{p1'}+\frac{q^2}{d_{1,1'}} & t_{s2} & t_{s1} +f_3(t)+if_4(t)  & 0 \\
t_{s2}^{*} & E_{p1}+f_1(t)+E_{p2'}+\frac{q^2}{d_{1,2'}} & 0 & t_{s1}+f_3(t)+if_4(t)  \\
t_{s1}^{*}+f_3(t)-if_4(t) & 0 & E_{p2}+f_2(t)+E_{p1'}+\frac{q^2}{d_{2,1'}} & t_{s2} \\
0   & t_{s1}^{*}+f_3(t)-if_4(t) & t_{s2}^{*} &  E_{p2}+f_2(t)+E_{p2'}+\frac{q^2}{d_{2,2'}}
%%%%%0 & t_{s2}^{*} & t_{s1} & 0 & 0 & E_{p2} + E_{p2'} \\
\end{pmatrix}.
\end{eqnarray*}
Let us investigate the equations of motion for 2-qubit system under influence of external charge particle. Let us assume that the quantum system is given as
\begin{eqnarray}
\ket{\psi(t)}=
\begin{pmatrix}
\gamma_1 \\
\gamma_2 \\
\gamma_3 \\
\gamma_4
\end{pmatrix}, |\gamma_1(t)|^2+|\gamma_2(t)|^2+|\gamma_3(t)|^2+|\gamma_4(t)|^2=1
\end{eqnarray}

We end up with 4 equation system given as
\begin{eqnarray}
\label{eqn}
[E_{p1}+f_1(t)) + E_{p1'}+\frac{q^2}{d_{1,1'}}]\gamma_1(t) + t_{s2}\gamma_2(t) + t_{s1}\gamma_3(t) +[f_3(t)+if_4(t)]\gamma_3(t)  = i \hbar \frac{d}{dt}\gamma_1(t), \\
t_{s2}^{*}\gamma_2(t) + [E_{p1}+f_1(t)+E_{p2'}+\frac{q^2}{d_{1,2'}}]\gamma_2(t)+ [t_{s1}+f_3(t)+if_4(t)]\gamma_4(t)=  i \hbar \frac{d}{dt}\gamma_2(t), \\
%%%%%[
( t_{s1}^{*}  +f_3(t)-if_4(t) )
\gamma_1(t)
+[ E_{p2}+f_2(t)+E_{p1'}+\frac{q^2}{d_{2,1'}}]\gamma_3(t) +
t_{s2}\gamma_4(t) =  i \hbar \frac{d}{dt}\gamma_3(t) \\
( t_{s1}^{*}+f_3(t)-if_4(t))\gamma_2 + t_{s2}^{*}\gamma_3(t) +(  E_{p2}+f_2(t)+E_{p2'}+\frac{q^2}{d_{2,2'}})\gamma_4(t)=i \hbar \frac{d}{dt}\gamma_4(t)
\end{eqnarray}
with  single proton movement in proximity of position based qubit generating
\begin{eqnarray}
f_1(t)= \sum_{k=1}^{Nprotons}V_1(k) \delta(t-t_1(k)),
f_2(t)= \sum_{k=1}^{Nprotons}V_2(k) \delta(t-t_1(k)),
f_3(t)= \sum_{k=1}^{Nprotons}(V_3(k)+iV_4(k))\delta(t-t_1(k)).
\end{eqnarray}
that in most simple version has the form
\begin{eqnarray}
f_1(t)= V_1 \delta(t-t_1),
f_2(t)= V_2 \delta(t-t_1),
f_3(t)= (V_3)+iV_4)\delta(t-t_1).
\end{eqnarray}
Applying operator $\int_{t_1-\delta t}^{t_1-\delta t}dt'$ to both sides of \ref{eqn} equation with very small $\delta t \rightarrow 0$ we obtain
4 algebraic relations as
\begin{eqnarray}
    \label{simple_equation6}
    i \hbar (\gamma_1(t_1^{+})-\gamma_1(t_1^{-}))=V_1\gamma_1(t_1^{+})+[V_3+iV_4]\gamma_3(t_1^{+}), \nonumber \\
    i \hbar (\gamma_2(t_1^{+})-\gamma_2(t_1^{-}))=V_2\gamma_2(t_1^{+})+ [V_3+iV_4]\gamma_4(t_1^{+}), \nonumber \\
    i \hbar (\gamma_3(t_1^{+})-\gamma_3(t_1^{-}))=[V_3-iV_4]\gamma_1(t_1^{+})+V_2\gamma_3(t_1^{+}),  \nonumber \\
    i \hbar (\gamma_4(t_1^{+})-\gamma_4(t_1^{-}))=[V_3-iV_4]\gamma_2(t_1^{+}) +V_2\gamma_4(t_1^{+}).
\end{eqnarray}
and is equivalent to the relation
\begin{eqnarray}
\frac{1}{i\hbar}
\begin{pmatrix}
i \hbar-V_1 & 0 & -[V_3+iV_4] & 0 \\
0 & i \hbar-V_1 & 0 & -[V_3+iV_4] \\
-[V_3-iV_4] & 0 & i \hbar-V_2 & 0 \\
0 & -[V_3-iV_4] & 0 & i \hbar-V_2 \\
\end{pmatrix}
\begin{pmatrix}
\gamma_1(t_1^{+})  \\
\gamma_2(t_1^{+})  \\
\gamma_3(t_1^{+})  \\
\gamma_4(t_1^{+})  \\
\end{pmatrix}=
\begin{pmatrix}
\gamma_1(t_1^{-})  \\
\gamma_2(t_1^{-})  \\
\gamma_3(t_1^{-})  \\
\gamma_4(t_1^{-})  \\
\end{pmatrix}
\end{eqnarray}
what brings the system of coupled quantum states of position based qubit after the passage of charged particle at time $t_1^{+}$ in dependence on the quantum state at time $t_1^{-}$ with condition $t_1^{+}=t_1^{-}+\Delta t$ for $\Delta t \rightarrow 0$ is  given in the algebraic form as
\begin{eqnarray}
\ket{\psi(t_1^{+})}=
\begin{pmatrix}
\gamma_1(t_1^{+})  \\
\gamma_2(t_1^{+})  \\
\gamma_3(t_1^{+})  \\
\gamma_4(t_1^{+})  \\
\end{pmatrix}=
\frac{1}{\hbar i}\int_{t_1^{-}}^{t_1^{+}} dt' \hat{H}(t') \ket{\psi(t_1^{-})}= \nonumber \\
=i\hbar
\begin{pmatrix}
i \hbar-V_1 & 0 & -[V_3+iV_4] & 0 \\
0 & i \hbar-V_1 & 0 & -[V_3+iV_4] \\
-[V_3-iV_4] & 0 & i \hbar-V_2 & 0 \\
0 & -[V_3-iV_4] & 0 & i \hbar-V_2 \\
\end{pmatrix}^{-1}
\begin{pmatrix}
\gamma_1(t_1^{-})  \\
\gamma_2(t_1^{-})  \\
\gamma_3(t_1^{-})  \\
\gamma_4(t_1^{-})  \\
\end{pmatrix}=\hat{M}
\begin{pmatrix}
\gamma_1(t_1^{-})  \\
\gamma_2(t_1^{-})  \\
\gamma_3(t_1^{-})  \\
\gamma_4(t_1^{-})  \\
\end{pmatrix}=\hat{M}\ket{\psi(t_1^{-})}= \nonumber \\ =
\begin{pmatrix}
M_{1,1} & M_{1,2} & M_{1,3} & M_{1,4} \\
M_{2,1} & M_{2,2} & M_{2,3} & M_{2,4} \\
M_{3,1} & M_{3,2} & M_{3,3} & M_{3,4} \\
M_{4,1} & M_{4,2} & M_{4,3} & M_{4,4} \\
\end{pmatrix}
\begin{pmatrix}
\gamma_1(t_1^{-})  \\
\gamma_2(t_1^{-})  \\
\gamma_3(t_1^{-})  \\
\gamma_4(t_1^{-})  \\
\end{pmatrix}= \nonumber \\ =
\begin{pmatrix}
M_{1,1} & M_{1,2} & M_{1,3} & M_{1,4} \\
M_{2,1} & M_{2,2} & M_{2,3} & M_{2,4} \\
M_{3,1} & M_{3,2} & M_{3,3} & M_{3,4} \\
M_{4,1} & M_{4,2} & M_{4,3} & M_{4,4} \\
\end{pmatrix}
[\gamma_1(t_1^{-})\ket{x_1}+\gamma_2(t_1^{-})\ket{x_2}+\gamma_3(t_1^{-})\ket{x_3}+\gamma_4(t_1^{-})\ket{x_4}]= \nonumber \\
=\hat{M}
(\ket{E_1}\bra{E_1}+\ket{E_2}\bra{E_2}+\ket{E_3}\bra{E_3}+\ket{E_4}\bra{E_4})[\gamma_1(t_1^{-})\ket{x_1}+\gamma_2(t_1^{-})\ket{x_2}+\gamma_3(t_1^{-})\ket{x_3}+\gamma_4(t_1^{-})\ket{x_4}]= \nonumber \\
=\hat{M}(\bra{E_1}\ket{x_1}\gamma_1(t_1^{-})+\bra{E_1}\ket{x_2}\gamma_2(t_1^{-})+\bra{E_1}\ket{x_3}\gamma_3(t_1^{-})+\bra{E_1}\ket{x_4}\gamma_4(t_1^{-}))\ket{E_1}+ \nonumber \\
+\hat{M}(\bra{E_2}\ket{x_1}\gamma_1(t_1^{-})+\bra{E_2}\ket{x_2}\gamma_2(t_1^{-})+\bra{E_2}\ket{x_3}\gamma_3(t_1^{-})+\bra{E_2}\ket{x_4}\gamma_4(t_1^{-}))\ket{E_2}+ \nonumber \\
+\hat{M}(\bra{E_3}\ket{x_1}\gamma_1(t_1^{-})+\bra{E_3}\ket{x_2}\gamma_2(t_1^{-})+\bra{E_3}\ket{x_3}\gamma_3(t_1^{-})+\bra{E_3}\ket{x_4}\gamma_4(t_1^{-}))\ket{E_3}+
\nonumber \\
+\hat{M}(\bra{E_4}\ket{x_1}\gamma_1(t_1^{-})+\bra{E_4}\ket{x_2}\gamma_2(t_1^{-})+\bra{E_4}\ket{x_3}\gamma_3(t_1^{-})+\bra{E_4}\ket{x_4}\gamma_4(t_1^{-}))\ket{E_4}=\nonumber \\
=\hat{M}
\begin{pmatrix}
\ket{E_1}\bra{E_1}\ket{x_1} & \ket{E_1}\bra{E_1}\ket{x_2} & \ket{E_1}\bra{E_1}\ket{x_3} & \ket{E_1}\bra{E_1}\ket{x_4} \\
\ket{E_2}\bra{E_2}\ket{x_1} & \ket{E_2}\bra{E_2}\ket{x_2} & \ket{E_2}\bra{E_2}\ket{x_3} & \ket{E_2}\bra{E_2}\ket{x_4} \\
\ket{E_3}\bra{E_3}\ket{x_1} & \ket{E_3}\bra{E_3}\ket{x_2} & \ket{E_3}\bra{E_3}\ket{x_3} & \ket{E_3}\bra{E_3}\ket{x_4} \\
\ket{E_4}\bra{E_4}\ket{x_1} & \ket{E_4}\bra{E_4}\ket{x_2} & \ket{E_4}\bra{E_4}\ket{x_3} & \ket{E_4}\bra{E_4}\ket{x_4} \\
\end{pmatrix}
\begin{pmatrix}
\gamma_1(t_1^{-})  \\
\gamma_2(t_1^{-})  \\
\gamma_3(t_1^{-})  \\
\gamma_4(t_1^{-})  \\
\end{pmatrix}= \nonumber \\
=\hat{M} \Bigg[  \ket{E_1}[
\begin{pmatrix}
\bra{E_1}\ket{x_1}, & \bra{E_1}\ket{x_2}, & \bra{E_1}\ket{x_3}, & \bra{E_1}\ket{x_4} \\
\end{pmatrix}
\begin{pmatrix}
\gamma_1(t_1^{-})  \\
\gamma_2(t_1^{-})  \\
\gamma_3(t_1^{-})  \\
\gamma_4(t_1^{-})  \\
\end{pmatrix}] \Bigg] + \nonumber \\
+\hat{M}\ket{E_2}[
\begin{pmatrix}
\bra{E_2}\Bigg[  \ket{x_1}, & \bra{E_2}\ket{x_2}, & \bra{E_2}\ket{x_3}, & \bra{E_2}\ket{x_4} \\
\end{pmatrix}
\begin{pmatrix}
\gamma_1(t_1^{-})  \\
\gamma_2(t_1^{-})  \\
\gamma_3(t_1^{-})  \\
\gamma_4(t_1^{-})  \\
\end{pmatrix}] \Bigg] + \nonumber \\
+\hat{M} \Bigg[  \ket{E_3}[
\begin{pmatrix}
\bra{E_3}\ket{x_1}, & \bra{E_3}\ket{x_2}, & \bra{E_3}\ket{x_3}, & \bra{E_3}\ket{x_4} \\
\end{pmatrix}
\begin{pmatrix}
\gamma_1(t_1^{-})  \\
\gamma_2(t_1^{-})  \\
\gamma_3(t_1^{-})  \\
\gamma_4(t_1^{-})  \\
\end{pmatrix}]\Bigg] + \nonumber \\
+\hat{M}\Bigg[ \ket{E_4}[
\begin{pmatrix}
\bra{E_4}\ket{x_1}, & \bra{E_4}\ket{x_2}, & \bra{E_4}\ket{x_3}, & \bra{E_4}\ket{x_4} \\
\end{pmatrix}
\begin{pmatrix}
\gamma_1(t_1^{-})  \\
\gamma_2(t_1^{-})  \\
\gamma_3(t_1^{-})  \\
\gamma_4(t_1^{-})  \\
\end{pmatrix}]\Bigg] = \nonumber \\ =
\begin{pmatrix}
M_{1,1} & M_{1,2} & M_{1,3} & M_{1,4} \\
M_{2,1} & M_{2,2} & M_{2,3} & M_{2,4} \\
M_{3,1} & M_{3,2} & M_{3,3} & M_{3,4} \\
M_{4,1} & M_{4,2} & M_{4,3} & M_{4,4} \\
\end{pmatrix}
\begin{pmatrix}
\bra{E_1}\ket{x_1}\gamma_1(t_1^{-})+\bra{E_1}\ket{x_2}\gamma_2(t_1^{-})+\bra{E_1}\ket{x_3}\gamma_3(t_1^{-})+\bra{E_1}\ket{x_4}\gamma_4(t_1^{-}) \nonumber \\
\bra{E_2}\ket{x_1}\gamma_1(t_1^{-})+\bra{E_2}\ket{x_2}\gamma_2(t_1^{-})+\bra{E_2}\ket{x_3}\gamma_3(t_1^{-})+\bra{E_2}\ket{x_4}\gamma_4(t_1^{-}) \nonumber \\
\bra{E_3}\ket{x_1}\gamma_1(t_1^{-})+\bra{E_3}\ket{x_2}\gamma_2(t_1^{-})+\bra{E_3}\ket{x_3}\gamma_3(t_1^{-})+\bra{E_3}\ket{x_4}\gamma_4(t_1^{-}) \nonumber \\
\bra{E_4}\ket{x_1}\gamma_1(t_1^{-})+\bra{E_4}\ket{x_2}\gamma_2(t_1^{-})+\bra{E_4}\ket{x_3}\gamma_3(t_1^{-})+\bra{E_4}\ket{x_4}\gamma_4(t_1^{-}) \nonumber \\
\end{pmatrix}
\end{eqnarray}
and matrix $\hat{M}$ can be rewritten as
\begin{eqnarray*}
\hat{M}=\frac{i\hbar}{(V_3^2+
   V_4)^2+(\hbar+i V_1) (\hbar+i V_2)} \times %%%\nonumber \\
\left(
\begin{array}{cccc}
 (V_2-i \hbar) & 0 & -(i V_4+V_3) & 0 \\
 0 & (V_2-i \hbar) & 0 & -(i V_4+V_3) \\
 -(-i V_4+V_3) & 0 & (V_1-i \hbar) & 0 \\
 0 & -(-i V_4+V_3) & 0 & (V_1-i \hbar) \\
\end{array}.
\right)= \nonumber \\
=\frac{1}{\hbar^4+\hbar^2 \left(V_1^2+V_2^2+2 \left(V_3^2+V_4^2\right)\right)+\left(-V_1 V_2+V_3^2+V_4^2\right)^2}
 \times [ (\hbar^2 (V_1 + V_2)) +i(\hbar (\hbar^2 - V_1 V_2 + V_3^2 + V_4^2))  ]\times  \nonumber \\ \times
\begin{pmatrix}
 (V_2-i \hbar) & 0 & -(i V_4+V_3) & 0 \\
 0 & (V_2-i \hbar) & 0 & -(i V_4+V_3) \\
 -(-i V_4+V_3) & 0 & (V_1-i \hbar) & 0 \\
 0 & -(-i V_4+V_3) & 0 & (V_1-i \hbar) \\
\end{pmatrix}= \nonumber \\
= \frac{1}{\hbar^4+\hbar^2 \left(V_1^2+V_2^2+2 \left(V_3^2+V_4^2\right)\right)+\left(-V_1 V_2+V_3^2+V_4^2\right)^2} \times`\nonumber \\
\times \Bigg[
\Big[ ( (\hbar^2 V_2(V_1 + V_2)) +(\hbar^2 (\hbar^2 - V_1 V_2 + V_3^2 + V_4^2)))  + i ( -\hbar^3(V_1+V_2)+V_2(\hbar (\hbar^2 - V_1 V_2 + V_3^2 + V_4^2))  )   \Big]
\begin{pmatrix}
1 & 0 & 0 & 0 \\
0 & 1 & 0 & 0 \\
0 & 0 & 0 & 0  \\
0 & 0 & 0 & 0
\end{pmatrix}+ \nonumber \\
+ \Big[ ( (\hbar^2 V_1(V_1 + V_2)) +(\hbar^2 (\hbar^2 - V_1 V_2 + V_3^2 + V_4^2)))  + i ( -\hbar^3(V_1+V_2)+V_1(\hbar (\hbar^2 - V_1 V_2 + V_3^2 + V_4^2))  )   \Big]
\begin{pmatrix}
0 & 0 & 0 & 0 \\
0 & 0 & 0 & 0 \\
0 & 0 & 1 & 0  \\
0 & 0 & 0 & 1
\end{pmatrix}+ \nonumber \\
- [ (\hbar^2 (V_1 + V_2)) +i(\hbar (\hbar^2 - V_1 V_2 + V_3^2 + V_4^2))  ] V_3
\begin{pmatrix}
0 & 0 & 1 & 0 \\
0 & 0 & 0 & 1 \\
1 & 0 & 0 & 0  \\
0 & 1 & 0 & 0
\end{pmatrix} + \nonumber \\
+ [ i(\hbar^2 (V_1 + V_2)) -(\hbar (\hbar^2 - V_1 V_2 + V_3^2 + V_4^2))  ] V_4
\begin{pmatrix}
0 & 0 & -1 & 0 \\
0 & 0 & 0 & -1 \\
+1 & 0 & 0 & 0  \\
0 & +1 & 0 & 0
\end{pmatrix} \Bigg] .
\end{eqnarray*}
One ends up with algebraic condition for the quantum state just after $t_1=t_1^{+}$ so we have the relation between quantum state at $t_1^{+}$ and $t_1^{-}$ expressed in the algebraic way as
\begin{eqnarray*}
c_{E1m}e^{i \phi_{E1m}}e^{\frac{E_1 t_1}{\hbar}}\ket{E_1}+c_{E2m}e^{i \phi_{E2m}}e^{\frac{E_2 t_1}{\hbar}}\ket{E_2}+c_{E3m}e^{\frac{E_3 t_1}{\hbar}}e^{i \phi_{E3m}}\ket{E_3}+c_{E4m}e^{i \phi_{E4}}e^{\frac{E_4 t_1}{\hbar}}\ket{E_4}=\nonumber \\
=
\begin{pmatrix}
\gamma_1(t_1^{+})  \\
\gamma_2(t_1^{+})  \\
\gamma_3(t_1^{+})  \\
\gamma_4(t_1^{+})  \\
\end{pmatrix}
= \frac{1}{\hbar^4+\hbar^2 \left(V_1^2+V_2^2+2 \left(V_3^2+V_4^2\right)\right)+\left(-V_1 V_2+V_3^2+V_4^2\right)^2} \times
\nonumber \\ \Bigg[
\Big[ ( (\hbar^2 V_2(V_1 + V_2)) +(\hbar^2 (\hbar^2 - V_1 V_2 + V_3^2 + V_4^2)))  + i ( -\hbar^3(V_1+V_2)+V_2(\hbar (\hbar^2 - V_1 V_2 + V_3^2 + V_4^2))  )   \Big]
\begin{pmatrix}
1 & 0 & 0 & 0 \\
0 & 1 & 0 & 0 \\
0 & 0 & 0 & 0  \\
0 & 0 & 0 & 0
\end{pmatrix}+ \nonumber \\
+ \Big[ ( (\hbar^2 V_1(V_1 + V_2)) +(\hbar^2 (\hbar^2 - V_1 V_2 + V_3^2 + V_4^2)))  + i ( -\hbar^3(V_1+V_2)+V_1(\hbar (\hbar^2 - V_1 V_2 + V_3^2 + V_4^2))  )   \Big]
\begin{pmatrix}
0 & 0 & 0 & 0 \\
0 & 0 & 0 & 0 \\
0 & 0 & 1 & 0  \\
0 & 0 & 0 & 1
\end{pmatrix}+ \nonumber \\
- [ (\hbar^2 (V_1 + V_2)) +i(\hbar (\hbar^2 - V_1 V_2 + V_3^2 + V_4^2))  ] V_3
\begin{pmatrix}
0 & 0 & 1 & 0 \\
0 & 0 & 0 & 1 \\
1 & 0 & 0 & 0  \\
0 & 1 & 0 & 0
\end{pmatrix} + \nonumber \\
+ [ i(\hbar^2 (V_1 + V_2)) -(\hbar (\hbar^2 - V_1 V_2 + V_3^2 + V_4^2))  ] V_4
\begin{pmatrix}
0 & 0 & -1 & 0 \\
0 & 0 & 0 & -1 \\
+1 & 0 & 0 & 0  \\
0 & +1 & 0 & 0
\end{pmatrix} \Bigg]
\begin{pmatrix}
\gamma_1(t_1^{-})  \\
\gamma_2(t_1^{-})  \\
\gamma_3(t_1^{-})  \\
\gamma_4(t_1^{-})  \\
\end{pmatrix}= \nonumber \\
= \frac{1}{\hbar^4+\hbar^2 \left(V_1^2+V_2^2+2 \left(V_3^2+V_4^2\right)\right)+\left(-V_1 V_2+V_3^2+V_4^2\right)^2} \times
\nonumber \\
\Bigg[
\begin{pmatrix}
+\Big[ ( (\hbar^2 V_2(V_1 + V_2)) +(\hbar^2 (\hbar^2 - V_1 V_2 + V_3^2 + V_4^2)))  + i ( -\hbar^3(V_1+V_2)+V_2(\hbar (\hbar^2 - V_1 V_2 + V_3^2 + V_4^2))  )   \Big]\gamma_1(t_1^{-}) \\
+\Big[ ( (\hbar^2 V_2(V_1 + V_2)) +(\hbar^2 (\hbar^2 - V_1 V_2 + V_3^2 + V_4^2)))  + i ( -\hbar^3(V_1+V_2)+V_2(\hbar (\hbar^2 - V_1 V_2 + V_3^2 + V_4^2))  )   \Big]\gamma_2(t_1^{-}) \\
+ \Big[ ( (\hbar^2 V_1(V_1 + V_2)) +(\hbar^2 (\hbar^2 - V_1 V_2 + V_3^2 + V_4^2)))  + i ( -\hbar^3(V_1+V_2)+V_1(\hbar (\hbar^2 - V_1 V_2 + V_3^2 + V_4^2))  )   \Big]\gamma_3(t_1^{-})  \\
+ \Big[ ( (\hbar^2 V_1(V_1 + V_2)) +(\hbar^2 (\hbar^2 - V_1 V_2 + V_3^2 + V_4^2)))  + i ( -\hbar^3(V_1+V_2)+V_1(\hbar (\hbar^2 - V_1 V_2 + V_3^2 + V_4^2))  )   \Big]\gamma_4(t_1^{-}) \\
\end{pmatrix} + \nonumber \\
\begin{pmatrix}
\Big[ -[(\hbar^2 (V_1 + V_2))V_3+(\hbar (\hbar^2 - V_1 V_2 + V_3^2 + V_4^2))V_4] - i [(\hbar^2 (V_1 + V_2))V_4+(\hbar (\hbar^2 - V_1 V_2 + V_3^2 + V_4^2))V_3]  \Big]\gamma_3(t_1^{-}) \\
\Big[ -[\hbar^2 (V_1 + V_2))V_3+(\hbar (\hbar^2 - V_1 V_2 + V_3^2 + V_4^2))V_4] - i [(\hbar^2 (V_1 + V_2))V_4+(\hbar (\hbar^2 - V_1 V_2 + V_3^2 + V_4^2))V_3]  \Big]\gamma_4(t_1^{-}) \\
\Big[ -[(\hbar (\hbar^2 - V_1 V_2 + V_3^2 + V_4^2))V_4 +(\hbar^2 (V_1 + V_2))V_3 ] + i [(\hbar^2 (V_1 + V_2))V_4-(\hbar (\hbar^2 - V_1 V_2 + V_3^2 + V_4^2))V_3]  \Big]\gamma_1(t_1^{-}) \\
\Big[ -[(\hbar (\hbar^2 - V_1 V_2 + V_3^2 + V_4^2))V_4 +(\hbar^2 (V_1 + V_2))V_3 ] + i [(\hbar^2 (V_1 + V_2))V_4-(\hbar (\hbar^2 - V_1 V_2 + V_3^2 + V_4^2))V_3]  \Big]\gamma_2(t_1^{-}) \\
\end{pmatrix}\Bigg]= \nonumber \\
=\hat{M}[c_{E1}e^{i \phi_{E1}}e^{\frac{E_1 t_1^{-}}{\hbar}}\ket{E_1}+c_{E2}e^{i \phi_{E2}}e^{\frac{E_2 t_1^{-}}{\hbar}}\ket{E_2}+c_{E3}e^{\frac{E_3 t_1^{-}}{\hbar}}e^{i \phi_{E3}}\ket{E_3}+c_{E4}e^{i \phi_{E4}}e^{\frac{E_4 t_1^{-}}{\hbar}}\ket{E_4}].
\end{eqnarray*}
Last equation implies 4 relations
\begin{eqnarray}
c_{E1m}e^{i \phi_{E1m}}e^{\frac{E_1 t_1}{\hbar}}=\bra{E_1}\hat{M}[c_{E1}e^{i \phi_{E1}}e^{\frac{E_1 t_1^{-}}{\hbar}}\ket{E_1}+c_{E2}e^{i \phi_{E2}}e^{\frac{E_2 t_1^{-}}{\hbar}}\ket{E_2}+c_{E3}e^{\frac{E_3 t_1^{-}}{\hbar}}e^{i \phi_{E3}}\ket{E_3}+c_{E4}e^{i \phi_{E4}}e^{\frac{E_4 t_1^{-}}{\hbar}}\ket{E_4}]= \nonumber \\ =\bra{E_1}\hat{M}
\begin{pmatrix}
\gamma_1(t_1^{-})  \\
\gamma_2(t_1^{-})  \\
\gamma_3(t_1^{-})  \\
\gamma_4(t_1^{-})  \\
\end{pmatrix}
, \nonumber \\
\end{eqnarray}
\begin{eqnarray}
c_{E2m}e^{i \phi_{E2m}}e^{\frac{E_2 t_1}{\hbar}}=\bra{E_2}\hat{M}[c_{E1}e^{i \phi_{E1}}e^{\frac{E_1 t_1^{-}}{\hbar}}\ket{E_1}+c_{E2}e^{i \phi_{E2}}e^{\frac{E_2 t_1^{-}}{\hbar}}\ket{E_2}+c_{E3}e^{\frac{E_3 t_1^{-}}{\hbar}}e^{i \phi_{E3}}\ket{E_3}+c_{E4}e^{i \phi_{E4}}e^{\frac{E_4 t_1^{-}}{\hbar}}\ket{E_4}]= \nonumber \\
=\bra{E_2}\hat{M}
\begin{pmatrix}
\gamma_1(t_1^{-})  \\
\gamma_2(t_1^{-})  \\
\gamma_3(t_1^{-})  \\
\gamma_4(t_1^{-})  \\
\end{pmatrix}
\end{eqnarray}
\begin{eqnarray}
c_{E3m}e^{i \phi_{E3m}}e^{\frac{E_3 t_1}{\hbar}}=\bra{E_3}\hat{M}[c_{E1}e^{i \phi_{E1}}e^{\frac{E_1 t_1^{-}}{\hbar}}\ket{E_1}+c_{E2}e^{i \phi_{E2}}e^{\frac{E_2 t_1^{-}}{\hbar}}\ket{E_2}+c_{E3}e^{\frac{E_3 t_1^{-}}{\hbar}}e^{i \phi_{E3}}\ket{E_3}+c_{E4}e^{i \phi_{E4}}e^{\frac{E_4 t_1^{-}}{\hbar}}\ket{E_4}]= \nonumber \\
=\bra{E_3}\hat{M}
\begin{pmatrix}
\gamma_1(t_1^{-})  \\
\gamma_2(t_1^{-})  \\
\gamma_3(t_1^{-})  \\
\gamma_4(t_1^{-})  \\
\end{pmatrix}, \nonumber \\
\end{eqnarray}
\begin{eqnarray}
c_{E4m}e^{i \phi_{E4m}}e^{\frac{E_4 t_1}{\hbar}}=\bra{E_4}\hat{M}[c_{E1}e^{i \phi_{E1}}e^{\frac{E_1 t_1^{-}}{\hbar}}\ket{E_1}+c_{E2}e^{i \phi_{E2}}e^{\frac{E_2 t_1^{-}}{\hbar}}\ket{E_2}+c_{E3}e^{\frac{E_3 t_1^{-}}{\hbar}}e^{i \phi_{E3}}\ket{E_3}+c_{E4}e^{i \phi_{E4}}e^{\frac{E_4 t_1^{-}}{\hbar}}\ket{E_4}]=\nonumber \\ =
\bra{E_4}\hat{M}
\begin{pmatrix}
\gamma_1(t_1^{-})  \\
\gamma_2(t_1^{-})  \\
\gamma_3(t_1^{-})  \\
\gamma_4(t_1^{-})  \\
\end{pmatrix}
\end{eqnarray}
The probability of occupancy of eigenergy $E_1$, $E_2$, $E_3$ and $E_4$ for interacting qubit system after measurement of charged particle passage is given by $|c_{E1m}|^2,$ $|c_{E2m}|^2$,$|c_{E3m}|^2$,$|c_{E4m}|^2$ and phase imprint of given eigenenergy state is given by factors $e^{i \phi_{E1m}}$
$ e^{i \phi_{2m}}$,$e^{i \phi_{E3m}}$,$e^{i \phi_{E4m}}$.
Let us consider the case of two symmetric qubits whose system is depicted at Fig. 2.

\begin{figure}
    \centering
    \includegraphics[scale=0.8]{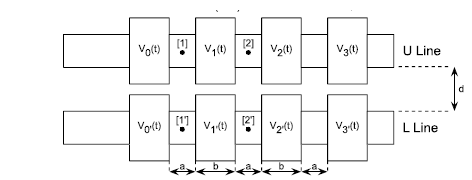}
    \includegraphics[scale=0.8]{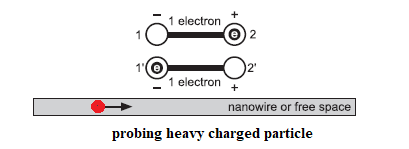}
\caption{Concept of two electrostatically interacting qubits and charged particle}
\end{figure}

%%%symmetric_qubits
We have the following Hamiltonian
\begin{eqnarray}
\hat{H}=
\left(
\begin{array}{cccc}
 E_{p1}+E_{p1'}+E_{c(1,1')} & i t_{2si}+t_{2sr} & i t_{1si}+t_{1sr} & 0 \\
 t_{2sr}-i t_{2si} & E_{p1}+E_{p2'}+E_{c(1,2')}  & 0 & i t_{1si}+t_{1sr} \\
 t_{1sr}-i t_{1si} & 0 & E_{p2}+E_{p1'}+E_{c(2,1')} & i t_{2si}+t_{2sr} \\
 0 & t_{1sr}-i t_{1si} & t_{2sr}-i t_{2si} & E_{p2}+E_{p2'}+E_{c(2,2')}  \\
\end{array}
\right)
\end{eqnarray}
that can be simplified by placement of 2 qubit system in the geometrical configuration giving the following electrostatic energies as $E_{c(1,1')}=E_{c(2,2')}=E_{c1}=\frac{q^2}{d}$ and $E_{c(2,1')}=E_{c(1,2')}=E_{c2}=\frac{q^2}{\sqrt{d^2+(a+b)^2}}$.
We set $E_{p2}=E_{p2'}=E_{p1}=E_{p1'}=E_p$ and we introduce $ E_{pp}=2E_p+\frac{q^2}{d}$ and
$ E_{pp1}=2E_p+\frac{q^2}{\sqrt{d^2+(a+b)^2}}$.

First simplfied Hamiltonian of 2 qubit system has the structure
\begin{eqnarray}
\hat{H}=
\left(
\begin{array}{cccc}
 E_{pp} & t_{s} & t_{s} & 0 \\
 t_{s} & E_{pp1} & 0 & t_{s} \\
 t_{s} & 0 & E_{pp1} & t_{s} \\
 0 & t_{s} & t_{s} & E_{pp} \\
\end{array}
\right)
\end{eqnarray}
and we assume that $t_s \in R$.
The eigenstates of simplified Hamiltonian are
\begin{eqnarray}
\ket{E_1}=
\begin{pmatrix}
-\frac{1}{\sqrt{2}} \\
0 \\ 0 \\ +\frac{1}{\sqrt{2}}
\end{pmatrix},
\ket{E_2}=
\begin{pmatrix}
0\\ -\frac{1}{\sqrt{2}} \\ \frac{1}{\sqrt{2}} \\ 0
\end{pmatrix}
\end{eqnarray}
and
\begin{eqnarray}
\ket{E_3}=
\begin{pmatrix}
 \frac{\sqrt{(E_{pp}-E_{pp1})^2+16 t_{s}^2}-E_{pp}+E_{pp1}}{\sqrt{2 \left(\sqrt{(E_{pp}-E_{pp1})^2+16 t_{s}^2}-E_{pp}+E_{pp1}\right)^2+32
   t_{s}^2}}, \\
-\frac{4 t_{s}}{\sqrt{2 \left(\sqrt{(E_{pp}-E_{pp1})^2+16 t_{s}^2}-E_{pp}+E_{pp1}\right)^2+32 t_{s}^2}}, \\
-\frac{4 t_{s}}{\sqrt{2
   \left(\sqrt{(E_{pp}-E_{pp1})^2+16 t_{s}^2}-E_{pp}+E_{pp1}\right)^2+32 t_{s}^2}}, \\
\frac{\sqrt{(E_{pp}-E_{pp1})^2+16 t_{s}^2}-E_{pp}+E_{pp1}}{\sqrt{2
   \left(\sqrt{(E_{pp}- E_{pp1})^2+16 t_{s}^2}-E_{pp}+E_{pp1}\right)^2+32 t_{s}^2}}
\end{pmatrix}
\end{eqnarray}
and
\begin{eqnarray}
\ket{E_4}=
\begin{pmatrix}
 \frac{\sqrt{(E_{pp}-E_{pp1})^2+16 t_{s}^2}+E_{pp}-E_{pp1}}{\sqrt{2 \left(\sqrt{(E_{pp}-E_{pp1})^2+16 t_{s}^2}+E_{pp}-E_{pp1}\right)^2+32
   \text{ts}^2}}, \\
\frac{4 t_{s}}{\sqrt{2 \left(\sqrt{(E_{pp}-E_{pp1})^2+16 _{ts}^2}+E_{pp}-E_{pp1}\right)^2+32 t_{s}^2}}, \\
\frac{4 t_{s}}{\sqrt{2
   \left(\sqrt{(E_{pp}-E_{pp1})^2+16 t_{s}^2}+E_{pp}-E_{pp1}\right)^2+32 t_{s}^2}} \\
\frac{\sqrt{(E_{pp}-E_{pp1})^2+16 t_{s}^2}+E_{pp}-E_{pp1}}{\sqrt{2
   \left(\sqrt{(E_{pp}-E_{pp1})^2+16 t_{s}^2}+E_{pp}-E_{pp1}\right)^2+32 t_{s}^2}}
\end{pmatrix}
\end{eqnarray}
We obtain
%\newline
%\tiny
\begin{eqnarray*}
e^{i\Phi_{E1m}}c_{E1m}=(i \hbar e^{i E_{pp} t_1} %\nonumber \\
 ( c_{E1} e^{i (\phi_{E1}-E_{pp}t_1)}(-2 i \hbar+V_1+V_2) \frac{1}{\sqrt{2}} \nonumber \\
+\frac{1}{2} [ \sqrt{2} c_{E2}e^{i (\phi_{E2}-E_{pp1}t_1)} (2 V_{3}+i (\hbar-1) V_{4}) \nonumber \\
%\end{eqnarray}
%\begin{eqnarray}
-[c_{E3}
   \exp \left(\frac{1}{2} i \left(t_1 \sqrt{(E_{pp}-E_{pp1})^2+16 t_{s}^2}-t_1
   (E_{pp}+E_{pp1})+2 \phi_{E3}\right)\right) \nonumber \\
\end{eqnarray*}
%\newline $ $
\begin{eqnarray*}
[E_{pp}^2 (V_{2}-V_{1})+4
   t_{s} \left(4 t_{s} (V_{2}-V_{1})+i (\hbar+1) V_{4}
   \sqrt{(E_{pp}-E_{pp1})^2+16 t_{s}^2}\right)+ \nonumber \\
+E_{pp} (V_{1}-V_{2})
   \left(\sqrt{(E_{pp}-E_{pp1})^2+16 t_{s}^2}+2 E_{pp1}\right)+E_{pp1}
   (V_{2}-V_{1}) \sqrt{(E_{pp}-E_{pp1})^2+16 t_{s}^2}+E_{pp1}^2
   (V_{2}-V_{1})]] \times
\end{eqnarray*}
%\newline
\begin{eqnarray*}
\frac{1}{\sqrt{(E_{pp}-E_{pp1})^2+16 t_{s}^2} \sqrt{16
   t_{s}^2-(E_{pp}-E_{pp1}) \left(\sqrt{(E_{pp}-E_{pp1})^2+16
   t_{s}^2}-E_{pp}+E_{pp1}\right)}}+ \nonumber \\
%%\end{eqnarray}
%%\begin{eqnarray}
+[c_{E4} e^{i \phi_{E4}-\frac{1}{2} i t_1 \left(\sqrt{(E_{pp}-E_{pp1})^2+16 t_{s}^2}+E_{pp}+E_{pp1}\right)} \times
\end{eqnarray*}
\begin{eqnarray}
 [E_{pp}^2 (V_{1}-V_{2})+4 t_{s} \left(4 t_{s} (V_{1}-V_{2})+i
   (\hbar+1) V_{4} \sqrt{(E_{pp}-E_{pp1})^2+16 t_{s}^2}\right)+ \nonumber \\
E_{pp}(V_{1}-V_{2}) \left(\sqrt{(E_{pp}-E_{pp1})^2+16 t_{s}^2}-2
   E_{pp1}\right)+ \nonumber \\
%\newline
%\begin{eqnarray}
+E_{pp1} (V_{2}-V_{1})
 \sqrt{(E_{pp}-E_{pp1})^2+16t_{s}^2}+E_{pp1}^2 (V_{1}-V_{2})]]\times \nonumber \\
\frac{1}{\sqrt{(E_{pp}-E_{pp1})^2+16 t_{s}^2}
\sqrt{(E_{pp}-E_{pp1}) \left(\sqrt{(E_{pp}-E_{pp1})^2+16
   t_{s}^2}+E_{pp}-E_{pp1}\right)+16 t_{s}^2}}]]) \nonumber \\
\times \frac{1}{\sqrt{2} \left(\hbar^2+i
   \hbar (V_{1}+V_{2}+V_{4} (V_{3}-i V_{4}))-V_{1} V_{2}+V_{3}
   (V_{3}-i V_{4})\right)}
\end{eqnarray}
\normalsize
The probability of occurrence of quantum state in energy $E_1$ after weak measurement is therefore equal to
$|e^{i\Phi_{E1m}}c_{E1m}|^2$.
and
\small
\newline $  $
\begin{eqnarray*}
  e^{i\Phi_{E2m}}c_{E2m}=
[i \hbar e^{i E_{pp1} t_1}[ \frac{1}{2} [\sqrt{2} c_{E1} e^{i
   (\phi_{E1}-E_{pp} t_1)} (2 V_{3}+i (\hbar-1) V_{4})+ \nonumber \\
   +[c_{E3}e^{\left(\frac{1}{2} i \left(t_1 \sqrt{(E_{pp}-E_{pp1})^2+16 t_s^2}-t_1
   (E_{pp}+E_{pp1})+2 \phi_{E3}\right)\right)} \nonumber \\
   (-i (\hbar+1)V_{4}
   (E_{pp}-E_{pp1}) \left(\sqrt{(E_{pp}-E_{pp1})^2+16
   t_s^2}-E_{pp}+E_{pp1}\right)+ \nonumber \\
   +4t_{s} (V_{2}-V_{1})
   \sqrt{(E_{pp}-E_{pp1})^2+16t_{s}^2}+16 i (\hbar+1)t_{s}^2
   V_{4})] \nonumber \\
   \times \frac{1}{\sqrt{(E_{pp}-E_{pp1})^2+16t_{s}^2} \sqrt{16
   t_{s}^2-(E_{pp}-E_{pp1}) \left(\sqrt{(E_{pp}-E_{pp1})^2+16
   t_{s}^2}-E_{pp}+E_{pp1}\right)}}+ \nonumber \\
%   +[c_{E4} e^{i \phi_{E4}-\frac{1}{2} i
%   t_1 \left(\sqrt{(E_{pp}-E_{pp1})^2+16t_{s}^2}+E_{pp}+E_{pp1}\right)} \left(i
%   (\hbar+1) V_{4} (E_{pp}-E_{pp1}) \left(\sqrt{(E_{pp}-E_{pp1})^2+16
%   t_{s}^2}+E_{pp}-E_{pp1}\right)+4 t_{s} (V_{1}-V_{2})
%   \sqrt{(E_{pp}-E_{pp1})^2+16t_{s}^2}+16 i (\hbar+1)t_{s}^2
%   V_{4}\right)] \nonumber \\
%%%%\end{eqnarray}
%$e^{i\Phi_{E2m}}c_{E2m}=
%[i \hbar e^{i E_{pp1} t_1}$ \newline
%%% $[ \frac{1}{2} [\sqrt{2} c_{E1} e^{i
%%%   (\phi_{E1}-E_{pp} t_1)} (2 V_{3}+i (\hbar-1) V_{4})+$
    %%%\newline
   %%% $+[c_{E3}e^{\left(\frac{1}{2} i \left(t_1 \sqrt{(E_{pp}-E_{pp1})^2+16 t_s^2}-t_1
   %%%(E_{pp}+E_{pp1})+2 \phi_{E3}\right)\right)}$ \newline
   %%$
   %%\left(-i (\hbar+1)V_{4}
   %%(E_{pp}-E_{pp1}) \left(\sqrt{(E_{pp}-E_{pp1})^2+16
   %%t_s^2}-E_{pp}+E_{pp1}\right)+4t_{s} (V_{2}-V_{1})
   %%\sqrt{(E_{pp}-E_{pp1})^2+16t_{s}^2}+16 i (\hbar+1)t_{s}^2
   %%V_{4}\right)]$
   %%%\newline
   %%$\times \frac{1}{\sqrt{(E_{pp}-E_{pp1})^2+16t_{s}^2} \sqrt{16
   %%t_{s}^2-(E_{pp}-E_{pp1}) \left(\sqrt{(E_{pp}-E_{pp1})^2+16
   %%t_{s}^2}-E_{pp}+E_{pp1}\right)}}+$ \newline
   %%%%$
   +(c_{E4} e^{i \phi_{E4}-\frac{1}{2} i
   t_1 \left(\sqrt{(E_{pp}-E_{pp1})^2+16t_{s}^2}+E_{pp}+E_{pp1}\right)} \times \nonumber \\
 %%%%%%  \begin{eqnarray}
  % +(c_{E4} e^{i \phi_{E4}-\frac{1}{2} i
  % t_1 \left(\sqrt{(E_{pp}-E_{pp1})^2+16t_{s}^2}+E_{pp}+E_{pp1}\right)} %\nonumber \\
 \times  [ i
   (\hbar+1) V_{4} (E_{pp}-E_{pp1}) ( \sqrt{(E_{pp}-E_{pp1})^2+16
   t_{s}^2}+E_{pp}-E_{pp1})+4 t_{s} (V_{1}-V_{2})
   \sqrt{(E_{pp}-E_{pp1})^2+16t_{s}^2}+16 i (\hbar+1)t_{s}^2
   V_{4}]) \nonumber \\ %$ \newline
   %\end{eqnarray}
    \times \frac{1}{\sqrt{(E_{pp}-E_{pp1})^2+16 t_{s}^2} \sqrt{(E_{pp}-E_{pp1})
   \left(\sqrt{(E_{pp}-E_{pp1})^2+16t_{s}^2}+E_{pp}-E_{pp1}\right)+16
   t_{s}^2}}]+\frac{c_{E2} e^{i (\phi_{E2}-E_{pp1} t_1)} (-2 i
   \hbar+V_{1}+V_{2})}{\sqrt{2}}]  ] \times  \nonumber \\
  % \end{eqnarray}
%%%%%\newline
 \times \frac{1}{\sqrt{2} \left(\hbar^2+i \hbar
   (V_1+V_{2}+V_{4} (V_{3}-i V_{4}))-V_{1} V_{2}+V_{3} (V_{3}-i
   V_{4})\right)} \nonumber \\
    \end{eqnarray*}
   \normalsize
   \newline
and
\newline $ $
\begin{eqnarray*}
  e^{i\Phi_{E3m}}c_{E3m}=
-[i \hbar e^{\frac{1}{2}it_1 \left(-\sqrt{(E_{pp}-E_{pp1})^2+16
   t_{s}^2}+E_{pp}+E_{pp1}\right)} \nonumber \\
   (\frac{1}{2} (\sqrt{2} e^{-it_1
   (E_{pp}+E_{pp1})}(c_{E2} e^{i (E_{pp}t_1+\phi_{E2})} \left(4
   t_{s} (V_{1}-V_{2})+i (\hbar+1)V_{4} \left(\sqrt{(E_{pp}-E_{pp1})^2+16
   t_{s}^2}-E_{pp}+E_{pp1}\right)\right) \nonumber \\
   -c_{E1} e^{i (E_{pp1}
   t_1+\phi_{E1})} ((V_{1}-V_{2}) (\sqrt{(E_{pp}-E_{pp1})^2+16
   t_{s}^2}-E_{pp}+E_{pp1})+4 i (\hbar+1) t_{s}V_{4}))+
\end{eqnarray*}
%$e^{i\Phi_{E3m}}c_{E3m}=
%-[i \hbar e^{\frac{1}{2}it_1 \left(-\sqrt{(E_{pp}-E_{pp1})^2+16
%   t_{s}^2}+E_{pp}+E_{pp1}\right)}$ \newline
   \small
    %$$
   %\newline
  % \begin{eqnarray
% $  -c_{E1} e^{i (E_{pp1}
%   t_1+\phi_{E1})} ((V_{1}-V_{2}) (\sqrt{(E_{pp}-E_{pp1})^2+16
%   t_{s}^2}-E_{pp}+E_{pp1})+4 i (\hbar+1) t_{s}V_{4}))+ $ \newline
$   +[2 i
   c_{E3} \sqrt{16 t_{s}^2-(E_{pp}-E_{pp1}) \left(\sqrt{(E_{pp}-E_{pp1})^2+16
   \text{ts}^2}-E_{pp}+E_{pp1}\right)} \exp \left(\frac{1}{2} i \left(t_1
   \sqrt{(E_{pp}-E_{pp1})^2+16t_{s}^2}-t_1 (E_{pp}+E_{pp1})+2
   \phi_{E3}\right)\right)$ %\nonumber \\ %\newline
   \begin{eqnarray}
   \left(2 \hbar \left(\sqrt{(E_{pp}-E_{pp1})^2+16t_{s}^2}-2
   t_{s} V_{4}\right)+i (V_{1}+V_{2}) \sqrt{(E_{pp}-E_{pp1})^2+16t_{s}^2}+4
   t_{s} (V_{4}+2 iV_{3})\right)]\frac{1}{\sqrt{(E_{pp}-E_{pp1})^2+16
   t_{s}^2}}) \nonumber \\ %\newline
     %\begin{eqnarray}
   -\frac{4 c_{E4} t_{s} (E_{pp}-E_{pp1}) (2 V_{3}+i (\hbar-1)
   V_{4}) \exp \left(\frac{1}{2} i \left(2 \phi_{E4}-t_1
   \left(\sqrt{(E_{pp}-E_{pp1})^2+16
   t_{s}^2}+E_{pp}+E_{pp1}\right)\right)\right)}{\sqrt{(E_{pp}-E_{pp1})
   \left(\sqrt{(E_{pp}-E_{pp1})^2+16 t_{s}^2}+E_{pp}-E_{pp1}\right)+16
   t_{s}^2}})] \times  \nonumber \\
   %\begin{eqnarray}
    \times \frac{1}{2 \sqrt{16 t_{s}^2-(E_{pp}-E_{pp1})
   \left(\sqrt{(E_{pp}-E_{pp1})^2+16t_{s}^2}-E_{pp}+E_{pp1}\right)}
   \left(\hbar^2+i \hbar (V_{1}+V_{2}+V_{4} (V_{3}-i V_{4}))-V_{1}
   V_{2}+V_{3} (V_{3}-i V_{4})\right)}
   \end{eqnarray}
\newline
\normalsize
and
%%%%\tiny
\small
%\newline
\begin{eqnarray*}
%$
e^{i\Phi_{E4m}}c_{E4m}=[i e^{\frac{1}{2} i t_1 \left(E_{pp}+E_{pp1}+\sqrt{(E_{pp}-E_{pp1})^2+16t_{s}^2}\right)} \hbar %%%\nonumber \\
(\left(E_{pp}-E_{pp1}+\sqrt{(E_{pp}-E_{pp1})^2+16t_{s}^2}\right) \times %%%%\nonumber \\
((\frac{e^{i (\phi_{E1}-E_{pp} t_1)}c_{E1}}{\sqrt{2}}+ \nonumber \\
%%%\end{eqnarray}
   %$
   %\newline
  % $ (\left(E_{pp}-E_{pp1}+\sqrt{(E_{pp}-E_{pp1})^2+16
  % t_{s}^2}\right) \times$
  % \newline
 %%  $ ((\frac{e^{i (\phi_{E1}-E_{pp} t_1)}
 %%  c_{E1}}{\sqrt{2}}+ $ \newline
   %%%$
   +[e^{-i (E_{pp}+E_{pp1})t_1} (e^{\frac{1}{2} i \left(2
   \phi_{E3}+t_1 \left(E_{pp}+E_{pp1}+\sqrt{(E_{pp}-E_{pp1})^2+16
   t_{s}^2}\right)\right)} \sqrt{16 t_{s}^2-(E_{pp}-E_{pp1})
   \left(-E_{pp}+E_{pp1}+\sqrt{(E_{pp}-E_{pp1})^2+16 t_{s}^2}\right)}
    c_{E3}+ \nonumber \\  %$ \newline
    %%\begin{eqnarray*}
    +c_{E4} e^{\frac{1}{2} i \left(2 \phi_{E4}+t_1
   \left(E_{pp}+E_{pp1}-\sqrt{(E_{pp}-E_{pp1})^2+16 t_{s}^2}\right)\right)}
   \sqrt{16
    t_{s}^2+(E_{pp}-E_{pp1}) \left(E_{pp}-E_{pp1}+\sqrt{(E_{pp}-E_{pp1})^2+16
    t_{s}^2}\right)})] \nonumber \\
    %%$ \newline $
    \times \frac{1}{2 \sqrt{(E_{pp}-E_{pp1})^2+16 t_{s}^2}})(V_{1}-i\hbar) \nonumber \\
   %% \end{eqnarray*}
   %% \newline $(V_{1}-i\hbar)+$ \newline
   %% \begin{eqnarray*}
   %%$
   +(-\frac{e^{i (\phi_{E2}-E_{pp1} t_1)} c_{E2}}{\sqrt{2}}+\frac{2
    c_{E4} e^{i \phi_{E4}-\frac{1}{2} i t_1
   \left(E_{pp}+E_{pp1}+\sqrt{(E_{pp}-E_{pp1})^2+16t_{s}^2}\right)} t_{s}}{\sqrt{16
    t_{s}^2+(E_{pp}-E_{pp1}) \left(E_{pp}-E_{pp1}+\sqrt{(E_{pp}-E_{pp1})^2+16
   t_{s}^2}\right)}} \nonumber \\
   -\frac{2 c_{E3} e^{\frac{1}{2} i \left(2 \phi_{E3}-(E_{pp}+E_{pp1})
   t_1+t_1 \sqrt{(E_{pp}-E_{pp1})^2+16 t_{s}^2}\right)} t_{s}}{\sqrt{16
   t_{s}^2-(E_{pp}-E_{pp1}) \left(-E_{pp}+E_{pp1}+\sqrt{(E_{pp}-E_{pp1})^2+16
   t_{s}^2}\right)}}) \nonumber \\
   %%$
    %%$
    (i V_{4}-V_{3}))+ \nonumber \\
    %%$ \newline
    %%\begin{eqnarray*}
    %$
    +4 t_{s} ((\frac{e^{i
   (\phi_{E2}-E_{pp1}t_1 )} c_{E2}}{\sqrt{2}}+\frac{2 c_{E4} e^{i
   \phi_{E4}-\frac{1}{2} i t_1 \left(E_{pp}+E_{pp1}+\sqrt{(E_{pp}-E_{pp1})^2+16
   t_{s}^2}\right)} t_{s}}{\sqrt{16 t_{s}^2+(E_{pp}-E_{pp1})
   \left(E_{pp}-E_{pp1}+\sqrt{(E_{pp}-E_{pp1})^2+16t_{s}^2}\right)}} \nonumber \\
   -\frac{2c_{E3}
   e^{\frac{1}{2} i \left(2 \phi_{E3}-(E_{pp}+E_{pp1}) t_1+t_1
   \sqrt{(E_{pp}-E_{pp1})^2+16 t_{s}^2}\right)} t_{s}}{\sqrt{16
   t_{s}^2-(E_{pp}-E_{pp1}) \left(-E_{pp}+E_{pp1}+\sqrt{(E_{pp}-E_{pp1})^2+16
   t_{s}^2}\right)}}) (V_1-i \hbar)+ \nonumber \\
   %$
   %\newline
   %\begin{eqnarray*}
   +((e^{-i (E_{pp}+E_{pp1})
   t_1} (e^{\frac{1}{2} i \left(2 \phi_{E3}+t_1
   \left(E_{pp}+E_{pp1}+\sqrt{(E_{pp}-E_{pp1})^2+16t_{s}^2}\right)\right)} \nonumber \\
   %\newline
   %%\begin{eqnarray*}
   \sqrt{16
   t_{s}^2-(E_{pp}-E_{pp1}) \left(-E_{pp}+E_{pp1}+\sqrt{(E_{pp}-E_{pp1})^2+16
   t_{s}^2}\right)} c_{E3}+ \nonumber \\
   %\newline
   %\begin{eqnarray*}
   %$
   +c_{E4} e^{\frac{1}{2} i \left(2 \phi_{E4}+t_1
   \left(E_{pp}+E_{pp1}-\sqrt{(E_{pp}-E_{pp1})^2+16t_{s}^2}\right)\right)} \sqrt{16
   t_{s}^2+(E_{pp}-E_{pp1}) \left(E_{pp}-E_{pp1}+\sqrt{(E_{pp}-E_{pp1})^2+16
   t_{s}^2}\right)})) \nonumber \\
   %$
   %\newline
   %\newline
   %\newline
   %\small
   % \begin{eqnarray*}
    \times \frac{1}{2 \sqrt{(E_{pp}-E_{pp1})^2+16t_{s}^2}}-\frac{c_{E1} e^{i
   (\phi_{E1}-E_{pp} t_1)}}{\sqrt{2}}) \nonumber \\
    %\end{eqnarray*}
   %\newline
   %$
   (i
   V_{4}-V_{3}))+\left(E_{pp}-E_{pp1}+\sqrt{(E_{pp}-E_{pp1})^2+16
   t_{s}^2}\right) (([e^{-i (E_{pp}+E_{pp1}) t_1} (e^{\frac{1}{2} i
   \left(2 \phi_{E3}+t_1(E_{pp}+E_{pp1}+\sqrt{(E_{pp}-E_{pp1})^2+16
   t_{s}^2})\right)} \times \nonumber \\
   %$
 %%    \end{eqnarray*}
 %%  \newline
  %% $
   \sqrt{16 t_{s}^2-(E_{pp}-E_{pp1})
   \left(-E_{pp}+E_{pp1}+\sqrt{(E_{pp}-E_{pp1})^2+16t_{s}^2}\right)}
   c_{E3}+ \nonumber \\
  %% $
 %%  \newline
   %QQQQ
  %% \newline
  %%  \begin{eqnarray*}
   +c_{E4} e^{\frac{1}{2} i \left(2 \phi_{E4}+t_1
   \left(E_{pp}+E_{pp1}-\sqrt{(E_{pp}-E_{pp1})^2+16 \text{ts}^2}\right)\right)} \sqrt{16
   t_{s}^2+(E_{pp}-E_{pp1}) \left(E_{pp}-E_{pp1}+\sqrt{(E_{pp}-E_{pp1})^2+16
   t_{s}^2}\right)})]\times \nonumber \\
   %%\newline
   \frac{1}{2 \sqrt{(E_{pp}-E_{pp1})^2+16t_{s}^2}}-\frac{\text{cE1} e^{i
   (\phi_{E1}-E_{pp} t_1)}}{\sqrt{2}}) (V_{2}-i \hbar) \nonumber \\
   %%%\newline
   %%%$
   %%%\begin{eqnarray*}
   -(\frac{e^{i
   (\phi_{E2}-E_{pp1} t_1)} c_{E2}}{\sqrt{2}}+\frac{2 c_{E4} e^{i
   \phi_{E4}-\frac{1}{2} i t_1\left(E_{pp}+E_{pp1}+\sqrt{(E_{pp}-E_{pp1})^2+16
    t_{s}^2}\right)} t_{s}}{\sqrt{16 t_{s}^2+(E_{pp}-E_{pp1})
   \left(E_{pp}-E_{pp1}+\sqrt{(E_{pp}-E_{pp1})^2+16 t_{s}^2}\right)}} \nonumber \\
   -\frac{2 \text{cE3}
   e^{\frac{1}{2} i \left(2 \phi_{E3}-(E_{pp}+E_{pp1}) t_1+t_1
   \sqrt{(E_{pp}-E_{pp1})^2+16t_{s}^2}\right)} t_{s}}{\sqrt{16
   t_{s}^2-(E_{pp}-E_{pp1}) \left(-E_{pp}+E_{pp1}+\sqrt{(E_{pp}-E_{pp1})^2+16
   t_{s}^2}\right)}}) (V_3+i \hbar V_{4}))+ \nonumber \\
   %%%\newline
  %%% QQQQ
  %%%% \newline
  %% \begin{eqnarray*}
   +4 t_{s}
   ((-\frac{e^{i (\phi_{E2}-E_{pp1} t_1)}c_{E2}}{\sqrt{2}}+\frac{2
   c_{E4} e^{i \phi_{E4}-\frac{1}{2} i t_1
   \left(E_{pp}+E_{pp1}+\sqrt{(E_{pp}-E_{pp1})^2+16 t_{s}^2}\right)} t_{s}}{\sqrt{16
   t_{s}^2+(E_{pp}-E_{pp1}) \left(E_{pp}-E_{pp1}+\sqrt{(E_{pp}-E_{pp1})^2+16
   t_{s}^2}\right)}} \nonumber \\
   %\end{eqnarray*}
   %%%\newline
   %%%$
   -\frac{2 c_{E3} e^{\frac{1}{2} i \left(2 \phi_{E3}-(E_{pp}+E_{pp1})
   t_1+t_1 \sqrt{(E_{pp}-E_{pp1})^2+16t_{s}^2}\right)}t_{s}}{\sqrt{16
   t_{s}^2-(E_{pp}-E_{pp1}) \left(-E_{pp}+E_{pp1}+\sqrt{(E_{pp}-E_{pp1})^2+16
   t_{s}^2}\right)}})(V_{2}-i\hbar) %%%%%\nonumber \\
   %%%$
   %%\newline
   %% $(V_{2}-i\hbar)$
   %%%\newline
   %%%%\begin{eqnarray*}
    -[\frac{e^{i (\phi_{E1}-E_{pp}
   t_1)} c_{E1}}{\sqrt{2}}+ \nonumber \\
   %%%\newline
   %%\begin{eqnarray*}
   +[e^{-i (E_{pp}+E_{pp1}) t_1}
   (e^{\frac{1}{2} i \left(2 \phi_{E3}+t_1
   \left(E_{pp}+E_{pp1}+\sqrt{(E_{pp}-E_{pp1})^2+16 t_{s}^2}\right)\right)} \sqrt{16
    t_{s}^2-(E_{pp}-E_{pp1}) \left(-E_{pp}+E_{pp1}+\sqrt{(E_{pp}-E_{pp1})^2+16
   t_{s}^2}\right)} c_{E3}+ \nonumber \\
   %%%\newline
   %\begin{eqnarray*}
    +c_{E4} e^{\frac{1}{2} i \left(2 \phi_{E4}+t_1
   \left(E_{pp}+E_{pp1}-\sqrt{(E_{pp}-E_{pp1})^2+16 t_{s}^2}\right)\right)} \nonumber \\  \sqrt{16
   t_{s}^2+(E_{pp}-E_{pp1}) \left(E_{pp}-E_{pp1}+\sqrt{(E_{pp}-E_{pp1})^2+16
   t_{s}^2}\right)})]\frac{1}{2 \sqrt{(E_{pp}-E_{pp1})^2+16 t_{s}^2}}] %%%%\nonumber \\
   %%\end{eqnarray*}
   %%\newline $
   (V_{3}+i\hbar V_{4})))] \nonumber \\
   %%$
%\newline
%%%\begin{eqnarray*}
\times \frac{1}{2 \sqrt{16 t_{s}^2+(E_{pp}-E_{pp1})
   \left(E_{pp}-E_{pp1}+\sqrt{(E_{pp}-E_{pp1})^2+16t_{s}^2}\right)}
   \left(\hbar^2+i (V_{1}+V_{2}+(V_{3}-iV_{4})V_{4}) \hbar-V_{1}
   V_{2}+V_{3} (V_{3}-i V_{4})\right)}
\end{eqnarray*}

   %\newline \newline
   \normalsize
Another step towards more general Hamiltonian for the system of symmetric coupling electrostatic qubits is when we assume the existence of complex values hopping constants $i t_{2si}+t_{2sr}$ and $i t_{1si}+t_{1sr}$ and in such case the Hamiltonian has the following form
\begin{eqnarray}
\hat{H}=
\left(
\begin{array}{cccc}
 E_{pp} & i t_{2si}+t_{2sr} & i t_{1si}+t_{1sr} & 0 \\
 t_{2sr}-i t_{2si} & E_{pp1} & 0 & i t_{1si}+t_{1sr} \\
 t_{1sr}-i t_{1si} & 0 & E_{pp1} & i t_{2si}+t_{2sr} \\
 0 & t_{1sr}-i t_{1si} & t_{2sr}-i t_{2si} & E_{pp} \\
\end{array}
\right)
\end{eqnarray}
\begin{landscape}
%%\begin{sidewaysfigure}
%%\begin{turn}{90}
We obtain 4 orthogonal eigenstates of the system
%\begin{landscape}
%%\begin{eqnarray}
%%\ket{E_1}=
%%\begin{pmatrix}
%a
%\end{pmatrix}
%\end{eqnarray}
\small
%%\newline
%%$\ket{E1}=$
%\newline
\tiny
\begin{eqnarray}
\ket{E1}=
\begin{pmatrix}
%\left\{
%\begin{array}{c}
\frac{(\text{t1si}-i \text{t1sr}) (\text{t2si}-i \text{t2sr}) \left(\sqrt{(\text{Epp}-\text{Epp1})^2+4 \left(-2 \sqrt{\left(\text{t1si}^2+\text{t1sr}^2\right)
   \left(\text{t2si}^2+\text{t2sr}^2\right)}+\text{t1si}^2+\text{t1sr}^2+\text{t2si}^2+\text{t2sr}^2\right)}-E_{pp}+E_{pp1}\right)}{2
   \sqrt{\left(t_{1si}^2+t_{1sr}^2\right) \left(t_{2si}^2+t_{2sr}^2\right)} \sqrt{4 \left(-2 \sqrt{\left(t_{1si}^2+t_{1sr}^2\right)
   \left(t_{2si}^2+t_{2sr}^2\right)}+t_{1si}^2+t_{1sr}^2+t_{2si}^2+t_{2sr}^2\right)-(E_{pp}-E_{pp1}) \left(\sqrt{(E_{pp}-E_{pp1})^2+4
   \left(-2 \sqrt{\left(t_{1si}^2+t_{1sr}^2\right)
   \left(t_{2si}^2+t_{2sr}^2\right)}+t_{1si}^2+t_{1sr}^2+t_{2si}^2+t_{2sr}^2\right)}-E_{pp}+E_{pp1}\right)}}, \nonumber
\\
\frac{(\text{t1sr}+i \text{t1si})
   \left(\sqrt{\left(\text{t1si}^2+\text{t1sr}^2\right) \left(\text{t2si}^2+\text{t2sr}^2\right)}-\text{t1si}^2-\text{t1sr}^2\right)}{\sqrt{\text{t1si}^2+\text{t1sr}^2} \sqrt{4
   \left(-2 \sqrt{\left(\text{t1si}^2+\text{t1sr}^2\right)
   \left(\text{t2si}^2+\text{t2sr}^2\right)}+\text{t1si}^2+\text{t1sr}^2+\text{t2si}^2+\text{t2sr}^2\right)-(\text{Epp}-\text{Epp1}) \left(\sqrt{(\text{Epp}-\text{Epp1})^2+4
   \left(-2 \sqrt{\left(\text{t1si}^2+\text{t1sr}^2\right)
   \left(\text{t2si}^2+\text{t2sr}^2\right)}+\text{t1si}^2+\text{t1sr}^2+\text{t2si}^2+\text{t2sr}^2\right)}-\text{Epp}+\text{Epp1}\right)}}, \nonumber  \\
\frac{(\text{t2sr}+i \text{t2si})
   \left(\sqrt{\left(\text{t1si}^2+\text{t1sr}^2\right) \left(\text{t2si}^2+\text{t2sr}^2\right)}-\text{t2si}^2-\text{t2sr}^2\right)}{\sqrt{\text{t2si}^2+\text{t2sr}^2} \sqrt{4
   \left(-2 \sqrt{\left(\text{t1si}^2+\text{t1sr}^2\right)
   \left(\text{t2si}^2+\text{t2sr}^2\right)}+\text{t1si}^2+\text{t1sr}^2+\text{t2si}^2+\text{t2sr}^2\right)-(\text{Epp}-\text{Epp1}) \left(\sqrt{(\text{Epp}-\text{Epp1})^2+4
   \left(-2 \sqrt{\left(\text{t1si}^2+\text{t1sr}^2\right)
   \left(\text{t2si}^2+\text{t2sr}^2\right)}+\text{t1si}^2+\text{t1sr}^2+\text{t2si}^2+\text{t2sr}^2\right)}-\text{Epp}+\text{Epp1}\right)}}, \nonumber  \\
\frac{\sqrt{(\text{Epp}-\text{Epp1})^2
   +4 \left(-2 \sqrt{\left(\text{t1si}^2+\text{t1sr}^2\right)
   \left(\text{t2si}^2+\text{t2sr}^2\right)}+\text{t1si}^2+\text{t1sr}^2+\text{t2si}^2+\text{t2sr}^2\right)}-\text{Epp}+\text{Epp1}}{2 \sqrt{4 \left(-2
   \sqrt{\left(\text{t1si}^2+\text{t1sr}^2\right) \left(\text{t2si}^2+\text{t2sr}^2\right)}+\text{t1si}^2+\text{t1sr}^2+\text{t2si}^2+\text{t2sr}^2\right)-(\text{Epp}-\text{Epp1})
   \left(\sqrt{(E_{pp}-E_{pp1})^2+4 \left(-2 \sqrt{\left(\text{t1si}^2+\text{t1sr}^2\right)
   \left(\text{t2si}^2+\text{t2sr}^2\right)}+t_{1si}^2+t_{1sr}^2+t_{2si}^2+t_{2sr}^2\right)}-E_{pp}+E_{pp1}\right)}}
%\end{array}{c}
%\right\}
\end{pmatrix}
\end{eqnarray}
%$
%%%$ $
\newline
%%%%$\ket{E2}=$
%%%\newline $ $
\begin{eqnarray}
\ket{E2}=
\begin{pmatrix}
%\left\{
%\begin{array}{c}
\frac{(\text{t1si}-i \text{t1sr}) (\text{t2si}-i \text{t2sr}) \left(\sqrt{(\text{Epp}-\text{Epp1})^2+4 \left(-2 \sqrt{\left(\text{t1si}^2+\text{t1sr}^2\right)
   \left(\text{t2si}^2+\text{t2sr}^2\right)}+\text{t1si}^2+\text{t1sr}^2+\text{t2si}^2+\text{t2sr}^2\right)}+\text{Epp}-\text{Epp1}\right)}{2
   \sqrt{\left(\text{t1si}^2+\text{t1sr}^2\right) \left(\text{t2si}^2+\text{t2sr}^2\right)} \sqrt{(\text{Epp}-\text{Epp1}) \left(\sqrt{(\text{Epp}-\text{Epp1})^2+4 \left(-2
   \sqrt{\left(\text{t1si}^2+\text{t1sr}^2\right)
   \left(\text{t2si}^2+\text{t2sr}^2\right)}+\text{t1si}^2+\text{t1sr}^2+\text{t2si}^2+\text{t2sr}^2\right)}+\text{Epp}-\text{Epp1}\right)+4 \left(-2
   \sqrt{\left(\text{t1si}^2+\text{t1sr}^2\right) \left(\text{t2si}^2+\text{t2sr}^2\right)}+\text{t1si}^2+\text{t1sr}^2+\text{t2si}^2+\text{t2sr}^2\right)}}, \nonumber \\
\frac{(\text{t1sr}+i
   \text{t1si}) \left(-\sqrt{\left(\text{t1si}^2+\text{t1sr}^2\right)
   \left(\text{t2si}^2+\text{t2sr}^2\right)}+\text{t1si}^2+\text{t1sr}^2\right)}{\sqrt{\text{t1si}^2+\text{t1sr}^2} \sqrt{(\text{Epp}-\text{Epp1})
   \left(\sqrt{(\text{Epp}-\text{Epp1})^2+4 \left(-2 \sqrt{\left(\text{t1si}^2+\text{t1sr}^2\right)
   \left(\text{t2si}^2+\text{t2sr}^2\right)}+\text{t1si}^2+\text{t1sr}^2+\text{t2si}^2+\text{t2sr}^2\right)}+\text{Epp}-\text{Epp1}\right)+4 \left(-2
   \sqrt{\left(\text{t1si}^2+\text{t1sr}^2\right) \left(\text{t2si}^2+\text{t2sr}^2\right)}+\text{t1si}^2+\text{t1sr}^2+\text{t2si}^2+\text{t2sr}^2\right)}},  \nonumber \\
\frac{(\text{t2sr}+i
   \text{t2si}) \left(-\sqrt{\left(\text{t1si}^2+\text{t1sr}^2\right)
   \left(\text{t2si}^2+\text{t2sr}^2\right)}+\text{t2si}^2+\text{t2sr}^2\right)}{\sqrt{\text{t2si}^2+\text{t2sr}^2} \sqrt{(\text{Epp}-\text{Epp1})
   \left(\sqrt{(\text{Epp}-\text{Epp1})^2+4 \left(-2 \sqrt{\left(\text{t1si}^2+\text{t1sr}^2\right)
   \left(\text{t2si}^2+\text{t2sr}^2\right)}+\text{t1si}^2+\text{t1sr}^2+\text{t2si}^2+\text{t2sr}^2\right)}+\text{Epp}-\text{Epp1}\right)+4 \left(-2
   \sqrt{\left(\text{t1si}^2+\text{t1sr}^2\right)
   \left(\text{t2si}^2+\text{t2sr}^2\right)}+\text{t1si}^2+\text{t1sr}^2+\text{t2si}^2+\text{t2sr}^2\right)}},   \\\frac{\sqrt{(\text{Epp}-\text{Epp1})^2+4 \left(-2
   \sqrt{\left(\text{t1si}^2+\text{t1sr}^2\right)
   \left(\text{t2si}^2+\text{t2sr}^2\right)}+\text{t1si}^2+\text{t1sr}^2+\text{t2si}^2+\text{t2sr}^2\right)}+\text{Epp}-\text{Epp1}}{2 \sqrt{(\text{Epp}-\text{Epp1})
   \left(\sqrt{(\text{Epp}-\text{Epp1})^2+4 \left(-2 \sqrt{\left(\text{t1si}^2+\text{t1sr}^2\right)
   \left(\text{t2si}^2+\text{t2sr}^2\right)}+\text{t1si}^2+\text{t1sr}^2+\text{t2si}^2+\text{t2sr}^2\right)}+\text{Epp}-\text{Epp1}\right)+4 \left(-2
   \sqrt{\left(\text{t1si}^2+\text{t1sr}^2\right) \left(\text{t2si}^2+\text{t2sr}^2\right)}+\text{t1si}^2+\text{t1sr}^2+\text{t2si}^2+\text{t2sr}^2\right)}}
%\end{array}{c}
\end{pmatrix}
\end{eqnarray}
%\right\}$
%$ $
%\newline
%%%$
%%%%\ket{E3}=
 %%%%$
%%%%%\newline
%$
%\left(
\begin{eqnarray}
\ket{E3}=
\begin{pmatrix}
 -\frac{(\text{t1si}-i \text{t1sr}) (\text{t2si}-i \text{t2sr}) \left(-\text{Epp}+\text{Epp1}+\sqrt{(\text{Epp}-\text{Epp1})^2+4
   \left(\text{t1si}^2+\text{t1sr}^2+\text{t2si}^2+\text{t2sr}^2+2 \sqrt{\left(\text{t1si}^2+\text{t1sr}^2\right) \left(\text{t2si}^2+\text{t2sr}^2\right)}\right)}\right)}{2
   \sqrt{\left(\text{t1si}^2+\text{t1sr}^2\right) \left(\text{t2si}^2+\text{t2sr}^2\right)} \sqrt{4 \left(\text{t1si}^2+\text{t1sr}^2+\text{t2si}^2+\text{t2sr}^2+2
   \sqrt{\left(\text{t1si}^2+\text{t1sr}^2\right) \left(\text{t2si}^2+\text{t2sr}^2\right)}\right)-(\text{Epp}-\text{Epp1})
   \left(-\text{Epp}+\text{Epp1}+\sqrt{(\text{Epp}-\text{Epp1})^2+4 \left(\text{t1si}^2+\text{t1sr}^2+\text{t2si}^2+\text{t2sr}^2+2 \sqrt{\left(\text{t1si}^2+\text{t1sr}^2\right)
   \left(\text{t2si}^2+\text{t2sr}^2\right)}\right)}\right)}} \\
 -\frac{i (\text{t1si}-i \text{t1sr}) \left(\text{t1si}^2+\text{t1sr}^2+\sqrt{\left(\text{t1si}^2+\text{t1sr}^2\right)
   \left(\text{t2si}^2+\text{t2sr}^2\right)}\right)}{\sqrt{\text{t1si}^2+\text{t1sr}^2} \sqrt{4 \left(\text{t1si}^2+\text{t1sr}^2+\text{t2si}^2+\text{t2sr}^2+2
   \sqrt{\left(\text{t1si}^2+\text{t1sr}^2\right) \left(\text{t2si}^2+\text{t2sr}^2\right)}\right)-(\text{Epp}-\text{Epp1})
   \left(-\text{Epp}+\text{Epp1}+\sqrt{(\text{Epp}-\text{Epp1})^2+4 \left(\text{t1si}^2+\text{t1sr}^2+\text{t2si}^2+\text{t2sr}^2+2 \sqrt{\left(\text{t1si}^2+\text{t1sr}^2\right)
   \left(\text{t2si}^2+\text{t2sr}^2\right)}\right)}\right)}} \\
 -\frac{i (\text{t2si}-i \text{t2sr}) \left(\text{t2si}^2+\text{t2sr}^2+\sqrt{\left(\text{t1si}^2+\text{t1sr}^2\right)
   \left(\text{t2si}^2+\text{t2sr}^2\right)}\right)}{\sqrt{\text{t2si}^2+\text{t2sr}^2} \sqrt{4 \left(\text{t1si}^2+\text{t1sr}^2+\text{t2si}^2+\text{t2sr}^2+2
   \sqrt{\left(\text{t1si}^2+\text{t1sr}^2\right) \left(\text{t2si}^2+\text{t2sr}^2\right)}\right)-(\text{Epp}-\text{Epp1})
   \left(-\text{Epp}+\text{Epp1}+\sqrt{(\text{Epp}-\text{Epp1})^2+4 \left(\text{t1si}^2+\text{t1sr}^2+\text{t2si}^2+\text{t2sr}^2+2 \sqrt{\left(\text{t1si}^2+\text{t1sr}^2\right)
   \left(\text{t2si}^2+\text{t2sr}^2\right)}\right)}\right)}} \\
 \frac{-\text{Epp}+\text{Epp1}+\sqrt{(\text{Epp}-\text{Epp1})^2+4 \left(\text{t1si}^2+\text{t1sr}^2+\text{t2si}^2+\text{t2sr}^2+2 \sqrt{\left(\text{t1si}^2+\text{t1sr}^2\right)
   \left(\text{t2si}^2+\text{t2sr}^2\right)}\right)}}{2 \sqrt{4 \left(\text{t1si}^2+\text{t1sr}^2+\text{t2si}^2+\text{t2sr}^2+2 \sqrt{\left(\text{t1si}^2+\text{t1sr}^2\right)
   \left(\text{t2si}^2+\text{t2sr}^2\right)}\right)-(\text{Epp}-\text{Epp1}) \left(-\text{Epp}+\text{Epp1}+\sqrt{(\text{Epp}-\text{Epp1})^2+4
   \left(\text{t1si}^2+\text{t1sr}^2+\text{t2si}^2+\text{t2sr}^2+2 \sqrt{\left(\text{t1si}^2+\text{t1sr}^2\right) \left(\text{t2si}^2+\text{t2sr}^2\right)}\right)}\right)}} \\
\end{pmatrix}
\end{eqnarray}

%%\right)
%%$
%%\newline
%%$\ket{E4}=$
%%%\newline
%$
%%\left\{
%%\begin{array}{c}
\begin{eqnarray}
\ket{E4}=
\begin{pmatrix}
%\left\{
-\frac{(t_{1si}-i t_{1sr}) (t_{2si}-i t_{2sr}) \left(\sqrt{(E_{pp}-E_{pp1})^2+4 \left(2 \sqrt{\left(t_{1si}^2+t_{1sr}^2\right)
   \left(t_{2si}^2+t_{2sr}^2\right)}+t_{1si}^2+t_{1sr}^2+t_{2si}^2+t_{2sr}^2\right)}+E_{pp}-E_{pp1}\right)}{2
   \sqrt{\left(t_{1si}^2+t_{1sr}^2\right) \left(t_{2si}^2+t_{2sr}^2\right)} \sqrt{(E_{pp}-E_{pp1}) \left(\sqrt{(E_{pp}-E_{pp1})^2+4 \left(2
   \sqrt{\left(t_{1si}^2+t_{1sr}^2\right)
   \left(t_{2si}^2+t_{2sr}^2\right)}+t_{1si}^2+t_{1sr}^2+t_{2si}^2+t_{2sr}^2\right)}+E_{pp}-E_{pp1}\right)+4 \left(2
   \sqrt{\left(t_{1si}^2+t_{1sr}^2\right) \left(t_{2si}^2+t_{2sr}^2\right)}+t_{1si}^2+t_{1sr}^2+t_{2si}^2+t_{2sr}^2\right)}} \\
,\frac{(\text{t1sr}+i
   \text{t1si}) \left(\sqrt{\left(\text{t1si}^2+\text{t1sr}^2\right)
   \left(\text{t2si}^2+\text{t2sr}^2\right)}+\text{t1si}^2+\text{t1sr}^2\right)}{\sqrt{\text{t1si}^2+\text{t1sr}^2} \sqrt{(\text{Epp}-\text{Epp1})
   \left(\sqrt{(\text{Epp}-\text{Epp1})^2+4 \left(2 \sqrt{\left(\text{t1si}^2+\text{t1sr}^2\right)
   \left(\text{t2si}^2+\text{t2sr}^2\right)}+\text{t1si}^2+\text{t1sr}^2+\text{t2si}^2+\text{t2sr}^2\right)}+\text{Epp}-\text{Epp1}\right)+4 \left(2
   \sqrt{\left(\text{t1si}^2+\text{t1sr}^2\right) \left(\text{t2si}^2+\text{t2sr}^2\right)}+\text{t1si}^2+\text{t1sr}^2+\text{t2si}^2+\text{t2sr}^2\right)}}, \\
\frac{(\text{t2sr}+i
   \text{t2si}) \left(\sqrt{\left(\text{t1si}^2+\text{t1sr}^2\right)
   \left(\text{t2si}^2+\text{t2sr}^2\right)}+\text{t2si}^2+\text{t2sr}^2\right)}{\sqrt{\text{t2si}^2+\text{t2sr}^2} \sqrt{(\text{Epp}-\text{Epp1})
   \left(\sqrt{(\text{Epp}-\text{Epp1})^2+4 \left(2 \sqrt{\left(\text{t1si}^2+\text{t1sr}^2\right)
   \left(\text{t2si}^2+\text{t2sr}^2\right)}+\text{t1si}^2+\text{t1sr}^2+\text{t2si}^2+\text{t2sr}^2\right)}+\text{Epp}-\text{Epp1}\right)+4 \left(2
   \sqrt{\left(\text{t1si}^2+\text{t1sr}^2\right)
   \left(t_{2si}^2+t_{2sr}^2\right)}+t_{1si}^2+t_{1sr}^2+t_{2si}^2+t_{2sr}^2\right)}} \\,\frac{\sqrt{(E_{pp}-E_{pp1})^2+4 \left(2
   \sqrt{\left(t_{1si}^2+t_{1sr}^2\right)
   \left(t_{2si}^2+t_{2sr}^2\right)}+t_{1si}^2+t_{1sr}^2+t_{2si}^2+t_{2sr}^2\right)}+E_{pp}-E_{pp1}}{2 \sqrt{(E_{pp}-E_{pp1})
   \left(\sqrt{(E_{pp}-E_{pp1})^2+4 \left(2 \sqrt{\left(t_{1si}^2+t_{1sr}^2\right)
   \left(t_{2si}^2+t_{2sr}^2\right)}+t_{1si}^2+t_{1sr}^2+t_{2si}^2+t_{2sr}^2\right)}+E_{pp}-E_{pp1}\right)+4 \left(2
   \sqrt{\left(t_{1si}^2+t_{1sr}^2\right) \left(t_{2si}^2+t_{2sr}^2\right)}+t_{1si}^2+t_{1sr}^2+t_{2si}^2+t_{2sr}^2\right)}}
%\right\}
\end{pmatrix}
\end{eqnarray}
%%\end{array}{c}
%%\right\}
%$
\normalsize
%%\end{landscape}
with 4 energy eigenvalues
%%\end{turn}
%%\end{sidewaysfigure}

\begin{equation}
E_1=\frac{1}{2} \left(-\sqrt{(E_{pp}-E_{pp1})^2+4 \left(-2 \sqrt{\left(t_{1si}^2+t_{1sr}^2\right)
   \left(t_{2si}^2+t_{2sr}^2\right)}+t_{1si}^2+t_{1sr}^2+t_{2si}^2+t_{2sr}^2\right)}+E_{pp}+E_{pp1}\right)
\end{equation}

\begin{equation}
E_2=\frac{1}{2} \left(\sqrt{(E_{pp}-E_{pp1})^2+4 \left(-2 \sqrt{\left(t_{1si}^2+t_{1sr}^2\right)
   \left(t_{2si}^2+t_{2sr}^2\right)}+t_{1si}^2+t_{1sr}^2+t_{2si}^2+t_{2sr}^2\right)}+E_{pp}+E_{pp1}\right)
\end{equation}

\begin{equation}
E_3=\frac{1}{2} \left(-\sqrt{(E_{pp}-E_{pp1})^2+4 \left(2 \sqrt{\left(t_{1si}^2+t_{1sr}^2\right)
   \left(t_{2si}^2+t_{2sr}^2\right)}+t_{1si}^2+t_{1sr}^2+t_{2si}^2+t_{2sr}^2\right)}+E_{pp}+E_{pp1}\right)
\end{equation}

\begin{equation}
E_4=\frac{1}{2} \left(\sqrt{(E_{pp}-E_{pp1})^2+4 \left(2 \sqrt{\left(t_{1si}^2+t_{1sr}^2\right)
   \left(t_{2si}^2+t_{2sr}^2\right)}+t_{1si}^2+t_{1sr}^2+t_{2si}^2+t_{2sr}^2\right)}+E_{pp}+E_{pp1}\right)
\end{equation}

\end{landscape}
\section{Concept of position dependent electrostatic roton qubit}
The idea is depicted in the Fig.4. Electron placed in the structure has two minima and two maxima in the effective confining potential. After one circulation the wavefunction of such electron is changed by $2 \Pi n$ phase and is coming back to its original value so we are dealing with periodic boundary conditons. We can express electron dynamics by effective Hamiltonian that is of the form
\begin{eqnarray}
H=
\begin{pmatrix}
E_{p1} & t_{21}  &  0         & t_{41} \\
t_{12} & E_{p2}  &  t_{32} &  0        \\
0          & t_{23}  &  E_{p3} & t_{43}  \\
t_{14}  & 0          & t_{34}  &  E_{p4} \\
\end{pmatrix}
\end{eqnarray}
Hermicity of Hamiltonian implies the condition $t_{kl}^{*}=t_{lk}$.  If four minima and four maxima of effective confining potential are the same the Hamiltonian reduced to the form
\begin{eqnarray}
H=
\begin{pmatrix}
E_{p1} & t_{s}  &  0         & t_{s} \\
t_{s} & E_{p2}  &  t_{s} &  0        \\
0          & t_{s}  &  E_{p1} & t_{s}  \\
t_{s}  & 0          & t_{s}  &  E_{p2} \\
\end{pmatrix}
\end{eqnarray}
After renormalization Hamiltonian can be simplified to be of the form
\begin{eqnarray}
H=
\begin{pmatrix}
E_{pq1} & 1  &  0         & 1 \\
1 & E_{pq2}  &  1 &  0        \\
0          & 1  &  E_{pq1} & 1  \\
1  & 0          & 1  &  E_{pq2} \\
\end{pmatrix}
\end{eqnarray}
The wavefunction of electron can be expressed in Wannier representation in the following form
\begin{eqnarray}
\ket{\psi}=
\begin{pmatrix}
\gamma_1(t) \\
\gamma_2(t) \\
\gamma_3(t) \\
\gamma_4(t) \\
\end{pmatrix}
\end{eqnarray}
where $|\gamma_1(t)|^2+ ..+|\gamma_4(t)|^2=1$  and
\begin{equation}
\begin{pmatrix}
1 \\
0 \\
0 \\
0 \\
\end{pmatrix}, .. ,
 \begin{pmatrix}
0 \\
0 \\
0 \\
1 \\
\end{pmatrix}
\end{equation}
 are denoting 4 orthonormal maximized localized wavefunctions (Wannier functions).
The system eigenstates and eigenergies are given as

The passing proton can affect the Hamiltonian in the following way given as $H=$ \small
\begin{eqnarray}
%%%H=\nonumber \\
\begin{pmatrix}
E_{p1}+V_1 \delta(t-t_1) & t_{21}+V_3 \delta(t-t_1)+ i V_4 \delta(t-t_1)   &  0         & t_{41}+ V_5 \delta(t-t_1) +iV_6 \delta(t-t_1) \\
t_{12} + V_3 \delta(t-t_1)- i V_4 \delta(t-t_1) & E_{p2}  &  t_{32} &  0        \\
0          & t_{23}  &  E_{p3} & t_{43} \\
t_{14}+ V_5 \delta(t-t_1) -iV_6 \delta(t-t_1)  & 0          & t_{34} &  E_{p4}  \\
\end{pmatrix}
\end{eqnarray}
Such Hamiltonian shall be referred to the situation given in Fig.3 and in Fig.4.
Let us assume symmetric position-based roton qubit having two minima (denoted by A) and two maxima (denoted by B) of electron confining potential that has periodic boundary condition. Then in first approximation we have semiconductor transmon Hamiltonian of the form
\begin{eqnarray}
H=\nonumber \\
\begin{pmatrix}
E_{A}+V_1 \delta(t-t_1) & t_{s}+V_3 \delta(t-t_1)+ i V_4 \delta(t-t_1)   &  0         & t_{s}+ V_5 \delta(t-t_1) +iV_6 \delta(t-t_1) \\
t_{s} + V_3 \delta(t-t_1)- i V_4 \delta(t-t_1) & E_{B}  &  t_{s} &  0        \\
0          & t_{s}  &  E_{A} & t_{s} \\
t_{s}+ V_5 \delta(t-t_1) -iV_6 \delta(t-t_1)  & 0          & t_{s} &  E_{B}  \\
\end{pmatrix}
\end{eqnarray}
Such Hamiltonian has eigenvalues as
\begin{eqnarray}
E_1=E_{p1}=E_A, E_2=E_{p2}=E_B, \nonumber \\
E_3=\frac{1}{2} \left(-\sqrt{ (E_{p1}-E_{p2})^2+16t_{s}^2}+E_{p1}+E_{p2}\right), \nonumber \\
E_4=\frac{1}{2} \left(+\sqrt{ (E_{p1}-E_{p2})^2+16t_{s}^2}+E_{p1}+E_{p2}\right).
\end{eqnarray}
%%%$E_1=E_A$ and $E_2=E_B$ and $E_3=\frac{1}{2} \left(-\sqrt{\text{Ep1}^2-2 \text{Ep1} \text{Ep2}+\text{Ep2}^2+16
%%   \text{ts}^2}+\text{Ep1}+\text{Ep2}\right)$ and $E_3=\frac{1}{2} \left(+\sqrt{\text{Ep1}^2-2 \text{Ep1} \text{Ep2}+\text{Ep2}^2+16
%%   \text{ts}^2}+\text{Ep1}+\text{Ep2}\right)$.
We have eigenvectors
   \begin{equation}
   \ket{E_1}= \frac{1}{\sqrt{2}}
    \begin{pmatrix}
    -1 \\
    0 \\
    +1 \\
    0 \\
    \end{pmatrix},
    \ket{E_2}= \frac{1}{\sqrt{2}}
    \begin{pmatrix}
    0 \\
    -1 \\
    0 \\
    +1 \\
    \end{pmatrix}.
   \end{equation}
 %%%%  $\ket{E_1}= \{-1,0,1,0\}$ and $\ket{E_2}=  \{0,-1,0,1\}$
   \begin{equation}
   \ket{E_3}=
   %%\left\{
   \frac{1}{\sqrt{2
   \left(\sqrt{(E_{p1}-E_{p2})^2+16t_{s}^2}+E_{p1}-E_{p2}\right)^2+32t_{s}^2}}
   \begin{pmatrix}
   -4 t_{s}, \\
   \sqrt{(E_{p1}-E_{p2})^2+16t_{s}^2}+E_{p1}-E_{p2} \\
    -4 t_{s}, \\
   \sqrt{(E_{p1}-E_{p2})^2+16t_{s}^2}+E_{p1}-E_{p2} \\
   \end{pmatrix}
   %%%\right\}
   %%\begin{pmatrix}
   %%-\frac{4 t_{s}}{\sqrt{(E_{p1}-E_{p2})^2+16 t_{s}^2}+(E_{p1}-E_{p2}) } \\
   %%1 \\
   %%-\frac{4 t_{s}}{\sqrt{(E_{p1}-E_{p2})^2+16t_{s}^2}+(E_{p1}-E_{p2})} \\
   %%1
   %%\end{pmatrix},
   \end{equation}
   \begin{equation}
   \ket{E_4}=
   \frac{1}{\sqrt{2 \left(\sqrt{(E_{p1}-E_{p2})^2+16
    t_{s}^2}-E_{p1}+E_{p2}\right)^2+32 t_{s}^2}}
   \begin{pmatrix}
   4t_{s} \\
   \sqrt{(E_{p1}-E_{p2})^2+16 t_{s}^2}-E_{p1}+E_{p2} \\
   4t_{s} \\
   \sqrt{(E_{p1}-E_{p2})^2+16 t_{s}^2}-E_{p1}+E_{p2} \\
   \end{pmatrix}.
   \end{equation}

  %% $\ket{E_3}= \left\{-\frac{4 \text{ts}}{\sqrt{\text{Ep1}^2-2 \text{Ep1} \text{Ep2}+\text{Ep2}^2+16
  %% \text{ts}^2}+\text{Ep1}-\text{Ep2}},1,-\frac{4 \text{ts}}{\sqrt{\text{Ep1}^2-2 \text{Ep1}
  %% \text{Ep2}+\text{Ep2}^2+16 \text{ts}^2}+\text{Ep1}-\text{Ep2}},1\right\}$ and
%%%  $ \ket{E_4}= \left\{\frac{4 \text{ts}}{\sqrt{\text{Ep1}^2-2 \text{Ep1} \text{Ep2}+\text{Ep2}^2+16
%%%   \text{ts}^2}-\text{Ep1}+\text{Ep2}},1,\frac{4 \text{ts}}{\sqrt{\text{Ep1}^2-2 \text{Ep1}
%%%   \text{Ep2}+\text{Ep2}^2+16 \text{ts}^2}-\text{Ep1}+\text{Ep2}},1\right\}$

It brings the following effective boundary condition for equation of motion at nearest presence of moving charged particle affecting position based qubit in the form
\begin{eqnarray}
V_1 \gamma_1(t_1^{+}) + (V_3 + i V_4) \gamma_2(t)+ (V_5+iV_6) \gamma_4(t_1^{+})=i \hbar (\gamma_1(t_1^{+})-\gamma_1(t_1^{-})) \nonumber \\
(V_3 - i V_4))\gamma_1(t_1^{+})=i \hbar (\gamma_2(t_1^{+})-\gamma_2(t_1^{-})) \nonumber  \\
(V_5-iV_6)  \gamma_1(t_1^{+}) =i \hbar (\gamma_4(t_1^{+})-\gamma_4(t_1^{-}))
\end{eqnarray}
We can express this boundary condition in the matrix form as
\begin{eqnarray}
\begin{pmatrix}
(V_1-i\hbar) & (V_3 + i V_4) & 0 &  (V_5+iV_6) \\
(V_3 - i V_4) & -i\hbar &  0 & 0 \\
0 & 0 & -i\hbar & 0 \\
(V_5-iV_6)  & 0 & 0 & -i\hbar \\
%(V_3 - i V_4)) & 0 &
\end{pmatrix}
\begin{pmatrix}
 \gamma_4(t_1^{+}) \\
 \gamma_2(t_1^{+}) \\
 \gamma_3(t_1^{+}) \\
 \gamma_4(t_1^{+}) \\
\end{pmatrix}=
-i\hbar
\begin{pmatrix}
 \gamma_1(t_1^{-}) \\
 \gamma_2(t_1^{-}) \\
 \gamma_3(t_1^{-}) \\
 \gamma_4(t_1^{-}) \\
\end{pmatrix}
\end{eqnarray}
Equivalently last matrix equation can be represented as
\begin{eqnarray}
\begin{pmatrix}
 \gamma_4(t_1^{+}) \\
 \gamma_2(t_1^{+}) \\
 \gamma_3(t_1^{+}) \\
 \gamma_4(t_1^{+}) \\
\end{pmatrix}=
-i\hbar
\begin{pmatrix}
(V_1-i\hbar) & (V_3 + i V_4) & 0 &  (V_5+iV_6) \\
(V_3 - i V_4) & -i\hbar &  0 & 0 \\
0 & 0 & -i\hbar & 0 \\
(V_5-iV_6)  & 0 & 0 & -i\hbar \\
%(V_3 - i V_4)) & 0 &
\end{pmatrix}^{-1}
\begin{pmatrix}
 \gamma_1(t_1^{-}) \\
 \gamma_2(t_1^{-}) \\
 \gamma_3(t_1^{-}) \\
 \gamma_4(t_1^{-}) \\
\end{pmatrix}=  %%%%\newline \\
 \hat{M}
 \begin{pmatrix}
 \gamma_1(t_1^{-}) \\
 \gamma_2(t_1^{-}) \\
 \gamma_3(t_1^{-}) \\
 \gamma_4(t_1^{-}) \\
\end{pmatrix}
\end{eqnarray}
and we obtain matrix M of the following structure %\newline
\begin{eqnarray}
M=
\left(
\begin{array}{cccc}
 \frac{\hbar^2}{\hbar^2+i V_{1}
   \hbar+V_{3}^2+V_{4}^2+V_{5}^2+V_{6}^2} & \frac{\hbar
   (V_{4}-i V_{3})}{\hbar^2+i V_{1}
   \hbar+V_{3}^2+V_{4}^2+V_{5}^2+V_{6}^2} & 0 & \frac{\hbar
   (V_{6}-iV_{5})}{\hbar^2+i V_{1}
   \hbar+V_{3}^2+V_{4}^2+V_{5}^2+V_{6}^2} \\
 -\frac{\hbar (i V_{3}+V_{4})}{\hbar^2+i V_{1}
   \hbar+V_{3}^2+V_{4}^2+V_{5}^2+V_{6}^2} &
   1-\frac{V_{3}^2+V_{4}^2}{\hbar^2+i V_{1}
   \hbar+V_{3}^2+V_{4}^2+V_{5}^2+V_{6}^2} & 0 & -\frac{(V_{3}-i
   V_{4}) (V_{5}+i V_{6})}{\hbar^2+i V_{1}
   \hbar+V_{3}^2+V_{4}^2+V_{5}^2+V_{6}^2} \\
 0 & 0 & 1 & 0 \\
 -\frac{\hbar (i V_{5}+V_{6})}{\hbar^2+i V_{1}
   \hbar+V_{3}^2+V_{4}^2+V_{5}^2+V_{6}^2} & -\frac{(V_{3}+i
   V_{4}) (V_{5}-i V_{6})}{\hbar^2+i V_{1}
   \hbar+V_{3}^2+V_{4}^2+V_{5}^2+V_{6}^2} & 0 & \frac{\hbar^2+i
   V_{1} \hbar+V_{3}^2+V_{4}^2}{\hbar^2+i V_{1}
   \hbar+V_{3}^2+V_{4}^2+V_{5}^2+V_{6}^2} \\
\end{array}
\right)
\end{eqnarray}
%%%$
\subsection{Case of symmetric semiconductor roton}
We can also consider the simplest version of semiconductor roton qubit. Such system can be approximated by 4 symmetric quantum dots on the square or by semciodncutor nanoring of radius R. In both cases the Hamiltonian can be given as
\begin{eqnarray}
\label{rotonH}
H=
\begin{pmatrix}
E_{p} & t_{s}e^{i \alpha}  &  0         & t_{s}e^{i \alpha} \\
t_{s}e^{-i \alpha} & E_{p}  &  t_{s}e^{i \alpha} &  0        \\
0          & t_{s}e^{-i \alpha}  &  E_{p} & t_{s}e^{i \alpha}  \\
t_{s}  & 0          & t_{s}e^{-i \alpha}  &  E_{p} \\
\end{pmatrix}
\end{eqnarray}
that has 4 energy eigenvalues
\begin{eqnarray}
E_1=  E_{p}-2 t_{s} |\sin \left(\frac{\alpha }{2}\right)| , %\nonumber \\
E_2=  E_{p}+2 t_{s} |\sin \left(\frac{\alpha }{2}\right)|, %\nonumber \\
E_3=  E_{p}-2 t_s |\cos (\frac{\alpha}{2} )| , %\nonumber \\
E_4=  E_{p}+2 t_s |\cos (\frac{\alpha}{2} )|.
\end{eqnarray}
and has 4 eigenstates
\begin{eqnarray}
\ket{E_1}=\frac{1}{2}\left(
\begin{array}{c}
 -i e^{ i \frac{3}{2} \alpha } \\
 -e^{i \alpha } \\
 i e^{i \frac{\alpha}{2} } \\
 1 \\
\end{array}
\right), %\nonumber \\
\ket{E_2}=\frac{1}{2}
\left(
\begin{array}{c}
 +i e^{i \frac{3}{2}\alpha } \\
 -e^{i \alpha } \\
 -i e^{i \frac{\alpha}{2} } \\
 1 \\
\end{array}
\right) \nonumber \\
\ket{E_3}=\frac{1}{2}
\left(
\begin{array}{c}
 -e^{i \frac{3}{2}\alpha } \\
 +e^{i \alpha } \\
 -e^{i \frac{\alpha}{2} } \\
 1 \\
\end{array}
\right),
\ket{E_4}=\frac{1}{2}
\left(
\begin{array}{c}
 e^{i \frac{3}{2}\alpha } \\
 e^{i \alpha } \\
 e^{i \frac{\alpha}{2} } \\
 1 \\
\end{array}
\right).
\end{eqnarray}
The presence of electron at nodes 1, 2, 3, 4 is expressed by vectors \newline
\begin{eqnarray}
\ket{x_1}=
\left(
\begin{array}{c}
1 \\
0 \\
0 \\
0 \\
\end{array}
\right),
\ket{x_2}=
\left(
\begin{array}{c}
0 \\
1 \\
0 \\
0 \\
\end{array}
\right),
\ket{x_3}=
\left(
\begin{array}{c}
0 \\
0 \\
1 \\
0 \\
\end{array}
\right),
\ket{x_4}=
\left(
\begin{array}{c}
0 \\
0 \\
0 \\
1 \\
\end{array}
\right), \nonumber \\
\ket{\psi(t)}=\gamma_1(t)\ket{x_1}+..+\gamma_1(t)\ket{x_4}, |\gamma_1(t)|^2+..+|\gamma_4(t)|^2=1.
\end{eqnarray}
In all energy eigenstates we see the systematic phase imprint across ring of semiconductor quantum dots what might correspond to the
placement of semiconductor ring to the external magnetic field. It is due to magnetic Aharonov-Bohm effect. Otherwise with case of placement of semiconducto ring of quantum dots in environment with zero magnetic field we have $\alpha=0$. Kinetic energy and momentum of quantum roton (semiconductor "transmon" qubit) is due to non-zero term $t_s$. Otherwise we are dealing with the case when electron position is localized among one between 4 nodes. If we refer to description of semiconductor roton qubit as ring of quantum dots of radius R we have the condition for single-value wavefunction after one circulation across ring/square expressed mathematically by condition
$2\pi k_n R = 2 \pi n$ that implies  $k_n= \frac{n}{R}$. This also implies that $t_s=n t_0$ and n is integer number. In the conducted considerations we have set n=1 so inside semiconductor ring we have one fluxon (one quantized flux of magnetic field is passing via interior of quantum ring so we have approximate mathematical condition for flux quantization $B a^2=\frac{\hbar^2}{e}n$ in case of 4 symmetric quantum dots on the square and with condition $2 \pi R^2 B = \frac{\hbar^2}{e}n$ in case of nanoring, where B is value of magnetic field in the center of square/nanoring that is preassumed to be constant in square/nanoring).  We also recognize that $4 \alpha = 2 \Pi n$ what brings $\alpha = \frac{\pi n}{2}$ and consequently $Sin(\frac{\alpha}{2})=+\frac{\sqrt{2}}{2},1,+\frac{\sqrt{2}}{2},0,-\frac{\sqrt{2}}{2},-1,-\frac{\sqrt{2}}{2},0,+\frac{\sqrt{2}}{2}, ..$ as well as .
$Cos(\frac{\alpha}{2})=0,-\frac{\sqrt{2}}{2},-1,-\frac{\sqrt{2}}{2},0,+\frac{\sqrt{2}}{2},1,+\frac{\sqrt{2}}{2},0,-\frac{\sqrt{2}}{2}, ..$ and therefore the spectrum of eigenergies $E_1, .., E_2$ can be tunned with magnetic field. This tunning however implies discrete eigenergies. We can also regulate the spectrum of eigenenergies by electric field as by tunning $E_p$ what we can do in the continuous manner (as chemical potential can be tunned in the continuous manner). Furthermore the Hamiltonian \ref{rotonH} can also refer to the square of 4 symmetric quantum dots with confining potential having 4 minima and 4 maxima. In such case we can tune both $t_s$ and $E_p$ in electrostatic way. It is also noticeable that the system of 4 symmetric quantum dots can be the source of magnetic field as when we are dealing with wavepacket of one electron moving in clockwise or anticlockwise direction and generating
one flux (or many) of magnetic field (fluxon given by $\frac{\hbar^2}{e}$). In such situation we have physical system that is similar to superconducting SQUID that can generate or detect one flux of magnetic field with occurrence of circulating non-dissipative electric current. In both cases we are assuming that quantum wavepacket is coherence along semiconductor or semiconductor ring. However it shall be underlined that in the superconductor the coherence time of wavepacket can be significantly bigger than in the case of semiconductor where decoherence time is in the range of ns. This is because superconductivity is the manifestation of macroscopic quantum effect that is thermodynamically supported so in first approximation non-dissipative circulating current in mesoscopic size SQUID (usually having bigger size that 300 nm and with value of quantized magnetic flux set to $\frac{\hbar^2}{2e}$, where factor 2 comes from the fact that Cooper pair participates in electrical transport) can last infinitely long time. This is not the case of semiconductor nanorings that has size usually below 300nm. We can preassume that before weak measurement of probing charge particle we have the quantum system given as
\begin{eqnarray}
  \ket{\psi(t_1)}=c_{E1} e^{\frac{E_1 t_1}{i \hbar}}e^{i \phi_{E1}}\ket{E_1}+..+c_{E4} e^{\frac{E_4 t_1}{i \hbar}}e^{i \phi_{E4}}\ket{E_4},
\end{eqnarray}
 where $|c_{E1}|^2$, .., $|c_{E4}|^2$ are the probabilities of occupancy of energy level $E_1$, .., $E_4$ by single electron in roton qubit and normalization condition holds $|c_{E1}|^2+..+|c_{E4}|^2=1$. After measurement at time $t_1$ by probing particle we have
 \begin{eqnarray}
  \ket{\psi(t_1^{+})}=c_{E1m} e^{\frac{E_1 t_1^{-}}{i \hbar}}e^{i \phi_{E1m}}\ket{E_1}+..+c_{E4m} e^{\frac{E_4 t_1^{-}}{i \hbar}}e^{i \phi_{E4m}}\ket{E_4}= \nonumber \\
  \hat{M}(c_{E1} e^{\frac{E_1 t_1^{-}}{i \hbar}}e^{i \phi_{E1}}\ket{E_1}+..+c_{E4} e^{\frac{E_4 t_1^{-}}{i \hbar}}e^{i \phi_{E4}}\ket{E_4}),
\end{eqnarray}
Again we obtain
 \begin{eqnarray}
  c_{E1m} e^{\frac{E_1 t_1^{+}}{i \hbar}}e^{i \phi_{E1m}}=
 \bra{E_1} \hat{M}(c_{E1} e^{\frac{E_1 t_1^{-}}{i \hbar}}e^{i \phi_{E1}}\ket{E_1}+..+c_{E4} e^{\frac{E_4 t_1^{-}}{i \hbar}}e^{i \phi_{E4}}\ket{E_4}), \nonumber  \\
 .. \nonumber \\
  c_{E4m} e^{\frac{E_4 t_1^{+}}{i \hbar}}e^{i \phi_{E4m}}=
 \bra{E_4} \hat{M}(c_{E1} e^{\frac{E_1 t_1^{-}}{i \hbar}}e^{i \phi_{E1}}\ket{E_1}+..+c_{E4} e^{\frac{E_4 t_1^{-}}{i \hbar}}e^{i \phi_{E4}}\ket{E_4}).
 \end{eqnarray}
 Consequently we obtain \newline \small
 $ c_{E1m}e^{i \phi_{E1m}}=[e^{i t_1\left(E_{p}-2 \sqrt{t_{s}^2 \sin
   ^2\left(\frac{\alpha }{2}\right)}\right)}$ \newline
 $  (e^{-iE_{p}
    t_1} (e^{i \left(\phi_{E1}+2 t_1
   \sqrt{t_{s}^2 \sin ^2\left(\frac{\alpha }{2}\right)}\right)}
   c_{E1}+c_{E4} e^{i \left(\phi_{E4}-2t_1
   \sqrt{t_{s}^2 \cos ^2\left(\frac{\alpha
   }{2}\right)}\right)}+c_{E3} e^{i \left(\phi_{E3}+\sqrt{2}
   t_1 \sqrt{t_{s}^2 (\cos (\alpha )+1)}\right)}+c_{E2}
   e^{i \left(\phi_{E2}-2 t_1 \sqrt{t_{s}^2 \sin
   ^2\left(\frac{\alpha }{2}\right)}\right)})$ \newline
$  \left(\hbar^2+iV_{1}
   \hbar+V_{3}^2+V_{4}^2\right)-e^{i (\alpha -E_{p}
   t_1 )} (-e^{i \left(\phi_{E1}+2t_1
   \sqrt{t_{s}^2 \sin ^2\left(\frac{\alpha }{2}\right)}\right)}
   c_{E1}+c_{E4} e^{i \left(\phi_{E4}-2 t_1
   \sqrt{t_{s}^2 \cos ^2\left(\frac{\alpha
   }{2}\right)}\right)}+c_{E3} e^{i \left(\phi_{E3}+\sqrt{2}
   t_1 \sqrt{t_{s}^2 (\cos (\alpha )+1)}\right)}$ \newline
   $-c_{E2}
   e^{i \left(\phi_{E2}-2t_1 \sqrt{t_{s}^2 \sin
   ^2\left(\frac{\alpha }{2}\right)}\right)}) (V_{3}+i
   V_{4}) (V_{5}-iV_{6})+$ %%%%%\newline
    $+(e^{i \left(\alpha
   +\phi_{E1}-E_{p}t_1+2t_1
   \sqrt{t_{s}^2 \sin ^2\left(\frac{\alpha }{2}\right)}\right)}
   \sqrt{-e^{i \alpha }}c_{E1}+$ \newline \newline
    $+e^{i \alpha } (e^{-i
   t_1 \left(E_{p}+\sqrt{2} \sqrt{t_{s}^2 (\cos (\alpha
   )+1)}\right)} \sqrt{e^{i \alpha }} \left(c_{E3} e^{i
   \left(\phi_{E3}+4t_1 \sqrt{t_{s}^2 \cos
   ^2\left(\frac{\alpha }{2}\right)}\right)}-c_{E4} e^{i
   \phi_{E4}}\right)$ \newline  $-c_{E2} e^{i
   \left(\phi_{E2}-t_1 \left(E_{p}+2 \sqrt{t_{s}^2
   \sin ^2\left(\frac{\alpha }{2}\right)}\right)\right)} \sqrt{-e^{i
   \alpha }})) \hbar (i
   V_{5}+V_{6})$  \newline  $-\sqrt{-e^{-i \alpha }} (e^{i
   \left(\phi_{E1}-E_{p}t_1+2t_1
   \sqrt{t_{s}^2 \sin ^2\left(\frac{\alpha }{2}\right)}\right)}
   \sqrt{-e^{i \alpha }}c_{E1}+$ \newline \newline $+e^{-it_1
   \left(E_{p}+\sqrt{2} \sqrt{t_{s}^2 (\cos (\alpha
   )+1)}\right)} \sqrt{e^{i \alpha }} \left(c_{E4} e^{i
   \phi_{E4}}-c_{E3} e^{i \left(\phi_{E3}+4t_1
   \sqrt{t_{s}^2 \cos ^2\left(\frac{\alpha
   }{2}\right)}\right)}\right)$ \newline $-c_{E2} e^{i
   \left(\phi_{E2}-t_1 \left(E_{p}+2 \sqrt{t_{s}^2
   \sin ^2\left(\frac{\alpha }{2}\right)}\right)\right)} \sqrt{-e^{i
   \alpha }}) \left(\hbar^2+iV_{1}
   \hbar+V_{3}^2+V_{4}^2+V_{5}^2+V_{6}^2\right)$ \newline $-e^
   {-i \alpha } ((e^{i \left(\alpha +\phi_{E1}-E_{p}
   t_1+2 t_1 \sqrt{t_{s}^2 \sin
   ^2\left(\frac{\alpha }{2}\right)}\right)} \sqrt{-e^{i \alpha }}
   c_{E1}+$ \newline $+e^{i \alpha } \left(e^{-it_1 \left(E_{p}+\sqrt{2} \sqrt{t_{s}^2 (\cos (\alpha
   )+1)}\right)} \sqrt{e^{i \alpha }} \left(c_{E3} e^{i
   \left(\phi_{E3}+4t_1\sqrt{t_{s}^2 \cos
   ^2\left(\frac{\alpha }{2}\right)}\right)}-c_{E4} e^{i
   \phi_{E4}}\right)-c_{E2} e^{i
   \left(\phi_{E2}-t_1 \left(E_{p}+2 \sqrt{t_{s}^2
   \sin ^2\left(\frac{\alpha }{2}\right)}\right)\right)} \sqrt{-e^{i
   \alpha }}\right)) \hbar (iV_{3}+V_{4})-e^{-i
   E_{p}t_1 } \left(e^{i \left(\phi_{E1}+2
   t_1 \sqrt{t_{s}^2 \sin ^2\left(\frac{\alpha
   }{2}\right)}\right)} c_{E1}+c_{E4} e^{i
   \left(\phi_{E4}-2t_1 \sqrt{t_{s}^2 \cos
   ^2\left(\frac{\alpha }{2}\right)}\right)}+c_{E3} e^{i
   \left(\phi_{E3}+\sqrt{2} t_1 \sqrt{t_{s}^2 (\cos
   (\alpha )+1)}\right)}+c_{E2} e^{i \left(\phi_{E2}-2
   t_1 \sqrt{t_{s}^2 \sin ^2\left(\frac{\alpha
   }{2}\right)}\right)}\right)$ \newline $ (V_{3}-iV_{4}) (V_{5}+i
   V_{6})+e^{i (\alpha -E_{p}t_1)}(-e^{i
   \left(\phi_{E1}+2t_1 \sqrt{t_{s}^2 \sin^2\left(\frac{\alpha }{2}\right)}\right)}c_{E1}+c_{E4} e^{i
   \left(\phi_{E4}-2t_1 \sqrt{t_{s}^2 \cos
   ^2\left(\frac{\alpha }{2}\right)}\right)}+c_{E3} e^{i
   \left(\phi_{E3}+\sqrt{2}t_1\sqrt{t_{s}^2 (\cos
   (\alpha )+1)}\right)}$ \newline $ -c_{E2} e^{i \left(\phi_{E2}-2
   t_1 \sqrt{t_{s}^2 \sin ^2\left(\frac{\alpha
   }{2}\right)}\right)}) \left(\hbar^2+iV_{1}
   \hbar+V_{5}^2+V_{6}^2\right))+$ \newline $+[e^{-2 i
   \alpha } \hbar (c_{E2} e^{i
   \left(\phi_{E2}-t_1 \left(E_{p}+2 \sqrt{t_{s}^2
   \sin ^2\left(\frac{\alpha }{2}\right)}\right)\right)} \left(e^{i
   \alpha } \left(-\sqrt{-e^{i \alpha }} \hbar-i
   V_{3}+V_{4}\right)+i V_{5}-V_{6}\right)+$  \newline  $+c_{E1}
   e^{i \left(\phi_{E1}-E_{p}t_1+2t_1
   \sqrt{t_{s}^2 \sin ^2\left(\frac{\alpha }{2}\right)}\right)}
   \left(e^{i \alpha } \left(\sqrt{-e^{i \alpha }} \hbar-i
   V_{3}+V_{4}\right)+i V_{5}-V_{6}\right)+e^{-i
   t_1\left(2 E_{p}+3 \sqrt{2} \sqrt{t_{s}^2 (\cos
   (\alpha )+1)}\right)}$ \newline $(c_{E3} e^{i
   \left(\phi_{E3}+t_1 \left(E_{p}+4 \sqrt{2}
   \sqrt{t_{s}^2 (\cos (\alpha )+1)}\right)\right)}
   \left(\hbar \left(e^{i \alpha }\right)^{\frac{3}{2}}+i \left(e^{i
   \alpha } (V_{3}+i V_{4})+V_{5}+i
   V_{6}\right)\right)+$  \newline $-c_{E4} e^{i
   \left(\phi_{E4}+t_1 \left(E_{p}+2 \sqrt{2}
   \sqrt{t_{s}^2 (\cos (\alpha )+1)}\right)\right)}
   \left(\hbar \left(e^{i \alpha }\right)^{\frac{3}{2}}+e^{i \alpha }
   (V_{4}-i V_{3})-i
   V_{5}+V_{6}\right)))]$
 %  \newline
    $\frac{1}{\sqrt{-e^{-i \alpha
   }}})]$ %\newline
    $\times \frac{1}{\hbar^2+i V_{1}
   \hbar+V_{3}^2+V_{4}^2+V_{5}^2+V_{6}^2}$ %\newline and
   \normalsize
\subsection{Case of asymmetric semiconductor quantum roton}
Symmetric quantum roton can be implemented with usage of concept of position-based qubits [4] and can be controlled both electrostatically or magnetically.
Let us consider slightly more general physical system of 2 symmetric quantum dots with localized energy set to $E_{p1}$ and two the same energy barriers with localized energy set to $E_{p2}$ such that $E_{p1}>>E_{p2} $ or at least $E_{p1}>E_{p2} $ (Condition $E_{p1} \neq E_{p2}$ defines asymmetric quantum roton). Similarly as before we are dealing with magnetic flux quantization in the interior of considered ring. Therefore we have that $\alpha=0$ is equivalent to the fact of absence of external magnetic field.  Assuming possible hopping given by term $t_s$ between nearest nodes via or over potential barrier we arrive to the Hamiltonian given as
\begin{eqnarray}
H=\left(
\begin{array}{cccc}
 E_{p1} & e^{i \alpha } t_{s} & 0 & e^{i \alpha } t_{s} \\
 e^{-i \alpha } t_{s} & E_{p2} & e^{i \alpha } t_{s} & 0 \\
 0 & e^{-i \alpha } t_{s} & E_{p1} & e^{i \alpha } t_{s} \\
 e^{-i \alpha } t_{s} & 0 & e^{-i \alpha } t_{s} & E_{p2} \\
\end{array}
\right)
\end{eqnarray}
that has 4 energy eigenvalues
\begin{eqnarray}
  E_1 &=& \frac{1}{2} \left(-\sqrt{16 t_{s}^2 \sin ^2\left(\frac{\alpha
   }{2}\right)+(E_{p1}-E_{p2})^2}+E_{p1}+E_{p2}\right) \\
  E_2 &=& \frac{1}{2} \left(+\sqrt{16 t_{s}^2 \sin ^2\left(\frac{\alpha
   }{2}\right)+(E_{p1}-E_{p2})^2}+E_{p1}+E_{p2}\right) \\
  E_3 &=& \frac{1}{2} \left(-\sqrt{16 t_{s}^2 \cos ^2\left(\frac{\alpha
   }{2}\right)+(E_{p1}-E_{p2})^2}+E_{p1}+E_{p2}\right) \\
  E_4 &=& \frac{1}{2} \left(+\sqrt{16 t_{s}^2 \cos ^2\left(\frac{\alpha
   }{2}\right)+(E_{p1}-E_{p2})^2}+E_{p1}+E_{p2}\right)
\end{eqnarray}
and we have 4 orthonormal eigenstates of the system \newline
\begin{eqnarray}
\ket{E_1}=\frac{1}{\sqrt{2 \left(+\sqrt{-8
   t_{s}^2 \cos (\alpha )+(E_{p1}-E_{p2})^2+8
   t_{s}^2}+(E_{p1}-E_{p2})\right)^2-16 t_{s}^2 (+\cos
   (\alpha )-1)}}\times \\
\begin{pmatrix}
2 e^{i \alpha } \left(-1+e^{i \alpha }\right)
   t_{s},\nonumber  \\
   -e^{i \alpha } \left(\sqrt{-8t_{s}^2 \cos
   (\alpha )+(E_{p1}-E_{p2})^2+8
   t_{s}^2}+(E_{p1}-E_{p2})\right),\nonumber  \\
   -2 \left(-1+e^{i \alpha }\right)
   t_{s}, \nonumber  \\
   \sqrt{-8 t_{s}^2 \cos (\alpha
   )+(E_{p1}-E_{p2})^2+8
   t_{s}^2}+E_{p1}-E_{p2}
   \end{pmatrix}
   \end{eqnarray}

   \begin{eqnarray}
 \ket{E_2}=\frac{1}{\sqrt{2 \left(-\sqrt{-8
   t_{s}^2 \cos (\alpha )+(E_{p1}-E_{p2})^2+8
   t_{s}^2}+(E_{p1}-E_{p2})\right)^2+16 t_{s}^2 (-\cos
   (\alpha )+1)}} \times \\
 \begin{pmatrix}
   2 e^{i \alpha } \left(-1+e^{i \alpha }\right)t_{s} %%%{\sqrt{2 \left(-\sqrt{-8 t_{s}^2 \cos (\alpha
  %%% )+(\text{Ep1}-\text{Ep2})^2+8
  %% \text{ts}^2}+\text{Ep1}-\text{Ep2}\right)^2+8 \text{ts}^2 (2-2 \cos
  %% (\alpha ))}},
   \nonumber \\
   -e^{i \alpha } \left(-\sqrt{-8 t_{s}^2 \cos
   (\alpha )+(E_{p1}-E_{p2})^2+8
   t_{s}^2}+E_{p1}-E_{p2}\right) %{\sqrt{2 \left(-\sqrt{-8
  % \text{ts}^2 \cos (\alpha )+(\text{Ep1}-\text{Ep2})^2+8
  % \text{ts}^2}+\text{Ep1}-\text{Ep2}\right)^2+8 \text{ts}^2 (2-2 \cos
  % (\alpha ))}}
   , \nonumber \\
   -2 \left(-1+e^{i \alpha }\right)
   t_{s}, \nonumber \\
   -\sqrt{-8t_{s}^2 \cos (\alpha
   )+(E_{p1}-E_{p2})^2+8
   t_{s}^2}+E_{p1}-E_{p2}
   \end{pmatrix}
   \end{eqnarray}

   \begin{eqnarray}
 \ket{E_3}=
 %  \begin{pmatrix}
%   -\frac{2 e^{i \alpha } \left(1+e^{i \alpha }\right)
%   \text{ts}}{\sqrt{2 \left(\sqrt{8 \text{ts}^2 \cos (\alpha
%   )+(\text{Ep1}-\text{Ep2})^2+8
%   \text{ts}^2}+\text{Ep1}-\text{Ep2}\right)^2+16 \text{ts}^2 (\cos
%   (\alpha )+1)}}, \nonumber \\
%   \frac{e^{i \alpha } \left(\sqrt{8 \text{ts}^2 \cos
%   (\alpha )+(\text{Ep1}-\text{Ep2})^2+8
%   \text{ts}^2}+\text{Ep1}-\text{Ep2}\right)}{\sqrt{2 \left(\sqrt{8
%   \text{ts}^2 \cos (\alpha )+(\text{Ep1}-\text{Ep2})^2+8
%   \text{ts}^2}+\text{Ep1}-\text{Ep2}\right)^2+16 \text{ts}^2 (\cos
%   (\alpha )+1)}},  \nonumber \\
%   -\frac{2 \left(1+e^{i \alpha }\right)
%   \text{ts}}{\sqrt{2 \left(\sqrt{8 \text{ts}^2 \cos (\alpha
%   )+(\text{Ep1}-\text{Ep2})^2+8
%   \text{ts}^2}+\text{Ep1}-\text{Ep2}\right)^2+16 \text{ts}^2 (\cos
%   (\alpha )+1)}},  \nonumber \\
%   \frac{\sqrt{8 \text{ts}^2 \cos (\alpha
%   )+(\text{Ep1}-\text{Ep2})^2+8
%   \text{ts}^2}+\text{Ep1}-\text{Ep2}}{\sqrt{2 \left(\sqrt{8 \text{ts}^2
%   \cos (\alpha )+(\text{Ep1}-\text{Ep2})^2+8
%   \text{ts}^2}+\text{Ep1}-\text{Ep2}\right)^2+16 \text{ts}^2 (\cos
%   (\alpha )+1)}}
%     \end{pmatrix}= \nonumber \\
     \frac{1}{\sqrt{2 \left(+\sqrt{8t_{s}^2 \cos (\alpha
   )+(E_{p1}-E_{p2})^2+8
   t_{s}^2}+(E_{p1}-E_{p2})\right)^2+16t_{s}^2 (+\cos
   (\alpha )+1)}} \times \\
   \begin{pmatrix}
   -2 e^{i \alpha } \left(1+e^{i \alpha }\right)
   t_{s}, \nonumber \\
   e^{i \alpha } \left(\sqrt{8 t_{s}^2 \cos
   (\alpha )+(E_{p1}-E_{p2})^2+8
   t_{s}^2}+E_{p1}-E_{p2}\right),  \nonumber \\
   -2 \left(1+e^{i \alpha }\right)
   t_{s},  \nonumber \\
   \sqrt{8 t_{s}^2 \cos (\alpha
   )+(E_{p1}-E_{p2})^2+8
   t_{s}^2}+E_{p1}-E_{p2}
     \end{pmatrix}
   \end{eqnarray}

   \begin{eqnarray}
    \ket{E_4}=\frac{1}{\sqrt{2 \left(-\sqrt{8 t_{s}^2 \cos (\alpha
   )+(E_{p1}-E_{p2})^2+8
   t_{s}^2}+(E_{p1}-E_{p2})\right)^2+16 t_{s}^2 (+\cos
   (\alpha )+1)}} \times  \\
     \begin{pmatrix}
     -2 e^{i \alpha } \left(1+e^{i \alpha }\right)
   t_{s}, \nonumber \\
   e^{i \alpha } \left(-\sqrt{8 t_{s}^2 \cos
   (\alpha )+(E_{p1}-E_{p2})^2+8
   t_{s}^2}+E_{p1}-E_{p2}\right),\nonumber \\
   -2 \left(1+e^{i \alpha }\right)
   t_{s},\nonumber \\
   -\sqrt{8 t_{s}^2 \cos (\alpha
   )+(E_{p1}-E_{p2})^2+8
   t_{s}^2}+E_{p1}-E_{p2}
    \end{pmatrix}= \nonumber \\
  =  \frac{1}{\sqrt{2 \left(-\sqrt{8  \cos (\alpha
   )+(\frac{E_{p1}-E_{p2}}{t_s})^2+8}+\frac{(E_{p1}-E_{p2})}{t_s}\right)^2+16 (+\cos
   (\alpha )+1)}} \times  \nonumber \\
     \begin{pmatrix}
     -2 e^{i \alpha } \left(1+e^{i \alpha }\right), \nonumber \\
   e^{i \alpha } \left(-\sqrt{8 \cos
   (\alpha )+(\frac{E_{p1}-E_{p2}}{t_s})^2+8}+\frac{E_{p1}-E_{p2}}{t_s}\right),\nonumber \\
   -2 \left(1+e^{i \alpha }\right),\nonumber \\
   -\sqrt{8  \cos (\alpha
   )+(\frac{E_{p1}-E_{p2}}{t_s})^2+8}+\frac{E_{p1}-E_{p2}}{t_s}
    \end{pmatrix}
   \end{eqnarray}
   It is remarkable to recognize that the occupancy at nodes 1, 2, 3 and 4 is function both of electric (expressed by factor $\frac{E_{p1}-E_{p2}}{t_s}$) and magnetic field (expressed by non-zero $\alpha$ coefficient ). Referring to the state $\ket{E_4}$ we have the same probabilities of occupancy nodes 1 and 3 set to
   \begin{equation}
   P_1=P_3=\frac{4((cos (\alpha)+1)^2+(\sin(\alpha))^2)}{\left(2 \left(-\sqrt{8  \cos (\alpha
   )+(\frac{E_{p1}-E_{p2}}{t_s})^2+8}+(\frac{E_{p1}-E_{p2}}{t_s})\right)^2+16 (+\cos
   (\alpha )+1)\right)}
   \end{equation}
   and probability of occupancy at nodes 2 and 4 nodes set to
   \begin{equation}
   P_2=P_4=\frac{ (-\sqrt{8  \cos (\alpha
   )+(\frac{E_{p1}-E_{p2}}{t_s})^2+8}+\frac{E_{p1}-E_{p2}}{t_s})^2}{\left(2 \left(-\sqrt{8  \cos (\alpha
   )+(\frac{E_{p1}-E_{p2}}{t_s})^2+8}+(\frac{E_{p1}-E_{p2}}{t_s})\right)^2+16 (+\cos
   (\alpha )+1)\right)}
   \end{equation}
   We use the property $(+\cos(\alpha )+1)=(+\cos(\frac{\alpha}{2})^2-\sin(\frac{\alpha}{2})^2+1)=+2\cos(\frac{\alpha}{2})^2 $ and the property
   $(-\cos(\alpha )+1)=(-\cos(\frac{\alpha}{2})^2+\sin(\frac{\alpha}{2})^2+1)=+2\sin(\frac{\alpha}{2})^2 $ and in such case of eigenstate $\ket{E_4}$ we obtain
   the ratio of probabilities
   \begin{eqnarray}
   \frac{P_1}{P_2}=\frac{P_3}{P_4}=\nonumber \\
   =\frac{4\sqrt{(cos (\alpha)+1)^2+(\sin(\alpha))^2}}{(-\sqrt{8  \cos (\alpha
   )+(\frac{E_{p1}-E_{p2}}{t_s})^2+8}+\frac{E_{p1}-E_{p2}}{t_s})^2}=\frac{4\sqrt{2(1+\cos(\alpha))}}{(-\sqrt{8(1+\cos (\alpha
   ))+(\frac{E_{p1}-E_{p2}}{t_s})^2}+\frac{E_{p1}-E_{p2}}{t_s})^2}= \nonumber \\
   =\frac{8|\cos(\frac{\alpha}{2})|}{(-\sqrt{16\cos (\frac{\alpha}{2}
   )^2+(\frac{E_{p1}-E_{p2}}{t_s})^2}+\frac{E_{p1}-E_{p2}}{t_s})^2}.
   \end{eqnarray}
In case of state $\ket{E_3}$ we have the ratio of probability occupancies given as
   \begin{eqnarray}
   \frac{P_1}{P_2}=\frac{P_3}{P_4}=\nonumber \\
   =\frac{4\sqrt{(cos (\alpha)+1)^2+(\sin(\alpha))^2}}{(+\sqrt{8  \cos (\alpha
   )+(\frac{E_{p1}-E_{p2}}{t_s})^2+8}+\frac{E_{p1}-E_{p2}}{t_s})^2}=\frac{4\sqrt{2(1+\cos(\alpha))}}{(\sqrt{8(1+\cos (\alpha
   ))+(\frac{E_{p1}-E_{p2}}{t_s})^2}+\frac{E_{p1}-E_{p2}}{t_s})^2}= \nonumber \\
   =\frac{8|\cos(\frac{\alpha}{2})|}{(+\sqrt{16\cos (\frac{\alpha}{2}
   )^2+(\frac{E_{p1}-E_{p2}}{t_s})^2}+\frac{E_{p1}-E_{p2}}{t_s})^2}.
   \end{eqnarray}
In case of state $\ket{E_2}$ we have the ratio of probability occupancies given as
      \begin{eqnarray}
   \frac{P_1}{P_2}=\frac{P_3}{P_4}=\nonumber \\
   =\frac{4\sqrt{(cos (\alpha)-1)^2+(\sin(\alpha))^2}}{(-\sqrt{-8\cos (\alpha
   )+(\frac{E_{p1}-E_{p2}}{t_s})^2+8}+\frac{E_{p1}-E_{p2}}{t_s})^2}=\frac{4\sqrt{2(1-\cos(\alpha))}}{(-\sqrt{8(1-\cos (\alpha
   ))+(\frac{E_{p1}-E_{p2}}{t_s})^2}+\frac{E_{p1}-E_{p2}}{t_s})^2}= \nonumber \\
   =\frac{8|\sin(\frac{\alpha}{2})|}{(-\sqrt{16\sin(\frac{\alpha}{2}
   )^2+(\frac{E_{p1}-E_{p2}}{t_s})^2}+\frac{E_{p1}-E_{p2}}{t_s})^2}.
   \end{eqnarray}
   In case of state $\ket{E_1}$ we have the ratio of probability occupancies given as
         \begin{eqnarray}
   \frac{P_1}{P_2}=\frac{P_3}{P_4}=\nonumber \\
   =\frac{4\sqrt{(cos (\alpha)-1)^2+(\sin(\alpha))^2}}{(+\sqrt{-8\cos (\alpha
   )+(\frac{E_{p1}-E_{p2}}{t_s})^2+8}+\frac{E_{p1}-E_{p2}}{t_s})^2}=\frac{4\sqrt{2(1-\cos(\alpha))}}{(+\sqrt{8(1-\cos (\alpha
   ))+(\frac{E_{p1}-E_{p2}}{t_s})^2}+\frac{E_{p1}-E_{p2}}{t_s})^2}= \nonumber \\
   =\frac{8|\sin(\frac{\alpha}{2})|}{(+\sqrt{16\sin(\frac{\alpha}{2}
   )^2+(\frac{E_{p1}-E_{p2}}{t_s})^2}+\frac{E_{p1}-E_{p2}}{t_s})^2}.
   \end{eqnarray}
   In general case when the quantum state is superposition of 4 energetic levels expressed by the equation
   \begin{equation}
   \ket{\psi(t)}=c_{E1}e^{i \phi_{E1}}e^{\frac{E_1 t}{i \hbar}}\ket{E_1}+c_{E2}e^{i \phi_{E2}}e^{\frac{E_2 t}{i \hbar}}\ket{E_2}+..+ c_{E4}e^{i \phi_{E4}}e^{\frac{E_4 t}{i \hbar}}\ket{E_4}.
   \end{equation}
  and probability of finding particle at node 1 is given by
   \begin{eqnarray}
 P_1(t)= |\bra{x_1} \ket{\psi(t)}|^2 =|c_{E1}e^{i \phi_{E1}}e^{\frac{E_1 t}{i \hbar}}\bra{x_1}\ket{E_1}+c_{E2}e^{i \phi_{E2}}e^{\frac{E_2 t}{i \hbar}}\bra{x_1}\ket{E_2}+..+ c_{E4}e^{i \phi_{E4}}e^{\frac{E_4 t}{i \hbar}}\bra{x_1}\ket{E_4}|^2 = \nonumber \\
 =\Bigg[\frac{c_{E1}e^{i \phi_{E1}}e^{\frac{E_1 t}{i \hbar}}2 e^{i \alpha } \left(-1+e^{i \alpha }\right)
   t_{s}}{\sqrt{2 \left(+\sqrt{-8
   t_{s}^2 \cos (\alpha )+(E_{p1}-E_{p2})^2+8
   t_{s}^2}+(E_{p1}-E_{p2})\right)^2-16 t_{s}^2 (+\cos
   (\alpha )-1)}}+ \nonumber \\
   +\frac{c_{E2}e^{i \phi_{E2}}e^{\frac{E_2 t}{i \hbar}}2t_s e^{i \alpha } \left(-1+e^{i \alpha }\right)}{\sqrt{2 \left(-\sqrt{-8
   t_{s}^2 \cos (\alpha )+(E_{p1}-E_{p2})^2+8
   t_{s}^2}+(E_{p1}-E_{p2})\right)^2+16 t_{s}^2 (-\cos
   (\alpha )+1)}} +  \nonumber \\
    +\frac{-c_{E3}e^{i \phi_{E3}}e^{\frac{E_3 t}{i \hbar}}2t_s e^{i \alpha } \left(1+e^{i \alpha }\right)}{\sqrt{2 \left(+\sqrt{8t_{s}^2 \cos (\alpha
   )+(E_{p1}-E_{p2})^2+8
   t_{s}^2}+(E_{p1}-E_{p2})\right)^2+16t_{s}^2 (+\cos
   (\alpha )+1)}}+ \nonumber \\
   +\frac{-c_{E4}e^{i \phi_{E4}}e^{\frac{E_4 t}{i \hbar}}2t_s e^{i \alpha } \left(1+e^{i \alpha }\right)}{\sqrt{2 \left(-\sqrt{8 t_{s}^2 \cos (\alpha
   )+(E_{p1}-E_{p2})^2+8
   t_{s}^2}+(E_{p1}-E_{p2})\right)^2+16 t_{s}^2 (+\cos
   (\alpha )+1)}}
   \Bigg] \nonumber \\
   \Bigg[\frac{c_{E1}e^{-i \phi_{E1}}e^{-\frac{E_1 t}{i \hbar}}2 e^{-i \alpha } \left(-1+e^{-i \alpha }\right)
   t_{s}}{\sqrt{2 \left(+\sqrt{-8
   t_{s}^2 \cos (\alpha )+(E_{p1}-E_{p2})^2+8
   t_{s}^2}+(E_{p1}-E_{p2})\right)^2-16 t_{s}^2 (+\cos
   (\alpha )-1)}}+ \nonumber \\
   +\frac{c_{E2}e^{-i \phi_{E2}}e^{-\frac{E_2 t}{i \hbar}}2t_s e^{-i \alpha } \left(-1+e^{-i \alpha }\right)}{\sqrt{2 \left(-\sqrt{-8
   t_{s}^2 \cos (\alpha )+(E_{p1}-E_{p2})^2+8
   t_{s}^2}+(E_{p1}-E_{p2})\right)^2+16 t_{s}^2 (-\cos
   (\alpha )+1)}} +  \nonumber \\
    +\frac{-c_{E3}e^{-i \phi_{E3}}e^{-\frac{E_3 t}{i \hbar}}2t_s e^{-i \alpha } \left(1+e^{-i \alpha }\right)}{\sqrt{2 \left(+\sqrt{8t_{s}^2 \cos (\alpha
   )+(E_{p1}-E_{p2})^2+8
   t_{s}^2}+(E_{p1}-E_{p2})\right)^2+16t_{s}^2 (+\cos
   (\alpha )+1)}}+ \nonumber \\
   +\frac{-c_{E4}e^{-i \phi_{E4}}e^{-\frac{E_4 t}{i \hbar}}2t_s e^{-i \alpha } \left(1+e^{-i \alpha }\right)}{\sqrt{2 \left(-\sqrt{8 t_{s}^2 \cos (\alpha
   )+(E_{p1}-E_{p2})^2+8
   t_{s}^2}+(E_{p1}-E_{p2})\right)^2+16 t_{s}^2 (+\cos
   (\alpha )+1)}}
   \Bigg]
  \end{eqnarray}
  and
   \begin{eqnarray}
 P_2(t)= |\bra{x_2} \ket{\psi(t)}|^2 =|c_{E1}e^{i \phi_{E1}}e^{\frac{E_1 t}{i \hbar}}\bra{x_2}\ket{E_1}+c_{E2}e^{i \phi_{E2}}e^{\frac{E_2 t}{i \hbar}}\bra{x_2}\ket{E_2}+..+ c_{E4}e^{i \phi_{E4}}e^{\frac{E_4 t}{i \hbar}}\bra{x_2}\ket{E_4}|^2 = \nonumber \\
 =\Bigg[\frac{c_{E1}e^{i \phi_{E1}}e^{\frac{E_1 t}{i \hbar}} (-1)e^{i \alpha }  \left( +\sqrt{-8 t_{s}^2 \cos (\alpha
   )+(E_{p1}-E_{p2})^2+8
   t_{s}^2}+E_{p1}-E_{p2}  \right)
   }{\sqrt{2 \left(+\sqrt{-8
   t_{s}^2 \cos (\alpha )+(E_{p1}-E_{p2})^2+8
   t_{s}^2}+(E_{p1}-E_{p2})\right)^2-16 t_{s}^2 (+\cos
   (\alpha )-1)}}+ \nonumber \\
   +\frac{c_{E2}e^{i \phi_{E2}}e^{\frac{E_2 t}{i \hbar}} (-1)e^{i \alpha } \left( -\sqrt{-8 t_{s}^2 \cos (\alpha
   )+(E_{p1}-E_{p2})^2+8
   t_{s}^2}+E_{p1}-E_{p2}  \right)    }{\sqrt{2 \left(-\sqrt{-8
   t_{s}^2 \cos (\alpha )+(E_{p1}-E_{p2})^2+8
   t_{s}^2}+(E_{p1}-E_{p2})\right)^2+16 t_{s}^2 (-\cos
   (\alpha )+1)}} +  \nonumber \\
    +\frac{c_{E3}e^{i \phi_{E3}}e^{\frac{E_3 t}{i \hbar}}e^{+i\alpha} \left( +\sqrt{-8 t_{s}^2 \cos (\alpha
   )+(E_{p1}-E_{p2})^2+8
   t_{s}^2}+E_{p1}-E_{p2}  \right)}{\sqrt{2 \left(+\sqrt{8t_{s}^2 \cos (\alpha
   )+(E_{p1}-E_{p2})^2+8
   t_{s}^2}+(E_{p1}-E_{p2})\right)^2+16t_{s}^2 (+\cos
   (\alpha )+1)}}+ \nonumber \\
   +\frac{c_{E4}e^{i \phi_{E4}}e^{\frac{E_4 t}{i \hbar}}e^{+i\alpha}   \left( -\sqrt{-8 t_{s}^2 \cos (\alpha
   )+(E_{p1}-E_{p2})^2+8
   t_{s}^2}+E_{p1}-E_{p2}   \right)}{\sqrt{2 \left(-\sqrt{8 t_{s}^2 \cos (\alpha
   )+(E_{p1}-E_{p2})^2+8
   t_{s}^2}+(E_{p1}-E_{p2})\right)^2+16 t_{s}^2 (+\cos
   (\alpha )+1)}}
   \Bigg] \nonumber \\
   \Bigg[\frac{c_{E1}e^{-i \phi_{E1}}e^{-\frac{E_1 t}{i \hbar}}(-1)e^{-i \alpha } \left( + \sqrt{-8 t_{s}^2 \cos (\alpha
   )+(E_{p1}-E_{p2})^2+8
   t_{s}^2}+E_{p1}-E_{p2} \right)}{\sqrt{2 \left(+\sqrt{-8
   t_{s}^2 \cos (\alpha )+(E_{p1}-E_{p2})^2+8
   t_{s}^2}+(E_{p1}-E_{p2})\right)^2-16 t_{s}^2 (+\cos
   (\alpha )-1)}}+ \nonumber \\
   +\frac{c_{E2}e^{-i \phi_{E2}}e^{-\frac{E_2 t}{i \hbar}} (-1)e^{-i \alpha } \left( -\sqrt{-8 t_{s}^2 \cos (\alpha
   )+(E_{p1}-E_{p2})^2+8
   t_{s}^2}+E_{p1}-E_{p2}  \right)}{\sqrt{2 \left(-\sqrt{-8
   t_{s}^2 \cos (\alpha )+(E_{p1}-E_{p2})^2+8
   t_{s}^2}+(E_{p1}-E_{p2})\right)^2+16 t_{s}^2 (-\cos
   (\alpha )+1)}} +  \nonumber \\
    +\frac{c_{E3}e^{-i \phi_{E3}}e^{-\frac{E_3 t}{i \hbar}}e^{-i\alpha} \left( +\sqrt{-8 t_{s}^2 \cos (\alpha
   )+(E_{p1}-E_{p2})^2+8
   t_{s}^2}+E_{p1}-E_{p2} \right)}{\sqrt{2 \left(+\sqrt{8t_{s}^2 \cos (\alpha
   )+(E_{p1}-E_{p2})^2+8
   t_{s}^2}+(E_{p1}-E_{p2})\right)^2+16t_{s}^2 (+\cos
   (\alpha )+1)}}+ \nonumber \\
   +\frac{c_{E4}e^{-i \phi_{E4}}e^{-\frac{E_4 t}{i \hbar}}e^{-i\alpha}  \left( -\sqrt{-8 t_{s}^2 \cos (\alpha
   )+(E_{p1}-E_{p2})^2+8
   t_{s}^2}+E_{p1}-E_{p2}  \right)}{\sqrt{2 \left(-\sqrt{8 t_{s}^2 \cos (\alpha
   )+(E_{p1}-E_{p2})^2+8
   t_{s}^2}+(E_{p1}-E_{p2})\right)^2+16 t_{s}^2 (+\cos
   (\alpha )+1)}}
   \Bigg]
  \end{eqnarray}
and
     \begin{eqnarray}
 P_3(t)= |\bra{x_3} \ket{\psi(t)}|^2 =|c_{E1}e^{i \phi_{E1}}e^{\frac{E_1 t}{i \hbar}}\bra{x_3}\ket{E_1}+c_{E2}e^{i \phi_{E2}}e^{\frac{E_2 t}{i \hbar}}\bra{x_3}\ket{E_2}+..+ c_{E4}e^{i \phi_{E4}}e^{\frac{E_4 t}{i \hbar}}\bra{x_3}\ket{E_4}|^2 = \nonumber \\
 =\Bigg[\frac{c_{E1}e^{i \phi_{E1}}e^{\frac{E_1 t}{i \hbar}}(-2)t_s \left(-1+e^{i \alpha }\right)
   t_{s}}{\sqrt{2 \left(+\sqrt{-8
   t_{s}^2 \cos (\alpha )+(E_{p1}-E_{p2})^2+8
   t_{s}^2}+(E_{p1}-E_{p2})\right)^2-16 t_{s}^2 (+\cos
   (\alpha )-1)}}+ \nonumber \\
   +\frac{c_{E2}e^{i \phi_{E2}}e^{\frac{E_2 t}{i \hbar}}  (-2)t_s \left(-1+e^{i \alpha }\right)    }{\sqrt{2 \left(-\sqrt{-8
   t_{s}^2 \cos (\alpha )+(E_{p1}-E_{p2})^2+8
   t_{s}^2}+(E_{p1}-E_{p2})\right)^2+16 t_{s}^2 (-\cos
   (\alpha )+1)}} +  \nonumber \\
    +\frac{c_{E3}e^{i \phi_{E3}}e^{\frac{E_3 t}{i \hbar}} (-2)t_s  \left(+1+e^{i \alpha }\right)}{\sqrt{2 \left(+\sqrt{8t_{s}^2 \cos (\alpha
   )+(E_{p1}-E_{p2})^2+8
   t_{s}^2}+(E_{p1}-E_{p2})\right)^2+16t_{s}^2 (+\cos
   (\alpha )+1)}}+ \nonumber \\
   +\frac{c_{E4}e^{i \phi_{E4}}e^{\frac{E_4 t}{i \hbar}}  (-2)t_s  \left(1+e^{i \alpha }\right)}{\sqrt{2 \left(-\sqrt{8 t_{s}^2 \cos (\alpha
   )+(E_{p1}-E_{p2})^2+8
   t_{s}^2}+(E_{p1}-E_{p2})\right)^2+16 t_{s}^2 (+\cos
   (\alpha )+1)}}
   \Bigg] \nonumber \\
   \Bigg[\frac{c_{E1}e^{-i \phi_{E1}}e^{-\frac{E_1 t}{i \hbar}}(-2) \left(-1+e^{-i \alpha }\right)
   t_{s}}{\sqrt{2 \left(+\sqrt{-8
   t_{s}^2 \cos (\alpha )+(E_{p1}-E_{p2})^2+8
   t_{s}^2}+(E_{p1}-E_{p2})\right)^2-16 t_{s}^2 (+\cos
   (\alpha )-1)}}+ \nonumber \\
   +\frac{c_{E2}e^{-i \phi_{E2}}e^{-\frac{E_2 t}{i \hbar}}(-2)t_s  \left(-1+e^{-i \alpha }\right)}{\sqrt{2 \left(-\sqrt{-8
   t_{s}^2 \cos (\alpha )+(E_{p1}-E_{p2})^2+8
   t_{s}^2}+(E_{p1}-E_{p2})\right)^2+16 t_{s}^2 (-\cos
   (\alpha )+1)}} +  \nonumber \\
    +\frac{c_{E3}e^{-i \phi_{E3}}e^{-\frac{E_3 t}{i \hbar}}(-2)t_s \left(1+e^{-i \alpha }\right)}{\sqrt{2 \left(+\sqrt{8t_{s}^2 \cos (\alpha
   )+(E_{p1}-E_{p2})^2+8
   t_{s}^2}+(E_{p1}-E_{p2})\right)^2+16t_{s}^2 (+\cos
   (\alpha )+1)}}+ \nonumber \\
   +\frac{c_{E4}e^{-i \phi_{E4}}e^{-\frac{E_4 t}{i \hbar}} (-2t_s)  \left(1+e^{-i \alpha }\right)}{\sqrt{2 \left(-\sqrt{8 t_{s}^2 \cos (\alpha
   )+(E_{p1}-E_{p2})^2+8
   t_{s}^2}+(E_{p1}-E_{p2})\right)^2+16 t_{s}^2 (+\cos
   (\alpha )+1)}}
   \Bigg]
  \end{eqnarray}
and

     \begin{eqnarray}
 P_4(t)= |\bra{x_4} \ket{\psi(t)}|^2 =|c_{E1}e^{i \phi_{E1}}e^{\frac{E_1 t}{i \hbar}}\bra{x_4}\ket{E_1}+c_{E2}e^{i \phi_{E2}}e^{\frac{E_2 t}{i \hbar}}\bra{x_4}\ket{E_2}+..+ c_{E4}e^{i \phi_{E4}}e^{\frac{E_4 t}{i \hbar}}\bra{x_4}\ket{E_4}|^2 = \nonumber \\
 =\Bigg[\frac{c_{E1}e^{i \phi_{E1}}e^{\frac{E_1 t}{i \hbar}} \left( +\sqrt{-8 t_{s}^2 \cos (\alpha
   )+(E_{p1}-E_{p2})^2+8
   t_{s}^2}+E_{p1}-E_{p2}  \right)
   }{\sqrt{2 \left(+\sqrt{-8
   t_{s}^2 \cos (\alpha )+(E_{p1}-E_{p2})^2+8
   t_{s}^2}+(E_{p1}-E_{p2})\right)^2-16 t_{s}^2 (+\cos
   (\alpha )-1)}}+ \nonumber \\
   +\frac{c_{E2}e^{i \phi_{E2}}e^{\frac{E_2 t}{i \hbar}}  \left( -\sqrt{-8 t_{s}^2 \cos (\alpha
   )+(E_{p1}-E_{p2})^2+8
   t_{s}^2}+E_{p1}-E_{p2}  \right)    }{\sqrt{2 \left(-\sqrt{-8
   t_{s}^2 \cos (\alpha )+(E_{p1}-E_{p2})^2+8
   t_{s}^2}+(E_{p1}-E_{p2})\right)^2+16 t_{s}^2 (-\cos
   (\alpha )+1)}} +  \nonumber \\
    +\frac{c_{E3}e^{i \phi_{E3}}e^{\frac{E_3 t}{i \hbar}} \left( +\sqrt{-8 t_{s}^2 \cos (\alpha
   )+(E_{p1}-E_{p2})^2+8
   t_{s}^2}+E_{p1}-E_{p2}  \right)}{\sqrt{2 \left(+\sqrt{8t_{s}^2 \cos (\alpha
   )+(E_{p1}-E_{p2})^2+8
   t_{s}^2}+(E_{p1}-E_{p2})\right)^2+16t_{s}^2 (+\cos
   (\alpha )+1)}}+ \nonumber \\
   +\frac{c_{E4}e^{i \phi_{E4}}e^{\frac{E_4 t}{i \hbar}}   \left( -\sqrt{-8 t_{s}^2 \cos (\alpha
   )+(E_{p1}-E_{p2})^2+8
   t_{s}^2}+E_{p1}-E_{p2}   \right)}{\sqrt{2 \left(-\sqrt{8 t_{s}^2 \cos (\alpha
   )+(E_{p1}-E_{p2})^2+8
   t_{s}^2}+(E_{p1}-E_{p2})\right)^2+16 t_{s}^2 (+\cos
   (\alpha )+1)}}
   \Bigg] \nonumber \\
   \Bigg[\frac{c_{E1}e^{-i \phi_{E1}}e^{-\frac{E_1 t}{i \hbar}} \left(  +\sqrt{-8 t_{s}^2 \cos (\alpha
   )+(E_{p1}-E_{p2})^2+8
   t_{s}^2}+E_{p1}-E_{p2} \right)}{\sqrt{2 \left(+\sqrt{-8
   t_{s}^2 \cos (\alpha )+(E_{p1}-E_{p2})^2+8
   t_{s}^2}+(E_{p1}-E_{p2})\right)^2-16 t_{s}^2 (+\cos
   (\alpha )-1)}}+ \nonumber \\
   +\frac{c_{E2}e^{-i \phi_{E2}}e^{-\frac{E_2 t}{i \hbar}} \left( -\sqrt{-8 t_{s}^2 \cos (\alpha
   )+(E_{p1}-E_{p2})^2+8
   t_{s}^2}+E_{p1}-E_{p2}  \right)}{\sqrt{2 \left(-\sqrt{-8
   t_{s}^2 \cos (\alpha )+(E_{p1}-E_{p2})^2+8
   t_{s}^2}+(E_{p1}-E_{p2})\right)^2+16 t_{s}^2 (-\cos
   (\alpha )+1)}} +  \nonumber \\
    +\frac{c_{E3}e^{-i \phi_{E3}}e^{-\frac{E_3 t}{i \hbar}} \left( +\sqrt{-8 t_{s}^2 \cos (\alpha
   )+(E_{p1}-E_{p2})^2+8
   t_{s}^2}+E_{p1}-E_{p2} \right)}{\sqrt{2 \left(+\sqrt{8t_{s}^2 \cos (\alpha
   )+(E_{p1}-E_{p2})^2+8
   t_{s}^2}+(E_{p1}-E_{p2})\right)^2+16t_{s}^2 (+\cos
   (\alpha )+1)}}+ \nonumber \\
   +\frac{c_{E4}e^{-i \phi_{E4}}e^{-\frac{E_4 t}{i \hbar}}  \left( -\sqrt{-8 t_{s}^2 \cos (\alpha
   )+(E_{p1}-E_{p2})^2+8
   t_{s}^2}+E_{p1}-E_{p2}  \right)}{\sqrt{2 \left(-\sqrt{8 t_{s}^2 \cos (\alpha
   )+(E_{p1}-E_{p2})^2+8
   t_{s}^2}+(E_{p1}-E_{p2})\right)^2+16 t_{s}^2 (+\cos
   (\alpha )+1)}}
   \Bigg]
  \end{eqnarray}

\normalsize
\begin{figure}
\centering
\includegraphics[scale=0.6]{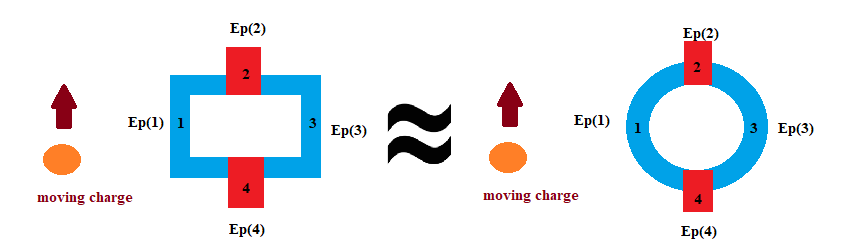}
\caption{Sensing of moving charge by the semiconductor transmon qubit (roton qubit). In first approximation only effect of externally moving charge at node 1 is accounted and hopping terms to the nearest neighbours are considered to be changed.. }
\end{figure}

\normalsize
\begin{figure}
\centering
\includegraphics[scale=0.6]{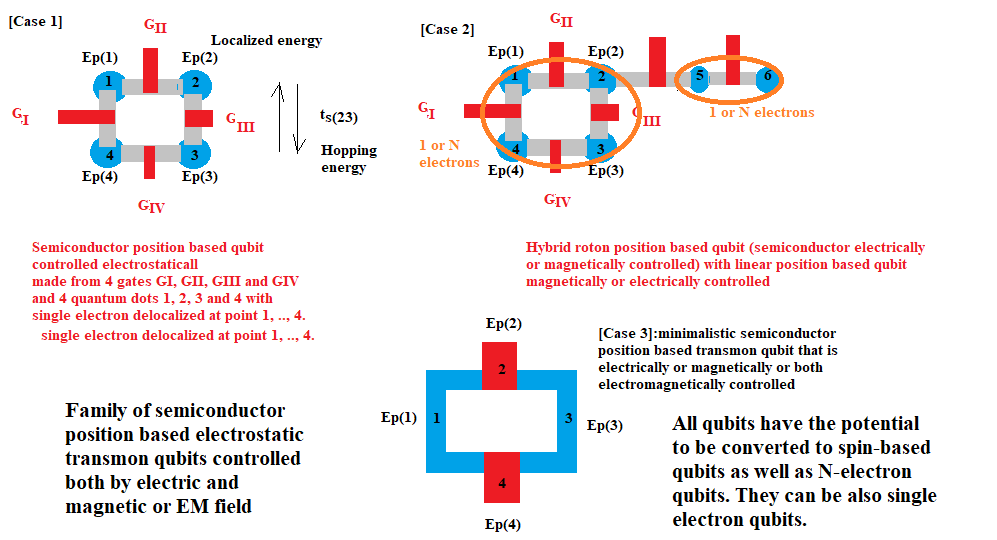}
\caption{Family of semiconductor position-based transmon qubits that can be single or many electron qubits as well as energy eigenvalue qubits or Wannier qubits or spin qubits. }
\end{figure}

\subsection{Action of phase rotating gate described analytically}

%%Let us consider the situation of single qubit from Fig.\ref{centralpicture} when we assume the following dependencies: $E_{p1}(t)=E_{p2}(t)=E_p=constant$ and $t_{s12}(t)=t_{s21}(t)=t_s(t)=constant_1$.
%In such
We inspect the derivation of formula for ration between $\psi_{2}(t)$ and $\psi_{1}(t)$ in case of situation $E_{p1}=E_{p2}=E_p=constant$ and with $t_s=constant \in R$.
The given assumptions that two system eigenergies are time independent. However we will assume that due to magnetic Aharonov-Bohm effect it is possible to have phase imprint on position based qubit between nodes 2 and 1. This phase imprint does not change the system energy but changes the system quantum state what has its importance in quantum information processing.  In the conducted considerations the probabilities of occupancy of eigenergies $E_1$ and $E_2$ are constant and determined by initial conditions and are $|\alpha_{E1}|^2$ and $|\alpha_{E2}|^2$. We have $|\alpha_{E2}(t)|=constant$, $|\alpha_{E2}(t)|=constant$,$\phi_{E2}(t)=constant$
and $\phi_{E1}(t)=constant$ .
The action of external source of vector potential is represented by phase imprints $\delta_1(t)$ and $\delta_2(t)$ at qubit nodes 1 and 2 and can be implemented by external conductor of electric current or charge moving in free space in the proximity to position-based qubit.
We obtain

\begin{figure}
\centering
\includegraphics[scale=1.0]{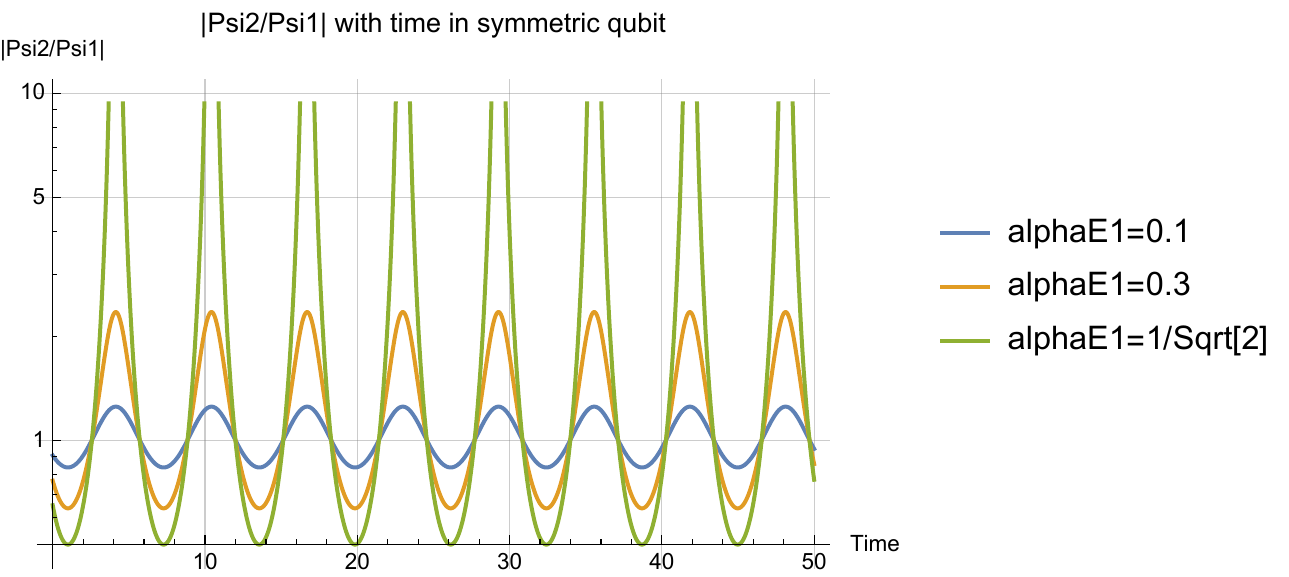} %{Phase_imprinting_gate.png}
\caption{Dependence of ratio $\frac{|\psi_2(t)|}{|\psi_1(t)|}$ in symmetric position dependent qubit with no external magnetic field.}
\label{Ratio}
\end{figure}

\begin{figure}
%\label{Psi2Psi1}
\centering
\includegraphics[scale=1.0]{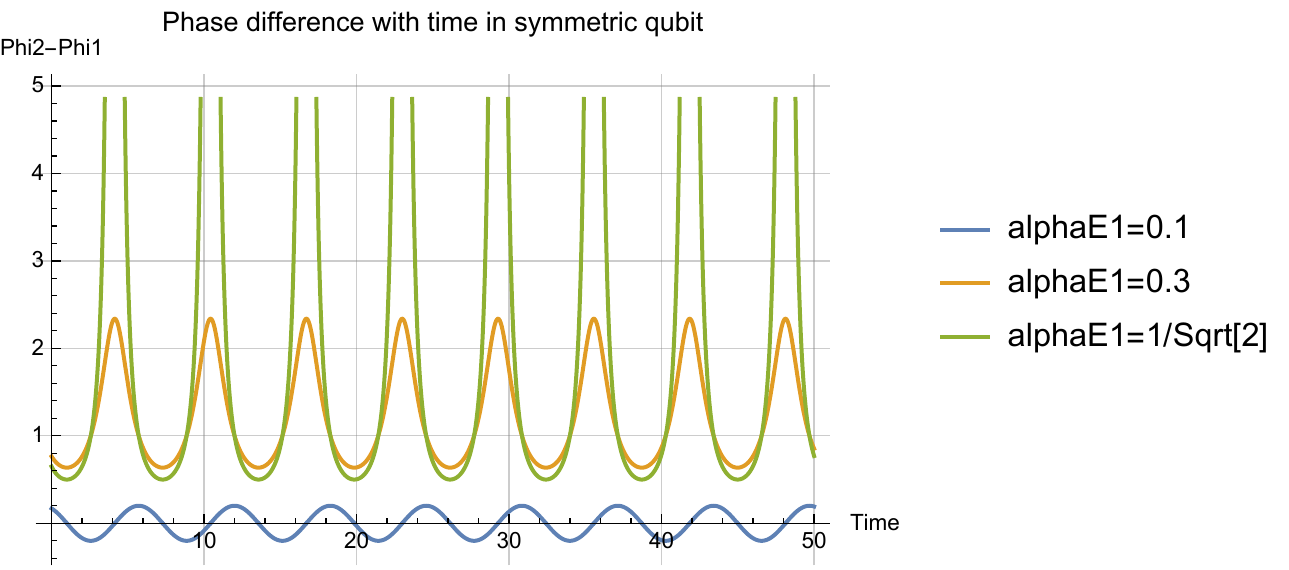} %{Phase_imprinting_gate.png}
\caption{Dependence of phase difference with time between nodes 2 and 1 in symmetric position dependent qubit with no external magnetic field.}
\label{Psi2Psi1}
\end{figure}

\begin{figure}
\centering
\includegraphics[scale=1.0]{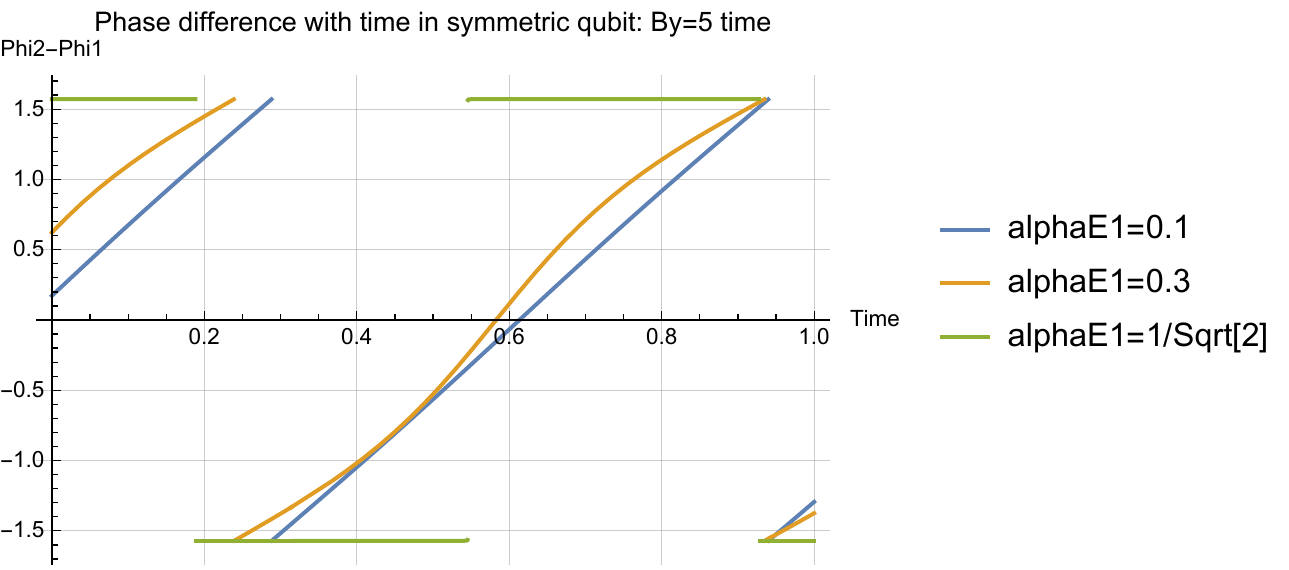} %{Phase_imprinting_gate.png}
\includegraphics[scale=1.0]{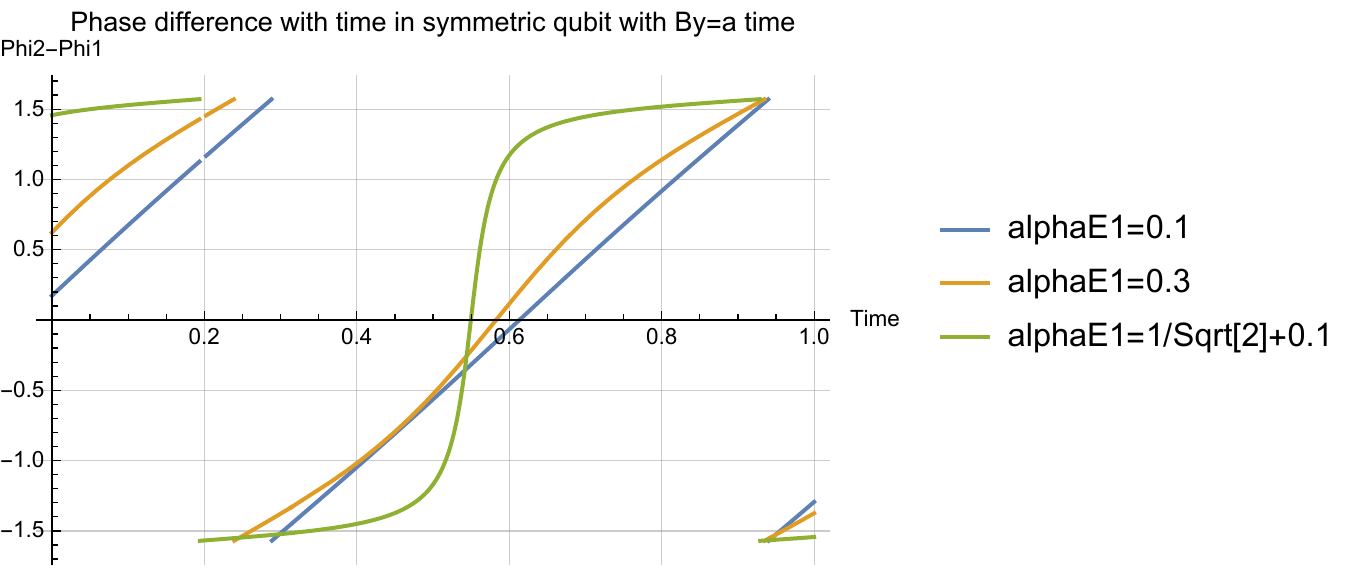}
\label{exotic}
%\caption{Dependence of phase difference with time between nodes 2 and 1 in symmetric position dependent qubit with time dependent external magnetic field.}
%\end{figure}
%\begin{figure}
\centering
\includegraphics[scale=1.0]{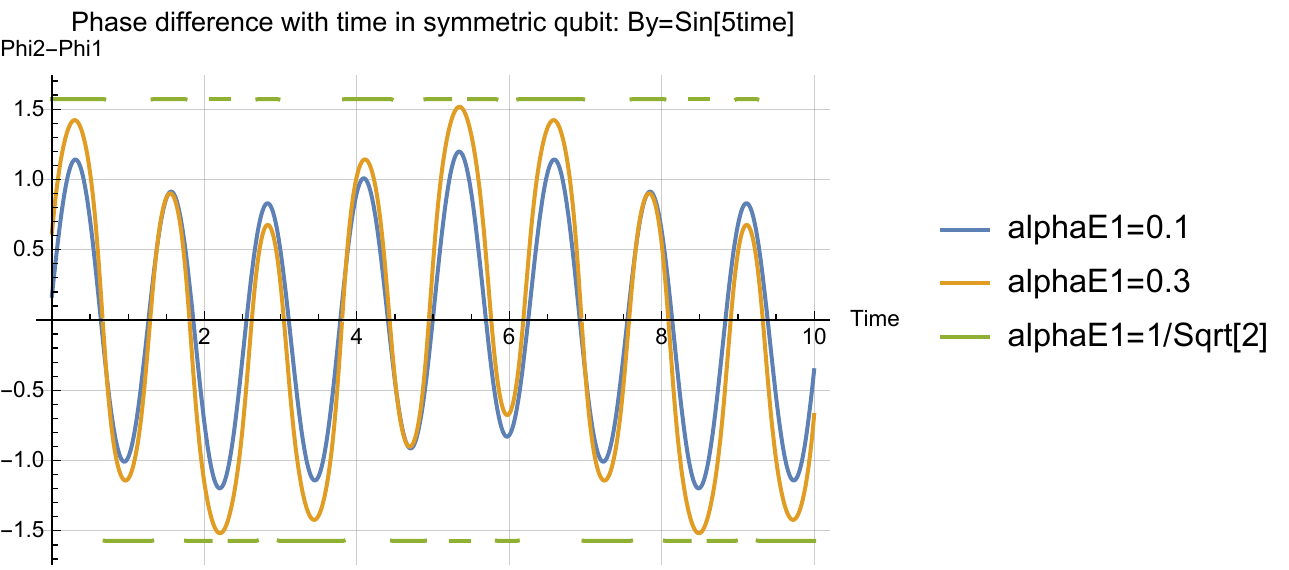} %{Phase_imprinting_gate.png}
\caption{Dependence of phase difference with time between nodes 2 and 1 in symmetric position dependent qubit with time dependent external magnetic field for the case of linear and sinusoidal dependence of external magnetic field affecting qubit on time.}
\label{exotic1}
\end{figure}

%1/2 (Sqrt[4 d^2 + (L - 2 x)^2] - Sqrt[4 d^2 + (L + 2 x)^2] +
 %  L Log[-L + Sqrt[4 d^2 + (L - 2 x)^2] + 2 x] +
  % L Log[L + 2 x + Sqrt[4 d^2 + (L + 2 x)^2]] +
   %2 x Log[(
    % L + 2 x + Sqrt[4 d^2 + (L + 2 x)^2])/(-L + Sqrt[
     % 4 d^2 + (L - 2 x)^2] + 2 x)])
\small
\begin{eqnarray}
\label{qq}
\frac{\psi_2(t)}{\psi_1(t)}=\frac{(-e^{\frac{-i}{\hbar}E_1t}e^{i \phi_{E1}(0)}|\alpha_{E1}|+e^{\frac{-i}{\hbar}E_2t}|\beta_{E2}|e^{i \phi_{E2}(0) })}{(+e^{\frac{-i}{\hbar}E_1t}e^{i \phi_{E1}(0)}|\alpha_{E1}|+e^{\frac{-i}{\hbar}E_2t}|\beta_{E2}|e^{i \phi_{E2}(0)})}
\frac{e^{i \delta_2(t)}}{e^{i \delta_1(t)}}=\frac{(-|\alpha_{E1}|+e^{\frac{-i}{\hbar}(E_2-E_1)t}|\beta_{E2}|e^{i (\phi_{E2}(0)-\phi_{E2}(0)) })}{(+|\alpha_{E1}|+e^{\frac{-i}{\hbar}(E_2-E_1))t}|\beta_{E2}|e^{i (\phi_{E2}(0)-\phi_{E1}(0)) })}e^{i(\delta_2(t)-\delta_1(t))}= \nonumber \\ =
\frac{(-|\alpha_{E1}|+e^{\frac{-i}{\hbar}(E_2-E_1)t}|\beta_{E2}|e^{i (\phi_{E2}(0)-\phi_{E2}(0)) })(+|\alpha_{E1}|+e^{\frac{+i}{\hbar}(E_2-E_1))t}|\beta_{E2}|e^{-i (\phi_{E2}(0)-\phi_{E1}(0)) })}{(+|\alpha_{E1}|+e^{\frac{-i}{\hbar}(E_2-E_1)t}|\beta_{E2}|e^{i (\phi_{E2}(0)-\phi_{E2}(0)) })(+|\alpha_{E1}|+e^{\frac{+i}{\hbar}(E_2-E_1))t}|\beta_{E2}|e^{-i (\phi_{E2}(0)-\phi_{E1}(0)) })} e^{i(\delta_2(t)-\delta_1(t))}= \nonumber \\
=\frac{(-|\alpha_{E1}|+e^{\frac{-i}{\hbar}(E_2-E_1)t}|\beta_{E2}|e^{i (\phi_{E2}(0)-\phi_{E2}(0)) })(+|\alpha_{E1}|+e^{\frac{+i}{\hbar}(E_2-E_1))t}|\beta_{E2}|e^{-i (\phi_{E2}(0)-\phi_{E1}(0)) })}{([+|\alpha_{E1}|+|\beta_{E2}|cos(\frac{(E_2-E_1)t}{\hbar}-(\phi_{E2}(0)-\phi_{E1}(0)))]^2+[|\beta_{E2}|sin(\frac{(E_2-E_1)t}{\hbar}-(\phi_{E2}(0)-\phi_{E1}(0)))]^2} e^{i(\delta_2(t)-\delta_1(t))}= \nonumber \\
=\frac{(-|\alpha_{E1}|+e^{\frac{-i}{\hbar}(E_2-E_1)t}|\beta_{E2}|e^{i (\phi_{E2}(0)-\phi_{E2}(0)) })(+|\alpha_{E1}|+e^{\frac{+i}{\hbar}(E_2-E_1))t}|\beta_{E2}|e^{-i (\phi_{E2}(0)-\phi_{E1}(0)) })}{1+2|\alpha_{E1}||\beta_{E2}|cos(\frac{(E_2-E_1)t}{\hbar}-(\phi_{E2}(0)-\phi_{E1}(0)))} e^{i(\delta_2(t)-\delta_1(t))}= \nonumber \\
=\frac{(|\beta_{E2}|^2-|\alpha_{E1}|^2+|\alpha_{E2}||\beta_{E2}|(e^{\frac{-i}{\hbar}(E_2-E_1)t}e^{i (\phi_{E2}(0)-\phi_{E2}(0)) }-e^{\frac{+i}{\hbar}(E_2-E_1)t}e^{-i (\phi_{E2}(0)-\phi_{E2}(0)) }))}{1+2|\alpha_{E1}||\beta_{E2}|cos(\frac{(E_2-E_1)t}{\hbar}-(\phi_{E2}(0)-\phi_{E1}(0)))} e^{i(\delta_2(t)-\delta_1(t))}= \nonumber  \\
=\frac{(|\beta_{E2}|^2-|\alpha_{E1}|^2)-2i|\alpha_{E2}||\beta_{E2}\sin[\hbar(E_2-E_1)t- (\phi_{E2}(0)-\phi_{E2}(0)) ]}{1+2|\alpha_{E1}||\beta_{E2}|cos(\frac{(E_2-E_1)t}{\hbar}-(\phi_{E2}(0)-\phi_{E1}(0)))} e^{i(\delta_2(t)-\delta_1(t))}= \nonumber \\
=\frac{(|\beta_{E2}|^2-|\alpha_{E1}|^2)\cos(\delta_{21}(t))+2\sin(\delta_{21}(t)|\alpha_{E2}||\beta_{E2}\sin[\hbar(E_2-E_1)t- (\phi_{E2}(0)-\phi_{E2}(0)) ]}{1+2|\alpha_{E1}||\beta_{E2}|cos(\frac{(E_2-E_1)t}{\hbar}-(\phi_{E2}(0)-\phi_{E1}(0)))} + \nonumber \\
+i\frac{(|\beta_{E2}|^2-|\alpha_{E1}|^2)\sin(\delta_{21}(t))-2\cos(\delta_{21}(t)|\alpha_{E2}||\beta_{E2}\sin[\hbar(E_2-E_1)t- (\phi_{E2}(0)-\phi_{E2}(0)) ]}{1+2|\alpha_{E1}||\beta_{E2}|cos(\frac{(E_2-E_1)t}{\hbar}-(\phi_{E2}(0)-\phi_{E1}(0)))}  = \nonumber \\
=\frac{1}{1+2|\alpha_{E1}||\beta_{E2}|cos(\frac{(E_2-E_1)t}{\hbar}-(\phi_{E2}(0)-\phi_{E1}(0)))}\times \nonumber \\ \times e^{+ i Arc_{Tan}\Bigg[\frac{(|\beta_{E2}|^2-|\alpha_{E1}|^2)\cos(\delta_{21}(t))+2\sin(\delta_{21}(t)|\alpha_{E2}||\beta_{E2}\sin[\hbar(E_2-E_1)t- (\phi_{E2}(0)-\phi_{E2}(0)) ]}{(|\beta_{E2}|^2-|\alpha_{E1}|^2)\sin(\delta_{21}(t))-2\cos(\delta_{21}(t)|\alpha_{E2}||\beta_{E2}\sin[\hbar(E_2-E_1)t- (\phi_{E2}(0)-\phi_{E2}(0)) ]}\Bigg]}= \nonumber \\
= \frac{|\psi_2(t)|}{|\psi_1(t)|}e^{i Arc_{Tan}\Bigg[\frac{(|\beta_{E2}|^2-|\alpha_{E1}|^2)\cos(\delta_{21}(t))+2\sin(\delta_{21}(t)|\alpha_{E2}||\beta_{E2}\sin[\hbar(E_2-E_1)t- (\phi_{E2}(0)-\phi_{E2}(0)) ]}{(|\beta_{E2}|^2-|\alpha_{E1}|^2)\sin(\delta_{21}(t))-2\cos(\delta_{21}(t)|\alpha_{E2}||\beta_{E2}\sin[\hbar(E_2-E_1)t- (\phi_{E2}(0)-\phi_{E2}(0)) ]}\Bigg]}=\nonumber \\ =\frac{1}{1+2|\alpha_{E1}||\beta_{E2}|cos(\frac{(E_2-E_1)t}{\hbar}-(\phi_{E2}(0)-\phi_{E1}(0)))}e^{i \phi(t)}
%{(+e^{\frac{-i}{\hbar}E_1t}|\alpha_{E1}|+e^{\frac{-i}{\hbar}E_2t}|\beta_{E2}|e^{i (\phi_{E2}(0)}-\phi_{E1}(0)}))}%e^{i (\delta_2(t)-\delta_1(t))}
%%\frac{e^{i \delta_2(t)}}{e^{i \delta_1(t)}}
%=
 %+\frac{i}{\hbar}E_2t})\alpha_{E1}\beta_{E2}}{(e^{-\frac{-i}{\hbar}E_1t})\alpha_{E1}+\frac{i}{\hbar}E_2t})\alpha_{E1}\beta_{E2}}=
\end{eqnarray}
\normalsize
and finally we obtain
\begin{equation}
\frac{|\psi_2(t)|}{|\psi_1(t)|}=\frac{1}{1+2|\alpha_{E1}||\beta_{E2}|cos(\frac{(E_2-E_1)t}{\hbar}-(\phi_{E2}(0)-\phi_{E1}(0)))},
\end{equation}
and
\begin{eqnarray}
\phi(t)=Arc_{Tan}\Bigg[\frac{(|\beta_{E2}|^2-|\alpha_{E1}|^2)\sin(\delta_{21}(t))-2\cos(\delta_{21}(t)|\alpha_{E1}||\beta_{E2}|\sin[\hbar(E_2-E_1)t- (\phi_{E2}(0)-\phi_{E1}(0)) ]}{(|\beta_{E2}|^2-|\alpha_{E1}|^2)\cos(\delta_{21}(t))+2\sin(\delta_{21}(t)|\alpha_{E1}||\beta_{E2}\sin[\hbar(E_2-E_1)t- (\phi_{E2}(0)-\phi_{E1}(0)) ]}\Bigg] = \nonumber \\
=Arc_{Tan}\Bigg[\frac{(|\beta_{E2}|^2-|\alpha_{E1}|^2)tan(\delta_{21}(t))-2|\alpha_{E1}||\beta_{E2}|\sin[\hbar(E_2-E_1)t- (\phi_{E2}(0)-\phi_{E1}(0)) ]}{(|\beta_{E2}|^2-|\alpha_{E1}|^2)+2tan(\delta_{21}(t)|\alpha_{E1}||\beta_{E2}\sin[\hbar(E_2-E_1)t- (\phi_{E2}(0)-\phi_{E1}(0)) ]}\Bigg]= \nonumber \\
=Arc_{Tan}\Bigg[\frac{tan(\delta_{21}(t))-2\frac{|\alpha_{E1}||\beta_{E2}|}{(|\beta_{E2}|^2-|\alpha_{E1}|^2)}\sin[\hbar(E_2-E_1)t- (\phi_{E2}(0)-\phi_{E1}(0)) ]}{1+2tan(\delta_{21}(t)\frac{|\alpha_{E1}||\beta_{E2}}{|\beta_{E2}|^2-|\alpha_{E1}|^2}\sin[\hbar(E_2-E_1)t- (\phi_{E2}(0)-\phi_{E1}(0)) ]}
\end{eqnarray}
Let us set $p=tan(\phi(t))$ and $s_1=2\frac{|\alpha_{E1}||\beta_{E2}|}{(|\beta_{E2}|^2-|\alpha_{E1}|^2)}$ and we obtain
\begin{equation}
tan(\delta_{21}(t))-s_1 sin(\phi_s(t)=p+s_1 sin(\phi_s(t)) tan(\delta_{21}(t))p
%\frac{1}{1+2|\alpha_{E1}||\beta_{E2}|cos(\frac{(E_2-E_1)t}{\hbar}-(\phi_{E2}(0)-\phi_{E1}(0)))},
\end{equation}
and therefore we have arrive to the expression
\begin{eqnarray}
tan(\delta_{21}(t))=\frac{p+s_1\sin(\phi_s(t))}{1-p s_1 \sin(\phi_s(t)) }=\frac{tan(\phi(t))+s_1\sin(\phi_s(t))}{1-tan(\phi(t)) s_1 \sin(\phi_s(t)) }=\frac{tan(\phi(t))+(2\frac{|\alpha_{E1}||\beta_{E2}|}{(|\beta_{E2}|^2-|\alpha_{E1}|^2)})\sin(\phi_s(t))}{1-tan(\phi(t))(2\frac{|\alpha_{E1}||\beta_{E2}|}{(|\beta_{E2}|^2-|\alpha_{E1}|^2)}) \sin(\phi_s(t)) }= \nonumber \\
\tan \Bigg[ \frac{e}{\hbar}\frac{\mu_0 I_p(t)}{4\pi} \Bigg[ (\sqrt{4 d^2 + ((a+b) - L)^2} - \sqrt{4 d^2 + ((a+b) + L)^2})  + \nonumber \\
+   \frac{(a+b)}{2} Log \Bigg[ \frac{ \left( (a+b) - \sqrt{4 d^2 + ((a+b) +L)^2} - L \right)  \left( -(a+b) - L + \sqrt{
        4 d^2 + ((a+b) + L)^2}  \right)  }{   \left( (a+b) + \sqrt{4 d^2 + (-(a+b) + L)^2} - L \right) \left( -(a+b) +
        L + \sqrt{ 4 d^2 + (-(a+b) + L)^2 )  } \right) } \Bigg] +
 \nonumber \\
+   \frac{L}{2}\times Log \Bigg[ \frac{  \left(  (  a+b) + \sqrt{4 d^2 + (+(a+b) + L)^2} - L \right) \left( +(a+b) - L + \sqrt{
         4 d^2 + (-(a+b) + L)^2}  \right) }{  \left( -(a+b) + \sqrt{4 d^2 + (-(a+b) + L)^2} + L\right) \left( -(a+b) -
          L - \sqrt{4 d^2 + (+(a+b) + L)^2} \right) }     \Bigg] \Bigg]  = \nonumber \\
=\frac{\tan(\phi(t))+2\frac{ \sqrt{|\alpha_{E1}(0)|^2-|\alpha_{E1}(0)|^4} }{\sqrt{(1-2\alpha_{E1}(0)^2)^2}}\sin(t \frac{E_2-E_1}{\hbar}- (\phi_{E2}(0)-\phi_{E1}(0)))}{1-\tan(\phi(t))2\frac{ \sqrt{|\alpha_{E1}(0)|^2-|\alpha_{E1}(0)|^4} }{\sqrt{(1-2\alpha_{E1}(0)^2)^2}}\sin(t \frac{E_2-E_1}{\hbar})- (\phi_{E2}(0)-\phi_{E1}(0))} .
%\frac{1}{1+2|\alpha_{E1}||\beta_{E2}|cos(\frac{(E_2-E_1)t}{\hbar}-(\phi_{E2}(0)-\phi_{E1}(0)))},
\end{eqnarray}
Finally we obtain the prescription for current function that generates any desired dynamics of $\phi(t)$ that is both continous or discontionus so we have
\begin{eqnarray}
%tan(\delta_{21}(t))=\frac{p+s_1\sin(\phi_s(t))}{1-p s_1 \sin(\phi_s(t)) }=\frac{tan(\phi(t))+s_1\sin(\phi_s(t))}{1-tan(\phi(t)) s_1 \sin(\phi_s(t)) }=\frac{tan(\phi(t))+(2\frac{|\alpha_{E1}||\beta_{E2}|}{(|\beta_{E2}|^2-|\alpha_{E1}|^2)})\sin(\phi_s(t))}{1-tan(\phi(t))(2\frac{|\alpha_{E1}||\beta_{E2}|}{(|\beta_{E2}|^2-|\alpha_{E1}|^2)}) \sin(\phi_s(t)) }= \nonumber \\
 I_p(t)=\frac{\hbar}{e}\frac{4\pi}{\mu_0}\Bigg[ (\sqrt{4 d^2 + ((a+b) - L)^2} - \sqrt{4 d^2 + ((a+b) + L)^2})  + \nonumber \\
+   \frac{(a+b)}{2} Log \Bigg[ \frac{ \left( (a+b) - \sqrt{4 d^2 + ((a+b) +L)^2} - L \right)  \left( -(a+b) - L + \sqrt{
        4 d^2 + ((a+b) + L)^2}  \right)  }{   \left( (a+b) + \sqrt{4 d^2 + (-(a+b) + L)^2} - L \right) \left( -(a+b) +
        L + \sqrt{ 4 d^2 + (-(a+b) + L)^2 )  } \right) } \Bigg] +
 \nonumber \\
+   \frac{L}{2}\times Log \Bigg[ \frac{  \left(  (  a+b) + \sqrt{4 d^2 + (+(a+b) + L)^2} - L \right) \left( +(a+b) - L + \sqrt{
         4 d^2 + (-(a+b) + L)^2}  \right) }{  \left( -(a+b) + \sqrt{4 d^2 + (-(a+b) + L)^2} + L\right) \left( -(a+b) -
          L - \sqrt{4 d^2 + (+(a+b) + L)^2} \right) }     \Bigg]^{-1}   \times \nonumber \\
\times Arc_{Tan}\Bigg[ \frac{\tan(\phi(t))+2\frac{ \sqrt{|\alpha_{E1}(0)|^2-|\alpha_{E1}(0)|^4} }{\sqrt{(1-2\alpha_{E1}(0)^2)^2}}\sin(t \frac{E_2-E_1}{\hbar}- (\phi_{E2}(0)-\phi_{E1}(0)))}{1-\tan(\phi(t))2\frac{ \sqrt{|\alpha_{E1}(0)|^2-|\alpha_{E1}(0)|^4} }{\sqrt{(1-2\alpha_{E1}(0)^2)^2}}\sin(t \frac{E_2-E_1}{\hbar})- (\phi_{E2}(0)-\phi_{E1}(0))}\Bigg] .
%\frac{1}{1+2|\alpha_{E1}||\beta_{E2}|cos(\frac{(E_2-E_1)t}{\hbar}-(\phi_{E2}(0)-\phi_{E1}(0)))},
\end{eqnarray}
On another hand knowing dynamis of external electric current imprinting phase drop on qubit we can can determine qubit effective phase dynamics
\tiny
\begin{eqnarray}
%tan(\delta_{21}(t))=\frac{p+s_1\sin(\phi_s(t))}{1-p s_1 \sin(\phi_s(t)) }=\frac{tan(\phi(t))+s_1\sin(\phi_s(t))}{1-tan(\phi(t)) s_1 \sin(\phi_s(t)) }=\frac{tan(\phi(t))+(2\frac{|\alpha_{E1}||\beta_{E2}|}{(|\beta_{E2}|^2-|\alpha_{E1}|^2)})\sin(\phi_s(t))}{1-tan(\phi(t))(2\frac{|\alpha_{E1}||\beta_{E2}|}{(|\beta_{E2}|^2-|\alpha_{E1}|^2)}) \sin(\phi_s(t)) }= \nonumber \\
%%%%(1-\tan(\phi(t))2\frac{ \sqrt{|\alpha_{E1}(0)|^2-|\alpha_{E1}(0)|^4} }{\sqrt{(1-2\alpha_{E1}(0)^2)^2}}\sin(t \frac{E_2-E_1}{\hbar})- (\phi_{E2}(0)-\phi_{E1}(0))) \times \nonumber \\
Tan \Bigg[-2\frac{ \sqrt{|\alpha_{E1}(0)|^2-|\alpha_{E1}(0)|^4} }{\sqrt{(1-2\alpha_{E1}(0)^2)^2}}\sin(t \frac{E_2-E_1}{\hbar}- (\phi_{E2}(0)-\phi_{E1}(0)))+\tan \Bigg[ \frac{e}{\hbar}\frac{\mu_0 I_p(t)}{4\pi} \Bigg[ (\sqrt{4 d^2 + ((a+b) - L)^2} - \sqrt{4 d^2 + ((a+b) + L)^2})  + \nonumber \\
+   \frac{(a+b)}{2} Log \Bigg[ \frac{ \left( (a+b) - \sqrt{4 d^2 + ((a+b) +L)^2} - L \right)  \left( -(a+b) - L + \sqrt{
        4 d^2 + ((a+b) + L)^2}  \right)  }{   \left( (a+b) + \sqrt{4 d^2 + (-(a+b) + L)^2} - L \right) \left( -(a+b) +
        L + \sqrt{ 4 d^2 + (-(a+b) + L)^2 )  } \right) } \Bigg] +
 \nonumber \\
+   \frac{L}{2}\times Log \Bigg[ \frac{  \left(  (  a+b) + \sqrt{4 d^2 + (+(a+b) + L)^2} - L \right) \left( +(a+b) - L + \sqrt{
         4 d^2 + (-(a+b) + L)^2}  \right) }{  \left( -(a+b) + \sqrt{4 d^2 + (-(a+b) + L)^2} + L\right) \left( -(a+b) -
          L - \sqrt{4 d^2 + (+(a+b) + L)^2} \right) }     \Bigg] \Bigg] \times \nonumber \\
\Bigg[ 1+ \Big[ 2\frac{ \sqrt{|\alpha_{E1}(0)|^2-|\alpha_{E1}(0)|^4} }{\sqrt{(1-2\alpha_{E1}(0)^2)^2}}\sin(t \frac{E_2-E_1}{\hbar}- (\phi_{E2}(0)-\phi_{E1}(0))) \Big] \times
\tan \Bigg[ \frac{e}{\hbar}\frac{\mu_0 I_p(t)}{4\pi}\Bigg[ (\sqrt{4 d^2 + ((a+b) - L)^2} - \sqrt{4 d^2 + ((a+b) + L)^2})  + \nonumber \\
+   \frac{(a+b)}{2} Log \Bigg[ \frac{ \left( (a+b) - \sqrt{4 d^2 + ((a+b) +L)^2} - L \right)  \left( -(a+b) - L + \sqrt{
        4 d^2 + ((a+b) + L)^2}  \right)  }{   \left( (a+b) + \sqrt{4 d^2 + (-(a+b) + L)^2} - L \right) \left( -(a+b) +
        L + \sqrt{ 4 d^2 + (-(a+b) + L)^2 )  } \right) } \Bigg] +
 \nonumber \\
+   \frac{L}{2}\times Log \Bigg[ \frac{  \left(  (  a+b) + \sqrt{4 d^2 + (+(a+b) + L)^2} - L \right) \left( +(a+b) - L + \sqrt{
         4 d^2 + (-(a+b) + L)^2}  \right) }{  \left( -(a+b) + \sqrt{4 d^2 + (-(a+b) + L)^2} + L\right) \left( -(a+b) -
          L - \sqrt{4 d^2 + (+(a+b) + L)^2} \right) }     \Bigg] \Bigg]
\Bigg]^{-1} \Bigg] %= \nonumber \\
=\phi(t)
%=\tan(\phi(t))+2\frac{ \sqrt{|\alpha_{E1}(0)|^2-|\alpha_{E1}(0)|^4} }{\sqrt{(1-2\alpha_{E1}(0)^2)^2}}\sin(t \frac{E_2-E_1}{\hbar}- (\phi_{E2}(0)-\phi_{E1}(0)))
%\frac{1}{1+2|\alpha_{E1}||\beta_{E2}|cos(\frac{(E_2-E_1)t}{\hbar}-(\phi_{E2}(0)-\phi_{E1}(0)))},
\end{eqnarray}
\normalsize
Evolution of dynamics of phase imprint on position based qubit can be both continous or discontinous as current flow can be continous as in case of circuits or discontinues [movement of charge in free space].
%We recognize that three frequencies are involved  $\omega_1=\frac{E_1}{\hbar}$, $\omega_{21m}=\frac{E_2-E_1}{\hbar}$,$\omega_{21p}=\frac{E_2+E_1}{\hbar}$  in the dynamics of phase difference of quantum state between nodes 2 and 1. We are usign sgin function as
%$sgn_{(\sin(\frac{(E_2-E_1) t}{\hbar}))}$ so it has 1 and -1 values for positive and negative values of $\sin\frac{(E_2-E_1) t}{\hbar}$ and 0 otherwise.
%What is more phase difference across position based qubit between nodes 1 and 2 is codepedent on the occupancy of the left and right node as given by last equation  in the case of time-independent Hamiltonian. Such situation is not taking place in most conventional qubits using energy eigenbases to encode information but takes place in position based semiconductor qubit. The ideal phase rotating gate implemented in position based qubit brings desired phase difference between wavefunctions at nodes 2 and 1 is not changing the occupancy of node 1 and 2. If we want to keep the occupancy from time t=0 we need to consider times $t_d=n\frac{2\pi \hbar}{E_2-E_1}$. At time t=0 phase difference was assumed to be 0.
Illlustration of phase difference evolution with time between 2 and 1 node in symmetric position depedent qubit with no external magnetic field is depicted in Fig.\ref{Ratio}.
Dependence of $\frac{|\psi_2(t)}{|\psi_1(t)|}$ o time in symmetric position depedent qubit with no external magnetic field is depicted in Fig.\ref{Psi2Psi1}. Evolution of phase difference in symmetric qubit under influence of time depedent magnetic field is given in Fig.\ref{exotic} and in Fig.\ref{exotic1}.
Linear effective phase imprint on time is reported as well as sinusoidal and step-like function. There is infnite class of phase imprinting functions that can be implemented.

%$\left\{
%%$-\frac{(\text{t1si}-i \text{t1sr}) (\text{t2si}-i \text{t2sr})}{2 \sqrt{\left(\text{t1si}^2+\text{t1sr}^2\right) \left(\text{t2si}^2+\text{t2sr}^2\right)}},$

%%%%$\frac{i
%%%%   \left(\sqrt{\left(\text{t1si}^2+\text{t1sr}^2\right) \left(\text{t2si}^2+\text{t2sr}^2\right)}+\text{t1si}^2+\text{t1sr}^2\right)}{2 (\text{t1si}+i \text{t1sr}) \sqrt{2
%%%%   \sqrt{\left(\text{t1si}^2+\text{t1sr}^2\right) \left(\text{t2si}^2+\text{t2sr}^2\right)}+\text{t1si}^2+\text{t1sr}^2+\text{t2si}^2+\text{t2sr}^2}},\frac{i
%%%%   \left(\sqrt{\left(\text{t1si}^2+\text{t1sr}^2\right) \left(\text{t2si}^2+\text{t2sr}^2\right)}+\text{t2si}^2+\text{t2sr}^2\right)}{2 (\text{t2si}+i \text{t2sr}) \sqrt{2
%%%%   \sqrt{\left(\text{t1si}^2+\text{t1sr}^2\right) \left(\text{t2si}^2+\text{t2sr}^2\right)}+\text{t1si}^2+\text{t1sr}^2+\text{t2si}^2+\text{t2sr}^2}},\frac{1}{2} $ %\right\}$
%$\frac {i   \left (\sqrt {\left (t1si^2 + t1sr^2 \right) \left (t2si^2 + t2sr^2 \right)} - t1si^2 - t1sr^2 \right)} {2 (t1si +
 %       i t1sr) \sqrt {-2\sqrt {\left (t1si^2 + t1sr^2 \right) \left (t2si^2 + t2sr^2 \right)} + t1si^2 + t1sr^2 + t2si^2 + t2sr^2}},$
%%$ -\frac {i\left (-\sqrt {\left (t1si^2 + t1sr^2 \right) \left (t2si^2 + t2sr^2 \right)} + t2si^2 + t2sr^2 \right)} {2 (t2si + i t2sr) \sqrt {-2\sqrt{\left (t1si^2 + t1sr^2 \right) \left (t2si^2 + t2sr^2 \right)} + t1si^2 + t1sr^2 + t2si^2 + t2sr^2}}, \frac {1}{2}$  %%\right\}$
\normalsize
\section{Conclusion}
%%Write your conclusion here.
I have shown the effect of transport of charged particle as proton in accelerator beam acting of single electron in position based qubit placed in the proximity of the accelerator. One can conclude that the beam of moving charged particles brings the change of occupancy of energetic levels in position electrostatic qubit and is inducing phase imprint across qubit.
In most general case one can expect that two level system represented by qubit will change its initial occupancy (as for example from 2) into N energy levels with phase imprint made on each eigenenergy level. However under assumption that the perturbing factor expressed by moving charge in accelerator beam is weak the conducted considerations are valid. Conducted considerations are also valid for the case of floating potential that is potential polarizing the qubit state. Therefore presented picture can be considered as phenomenological model of noise for electrostatic qubit that provides the description for qualitative and quantitative assessment of noise on two kinds of decoherence times commonly known as $T_1$ and $T_2$.
The presented results were presented at the seminar [5]. In such way one can account for very complicated electromagnetic environment in which position electrostatic semiconductor qubit is placed. In particular one can trace the decay of quantum information encoded in the qubit. One also expects that in the situation of 2 electrostatically qubits the passage of external charged particles is changing the quantum entanglement between qubits and anticorrelation function characterising two interacting qubits. Part of this work was presented in [3] and in [6, 14, 15].  The results can be extended quite straightforward to the more complicated structures by the mathematical framework given in [6], [7] and [8-15]. Particular attention shall be paid to the structures depicted in Fig.3.
\begin{figure}
\centering
\includegraphics[scale=2.2]{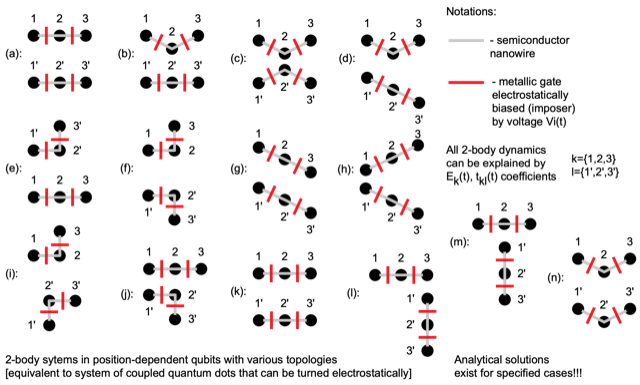}
\caption{Zoo of basic position dependent qubit topologies that could be used for beam diagnostics.}
\includegraphics[scale=0.5]{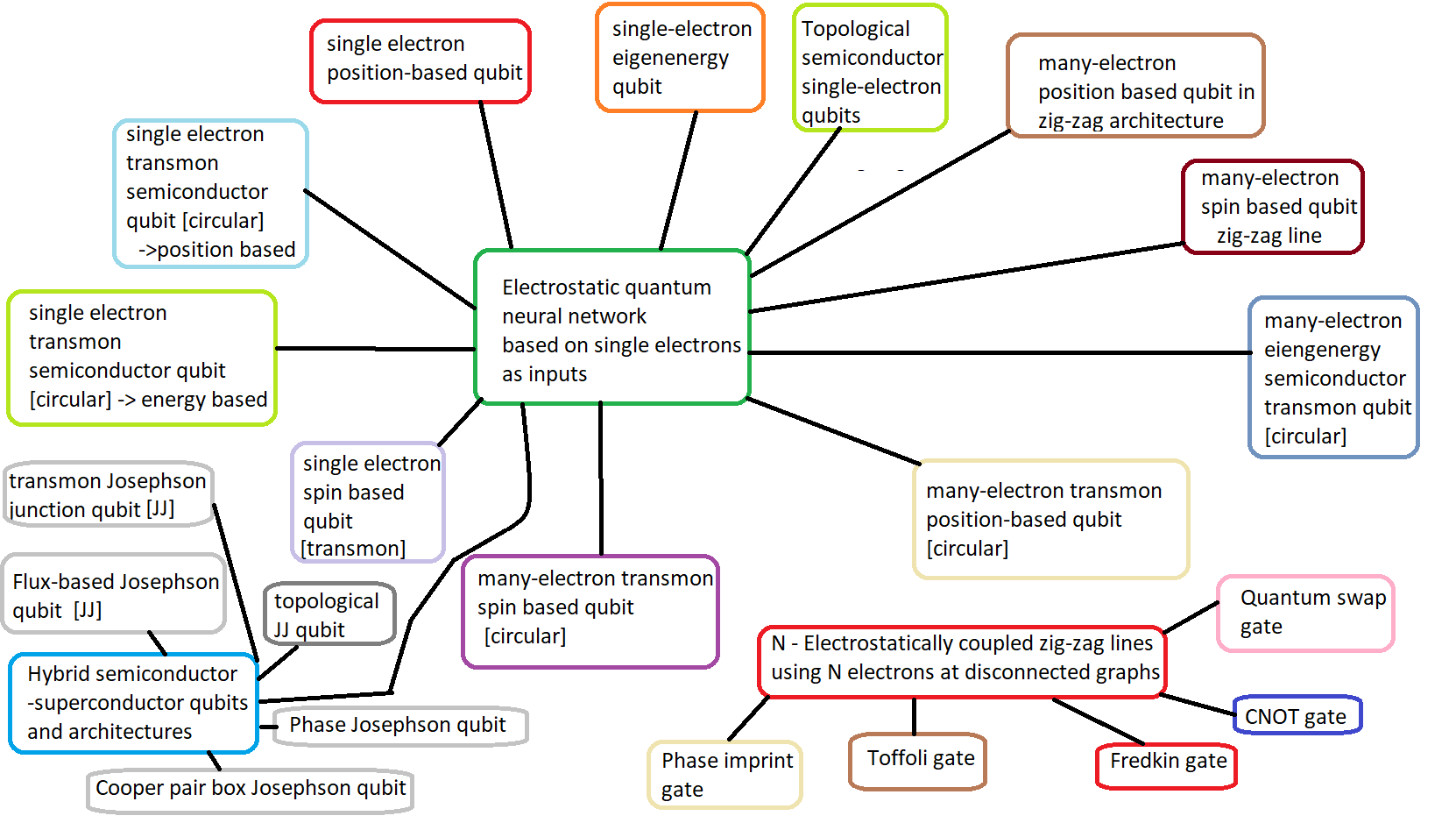}
\caption{Additional basic position dependent qubit topologies that could be used for beam diagnostics in quantum sensing and in quantum processing of obtained information. Central structure is matrix of semiconductor quantum dots implementing quantum neural network. }
\end{figure}
\newpage
\section{References}
[1]. T.Fujisawa, T.Hayashi, Y.Hirayama, "Electron counting of single-electron tunneling current",
Applied Physical Letters, Vol. 84, 2343, 2004, \href{https://doi.org/10.1063/1.1691491}{( https://doi.org/10.1063/1.1691491 )} \newline \newline
[2]. A.Bednorz, K.Franke, W.Belzig, "Noninvasiveness and time symmetry of weak measurements", \newline
New Journal of Physics, Volume 15, 2013, \newline \href{https://iopscience.iop.org/article/10.1088/1367-2630/15/2/023043}{ ( https://iopscience.iop.org/article/10.1088/1367-2630/15/2/023043 ) } \newline \newline
[3]. K.Pomorski, P.Giounanlis; E.Blokhina; D.Leipold; R.Staszewski, "Analytic view on coupled single electron lines", Semiconductor Science and Technology, Volume 34, Number 12, 2019, \newline \href{https://iopscience.iop.org/article/10.1088/1361-6641/ab4f40/meta}{ ( https://iopscience.iop.org/article/10.1088/1361-6641/ab4f40/meta )} \newline \newline
[4].K.Pomorski, P.Giounanlis, E.Blokhina, D.Leipold, P.Peczkowki, R.Staszewski, "From two types of electrostatic position-dependent semiconductor qubits to quantum universal gates and hybrid semiconductor-superconducting quantum computer", Spie, Proceedings Volume 11054, Superconductivity and Particle Accelerators 2018; 110540M, published on 2019, \href{https://www.spiedigitallibrary.org/conference-proceedings-of-spie/11054/110540M/From-two-types-of-electrostatic-position-dependent-semiconductor-qubits-to/10.1117/12.2525217.short}{(https://www.spiedigitallibrary.org/conference-proceedings-of-spie/11054/110540M/From-two-types-of-electrostatic-position-dependent-semiconductor-qubits-to/10.1117/12.2525217.short)}. \newline \newline
[5]. K.Pomorski, "Detection of moving charge by position dependent qubits", CERN 2020 -Dublin UCD Webinar 23-January-2020, \href{https://indico.cern.ch/event/876476/contributions/3693568/}{(https://indico.cern.ch/event/876476/contributions/3693568/)}, \newline
\href{https://vidyoreplay.cern.ch/replay/showRecordingExternal.html?key=6lP8xJZuDcPToTD}{(https://vidyoreplay.cern.ch/replay/showRecordingExternal.html?key=6lP8xJZuDcPToTD)}. \newline \newline
[6]. K.Pomorski, R.Staszewski, "Analytical Solutions for N-Electron Interacting System Confined in Graph of Coupled Electrostatic Semiconductor and Superconducting Quantum Dots in Tight-Binding Model with Focus on Quantum Information Processing", ArxiV:1907.02094, 2019, \newline \url{https://arxiv.org/abs/1907.03180}{(https://arxiv.org/abs/1907.03180)}. \newline \newline
[7]. K.Pomorski, R.B.Staszewski, “Towards quantum internet and non-local communication in position based qubits”, 2019, Arxiv: 1911.02094 \href{https://arxiv.org/abs/1911.02094}{(https://arxiv.org/abs/1911.02094)} \newline \newline
[8]. I.Bashir, M.Asker, C.Cetintepe, D.Leipold, A.Esmailiyan, H.Wang, T.Siriburanon, P.Giounanlis, E.Blokhina,. K.Pomorski, R.B.Staszewski, "Mixed-Signal Control Core for a Fully Integrated Semiconductor Quantum Computer System-on-Chip", Sep 2019 ESSCIRC 2019 - IEEE 45th European Solid State Circuits Conference (ESSCIRC), \href{https://ieeexplore.ieee.org/abstract/document/8902885/}{( https://ieeexplore.ieee.org/abstract/document/8902885/ )}. \newline \newline
[9]. P.Giounanlis, E.Blokhina, K.Pomorski, D.Leipold, R.Staszewski, "Modeling of Semiconductor Electrostatic Qubits Realized Through Coupled Quantum Dots", IEEE Open Access, 2019, 10.1109/ACCESS.2019.2909489, \href{https://ieeexplore.ieee.org/stamp/stamp.jsp?arnumber=8681511}{(https://ieeexplore.ieee.org/stamp/stamp.jsp?arnumber=8681511)}, \newline \newline
[10]. K.Pomorski, "Quantum information processing, quantum communication and quantum Artificial Intelligence in semiconductor quantum computer" , FQIS 2020-International Workshop for Young Researchers on the Future of Quantum Science and Technology, 3-6 February 2020, \url{https://fqst2020.com/}{(https://fqst2020.com/)} \newline \newline \newline
[11]. K.Pomorski, R.Staszewski et al. , "Programmable quantum matter in semiconductor", Seminar in Dublin Institute of Advanced Studies, 24-09-2019,
\href{https://www.dias.ie/2019/09/24/tuesday-1st-october-programmable-quantum-matter-in-semiconductor-electronics/}{( https://www.dias.ie/2019/09/24/tuesday-1st-october-programmable-quantum-matter-in-semiconductor-electronics/  )} \newline \newline
[12]. K.Pomorski, P.Giounanlist, E.Blokhina, R.B.Staszewski, "ISIF 2019 conference on Integrated Functionalities", Dublin, 12 August 2019.
\newline \newline
[13]. K.Pomorski,  P.Giounanlist, E.Blokhina, R.B.Staszewski,"Modeling quantum universal gates in semiconductor CMOS", 17 Jan 2019, Scalable Hardware Platforms for Quantum Computing Proc. of Scalable Hardware Platforms for Quantum Computing, Physikzentrum, Bad Honnef, Germany.
\newline \newline
[14]. K.Pomorski, “Analytic view on N body interaction in electrostatic quantum gates and decoherence effects in tight-binding model”, ArXiv: 1912.01205, 2019  \href{https://arxiv.org/abs/1912.01205}{(https://arxiv.org/abs/1912.01205)}, \newline \newline
[15]. K.Pomorski, "Analytical view on tunnable electrostatic quantum swap gate in tight-binding model", ArXiv: 2001.02513, 2020
 \href{https://arxiv.org/pdf/2001.02513.pdf }{(https://arxiv.org/pdf/2001.02513.pdf)}.
%%[13].
\end{document}